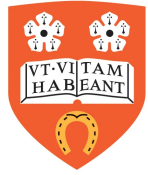

# The metal content of hot DA white dwarf spectra

# Nathan James Dickinson

**Supervisor:**
Martin Barstow

A thesis submitted for the degree of
Doctor of Philosophy
at the University of Leicester

March 2012

# Declaration

I hereby declare that no part of this thesis has been previously submitted to this or any other University as part of the requirement for a higher degree. The work described herein was conducted by the undersigned, except for contributions from colleagues as acknowledged in the text.

Nathan James Dickinson
March 2012

# The metal content of hot DA white dwarf spectra

## Nathan James Dickinson

## ABSTRACT

In this thesis, a study of the high ionisation-stage metal absorption features in the spectra of hot DA white dwarfs is presented. Metals are present in the photospheres of such stars due to radiative levitation (Chayer et al. 1994, 1995; Chayer Fontaine & Wesemael 1995). However, studies of the patterns between metal abundance and $T_{eff}$ show that, though the broad patterns predicted are seen, individual abundance measurements often do not reflect the predictions of radiative levitation theory (e.g. Barstow et al. 2003b). In this thesis, an analysis of the nitrogen abundance in three stars is performed, where a highly abundant layer of nitrogen was thought to reside at the top of the photospheres of the stars. The nitrogen abundance and distribution in these DAs is found to be homogeneous and of an abundance in keeping with stars of higher $T_{eff}$.

The accretion of metals from circumstellar discs has been shown to be the source of photospheric metals in DAs with $T_{eff}$ < 25,000 K (e.g. Zuckerman et al. 2003), where gravitational diffusion dominates (Koester & Wilken, 2006). In some cases, gaseous components are seen at such discs (e.g. SDSS 122859.93+104032.9; Gänsicke et al. 2006). A survey is made of a sample of hot (19,000 K < $T_{eff}$ < 51,000 K) DAs, where similar accretion may explain the inability of radiative levitation theory alone to account for the detected photospheric metal abundances. No circumstellar gas discs are found, though accretion from as yet undetected circumstellar sources remains an attractive explanation of the photospheric abundances of the stars.

Circumstellar absorption is seen in the UV spectra of some hot DA stars (Holberg et al. 1998; Bannister et al. 2003). Sources suggested for this material include circumstellar discs, the ionisation of the ISM, stellar mass loss and planetary nebulae. A re-analysis of this absorption is presented, using a technique that for the first time allows proper modelling of the circumstellar absorption features, and provides column densities for all components. The ionisation of circumstellar discs or planetesimals, the ionisation of the ISM and the ionisation of mass lost by binary companions are put forward as the origin for this circumstellar material.

# Publications

A significant amount of work contained in this thesis has been published in the following papers:

"*On the Origin of Metals in Some Hot White Dwarf Photospheres*,"
Burleigh M.R., Barstow M.A., Farihi J., Bannister N.P., Dickinson N.J., Steele P.R., Dobbie P.D., Faedi F., Gänsicke B.T., 2010, in '*17th European White Dwarf Workshop*,' eds. Werner K., Rauch T., AIP Conference Proceedings, Vol. 1273, p. 473

"*On the Origin of Metals in Some Hot White Dwarf Photospheres*,"
Burleigh M.R., Barstow M.A., Farihi J., Bannister N.P., Dickinson N.J., Steele P.R., Dobbie P.D., Faedi F., Gänsicke B.T., 2011 in '*Planetary Systems Beyond the Main Sequence*', eds. Schuh S., Dreschel H., Heber U., AIP Conference Proceedings, Vol. 1331, p. 289

"*On the Origin of Metals in Some Hot White Dwarf Photospheres*,"
Burleigh M.R., Barstow M.A., Farihi J., Bannister N.P., Dickinson N.J., Steele P.R., Dobbie P.D., Faedi F., Gänsicke B.T., 2012, MNRAS, *in preparation*

"*The Stratification of Metals in Hot DA White Dwarfs Atmospheres*,"
Dickinson N.J., Barstow M.A., Hubeny I., 2010, in '*17th European White Dwarf Workshop*,' eds. Werner K., Rauch T., AIP Conference Proceedings, Vol. 1273, p. 400

"*The distribution of metals in hot DA white dwarfs*,"
Dickinson N.J., Barstow M.A., Hubeny I., 2012, MNRAS, 421, 3222

"*The origin of circumstellar features in the spectra of hot DA white dwarfs*,"
Dickinson N.J., Barstow M.A., Welsh B.Y., Burleigh M., Farihi J., Redfield S., Unglaub K., MNRAS, 2012, *in press*

# Acknowledgements

The work contained in this thesis could not have been possible without the help of many other people. Obviously, I have a lot to thank Martin for, for helping me become a (good?) scientist, opening my eyes to the fascinating area that is white dwarf astronomy, and for fully supporting me along the way. The other staff in the white dwarf group at Leicester also shoulder some blame; thanks to Matt for a great observing experience in La Palma, and for always knowing where to drink, no matter what city or what country we happen to be in ("this beer tastes like bacon!"); Sarah has been a great source of help, and has never made me feel stupid, even when I've asked stupid questions; thanks to Jay for both showing me around Hawaii and for all the help he's given me in the later study of this thesis. Thanks also to Paul, Dave, Katherine and Simon. All of these people, plus those who I have met on my travels, have shown me that a career in science is truly worthwhile.

I'd also like to thank Ivan Hubeny for his help in getting my white dwarf modelling skills to a decent standard. Barry Welsh has contributed a lot to my understanding of the ISM and has opened my eyes to wealth of possibilities that exist in reality TV, should my astronomy career not work out. Similarly, Seth Redfield and Klaus Unglaub have been important in shaping my understanding of the LISM and white dwarf mass loss. The quality of the work here would be far from what it is now without the guidance and help of all. I'd also like to thank the other PhD students in the X-ray and Observational Astronomy and Theoretical Astrophysics Groups, for being good sports during teatime rants.

I would never have achieved what I have done, had I not had the support of my parents, Andrew and Joy, from day one; they made me realise there is no limit to the reward of hard work, and that you can do anything you want to. Thank you. Credit should also be given to my brother, Luke; no matter how sincerely I try to explain what I do, I'll always get a laugh ("so what do you *actually* do all day?") Last, but by no means least, I'd like to thank Sophia, for enduring the non-existent weekends and evenings, and the tantrums when things don't work. Together, anything is possible.

*For Sophia*

# Contents.

# List of figures.

# List of tables.

# *Chapter 1.*

# Introduction.

## 1.1. White dwarfs - an overview.

After stars like the Sun have finished their nuclear burning and have expelled their outer layers, the degenerate, cooling stellar core remains. This remnant core is a white dwarf. White dwarfs typically have radii similar to that of the Earth, and have masses around 0.6 solar masses ($M_\odot$). This small radius gives these objects extraordinarily high densities, making them astronomical 'compact objects,' a title shared only with neutron stars and stellar mass black holes. Being the evolutionary end point of stars with initial masses $\leq 8\ M_\odot$ (e.g. Weidemann 1987; Casewell et al. 2009), approximately 98% of stars will end their lives as white dwarfs (e.g. Wood 1992).

White dwarfs have a variety of applications in astronomical studies. As the end states of the evolution of the majority of stars, white dwarfs are of massive importance to our understanding of the life cycle of most of the stars in the Universe. Signatures of processes that occurred in earlier phases of the star's life are exhibited by the white dwarf, and thus these objects allow us to understand not just the end state of stellar evolution they represent, but also previous stages of the star's life. The pollution of white dwarfs by circumstellar planetary debris discs (discussed in more detail in Section 1.7.3) can be used to probe the end points of planetary system evolution, and indeed the ultimate fate of the Sun and Solar System.

Although the subject of this thesis is the metal features in white dwarf spectra, broadly speaking white dwarfs are thought of as having 'pristine' atmospheres. Due to their exceedingly high surface gravity (log $g$ ~7–9), the early view was that all elements but the lightest atmospheric constituent (i.e. hydrogen and/or helium) diffuse downward to leave a smooth continuum spectrum with no spectral features other than H or He (e.g. Schatzmann 1958). By examining interstellar medium (ISM) absorption features superimposed over the white dwarf continuum, these objects are often used to probe the structure and ionisation state of the ISM (e.g. Barstow et al. 2010; Redfield & Linsky, 2002, 2004a,b, 2008). Given their simple spectral form, white dwarfs are also often used as calibration targets for astronomical observations (e.g. Bohlin et al. 2011; Bohlin, Dickinson & Calzetti, 2001).

Being among the oldest stellar objects, white dwarfs can be used to determine the age of stellar populations, and as such can be used to provide an age estimate for the Galactic disc (e.g. Fontaine, Brassard & Bergeron, 2001). Given the use of supernovae as 'standard candles' in extragalactic astronomy, better understanding of white dwarfs as possible Type Ia supernova progenitors (e.g. Maxted et al. 2000) will allow a better understanding of the cosmological distance ladder and dark energy (it must be stressed that although white dwarfs are thought to be viable Type Ia supernova progenitors, a detailed understanding of the precise mechanism(s) is still a work in progress). White dwarfs are interesting not just to the astronomer; the interior of a white dwarf contains matter of a phenomenal density, making these objects unique physical laboratories.

## 1.2. The discovery of white dwarfs.

William Herschel was the first person to observe a white dwarf, discovering the star 40 Eridani (Eri) B (WD 0413–074) in 1783. However, he did not immediately notice anything untoward about the star. Almost fifty years later, in 1834, Friedrich Bessel noted the oscillating proper motion of Sirius (the dog star), hinting that a binary companion was present; a similar observation was made of the star Procyon (Bessel, 1844). Using subsequent observations, Peters first calculated the orbital elements of the Sirius binary in 1851 (Peters, 1851). Some 11 years later, Clarke first imaged Sirius B (WD 0642–166; Figure 1.1) using the 18.5 inch refracting telescope he was testing for the Dearborn Observatory, at the time the largest telescope in the USA. Bond officially reported this discovery later that year, after having observed the star himself at the Observatory of Harvard College (Bond, 1862).

The exotic and enigmatic nature of white dwarf stars began to emerge shortly after the discovery of Sirius B. The Russian astronomer Otto Struve determined that Sirius B has a mass of around half that of Sirius A, and reasoned that given they are at the same distance, Sirius B ought to be a $1^{st}$ magnitude star with a radius 80% of that of Sirius A, if both stars are made of the same material (Struve, 1866). Since Sirius B was found to be an $8^{th}$ magnitude star with a diminutive radius, Struve postulated that Sirius A and B are of "a very different physical constitution."

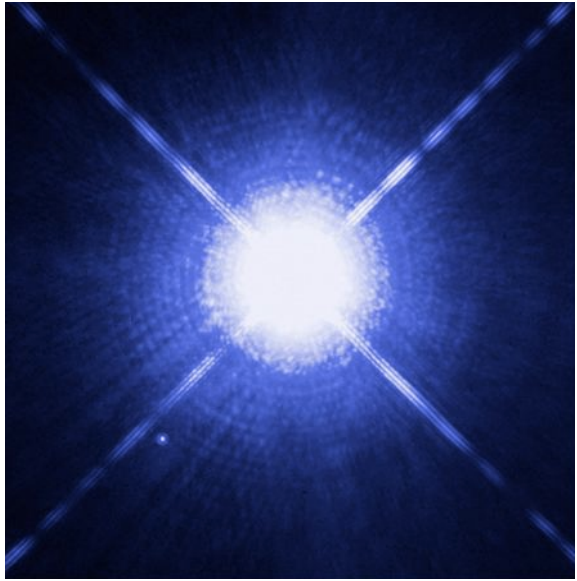

**Figure 1.1.** An image of the Sirius binary system, taken using the Hubble Space Telescope's Wide Field Planetary Camera 2. The white dwarf, Sirius B, is the smaller of the two stars, to the lower left of the larger star, Sirius A. Image credit: NASA, ESA, H. Bond (STScI) and M. Barstow (University of Leicester).

Much later, in 1915, Adams obtained a spectrum of Sirius B at the Mount Wilson Observatory. The star was found to have an effective temperature ($T_{eff}$) around 29,000 K, almost three times greater than Sirius A, though it was almost 1,000 times less luminous (Adams, 1915). Spectra of 40 Eri B and Procyon B (WD 0736+053) revealed a similar situation in those systems. Physical theories at the time required that hotter stars had a far greater luminosity than cooler stars. The only explanation for the observations was that the anomalous stars had radii approaching that of the Earth, giving them densities $10^5$–$10^6$ times greater than that of the Sun. No contemporary physical theory could explain such an object, since according to the astrophysics of the time such stars should undergo a gravitational collapse.

The first isolated white dwarf was discovered a few years later, by van Maanen (van Maanen's Star, WD 0046+051), in 1917 (van Maanen, 1917). Though originally thought to be an F0 star (van Maanen, 1917), the star is actually of the DZ class, and was therefore the first metal polluted white dwarf to be discovered. The source of these metals is now understood to be due to the accretion of disrupted planetary debris by the white dwarf (see Section 1.7.3 for a detailed discussion of this phenomenon). The term white dwarf was not used until 1922, when it was coined by Luyten to describe the white colour and small radii of these stars.

## 1.3. The structure of white dwarfs.

The matter making up these stars remained a mystery until the advent of quantum mechanics, and Fermi and Dirac's statistical theory of an electron gas in 1926. According to this theory, the behaviour of individual electrons inside an atom is governed by quantum mechanics, and the electrons will only occupy discrete energy levels. In materials with closely spaced atoms, the most loosely bound electrons can move freely, and are considered to form a gas; this electron gas is responsible for the conduction of heat and electricity in metals. However, the electron energy levels remain quantised, and each electron will occupy the lowest energy level available to it up to a maximum energy limit (the 'Fermi energy'). Any states with the same energy are 'degenerate,' and a gas such as this is said to be degenerate since all available electron states are occupied. The 'Pauli Exclusion Principle' prevents any two electrons sharing the exact same energy state.

In the extreme pressure environment of a white dwarf, the ions are compressed so closely that their quantised electron energy structures are broken down; an electron gas is again present. According to the Heisenberg Uncertainty Principle, $\Delta x \Delta p \geq h/4\pi$ (where $\Delta x$ is the position uncertainty and $\Delta p$ is the momentum uncertainty of a given electron, and $h$ is Planck's constant), so that as this electron gas becomes compressed $\Delta x$ decreases, raising $\Delta p$. This in turn raises the pressure of the gas. Since, according to the Pauli Exclusion principle, only two electrons (with opposite spin values) can occupy a given position-momentum phase cell, such compression is resisted by this 'degeneracy pressure.' In 1926, Fowler showed that it is this outward degeneracy pressure that provides a resistance to the gravitational collapse of a white dwarf.

A mere five years later, the noted Indian physicist Subrahmanyan Chandrasekhar (nephew of the Nobel Prize winning C.V. Raman, after whom Raman scattering is named) combined quantum mechanics with the theory of relativity and derived the equations describing the structure of white dwarfs. He predicted both the white dwarf mass-radius relation ($R \propto M^{-1/3}$, where $R$ and $M$ are the white dwarf radius and mass, respectively) and that at some mass limit (1.4 $M_\odot$, the Chandrasekhar mass), the outward degeneracy pressure in a non-rotating white dwarf would be insufficient to support the weight of the star, and that the star would undergo a violent gravitational collapse. It is a point of historical interest that although many of Chandrasekhar's contemporaries (including Bohr, Pauli and Fowler) agreed with his theory, they did not initially publicly support him as the eminent British astrophysicist Sir Arthur Eddington harshly rejected Chandrasekhar's work, stating that he thought "there should be a law of Nature to prevent a star from behaving in this absurd way" (Meeting of the Royal Astronomical Society, 11$^{th}$

January 1935, as reported in the Observatory, 1935). Chandrasekhar won the Nobel Prize in 1983 for this work, died in 1995, and the Chandra X-ray observatory (launched July 23$^{rd}$, 1999) is named in his honour.

A white dwarf is commonly composed of a carbon and oxygen ions (leftover from the helium burning processes in the progenitor star) and degenerate electron plasma core. Typically, 99.99% of the mass of the object is contained within this core. In cooler white dwarfs, with 4,000 K < $T_{eff}$ < 10,000 K, the cores of these stars crystallise. Above this core lies a thin non-degenerate atmosphere, composed mainly of hydrogen and/or helium.

## 1.4. White dwarf classification.

As far back as the Father Angelo Secchi's attempts at stellar classification in the 1860s, stars have been categorised using their spectral characteristics, culminating in the Harvard classification system used today. White dwarfs are no exception. Attempts to classify white dwarfs were begun by Kuiper (1941). Luyten (1952) found that white dwarfs occupied a continuous, lower luminosity strip parallel to the main sequence on the Hertzsprung-Russell diagram (Figure 1.2), leading to the white dwarf classification system devised by Greenstein (1960); the prefix "D" (for degenerate), followed by the Harvard spectral class for the star. Using this system, the hydrogen dominated objects occupied the DA class, while the mainly helium rich stars occupied the DB, DC, DF, DG, DK and DM classes. However, as white dwarf studies evolved and many hybrid objects were discovered, the classification system became unworkable; the system also gave no indication of a white dwarf's $T_{eff}$, which can vary from a few thousand to ~150,000K.

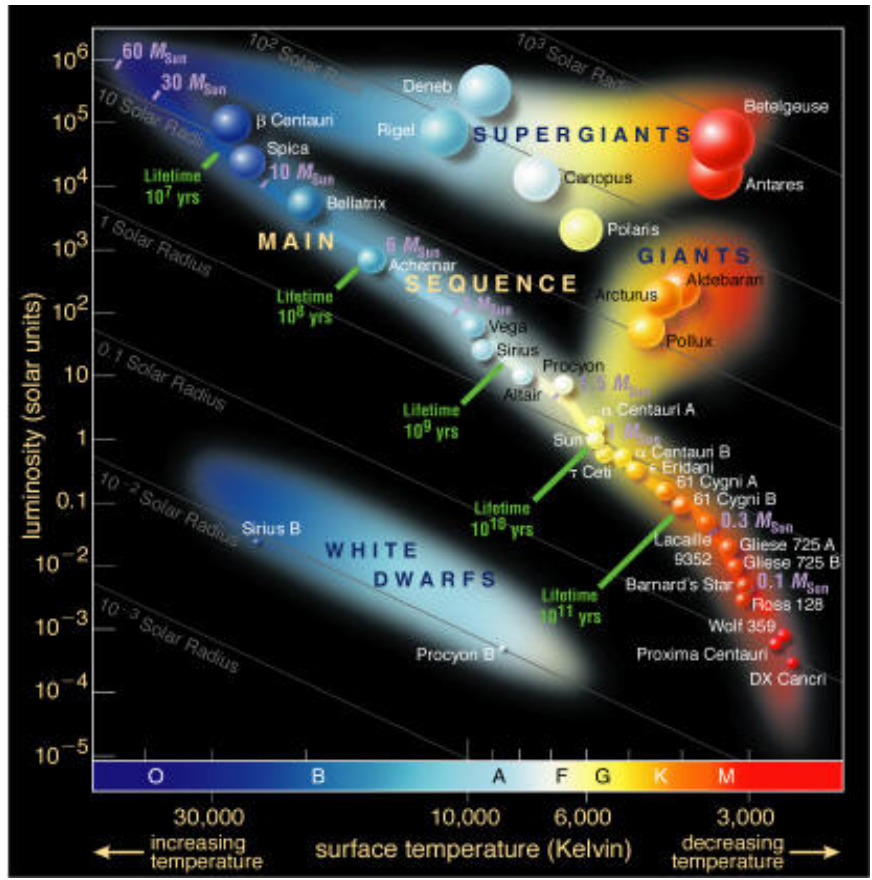

**Figure 1.2.** The Hertzsprung-Russell diagram, including the positions of Sirius B and Procyon B (image from http://www.daviddarling.info/encyclopedia/H/HRdiag.html).

This led Sion et al. (1983) to introduce the classification system for white dwarfs that is still in use today. The new system retains the 'D' to signify the degenerate nature of white dwarfs, followed by a symbol signifying the main atmospheric constituent (more detail is given in Table 1.1). Though often dropped when quoting a white dwarf's class, a temperature index is also used, equal to 50,400 K divided by the $T_{eff}$ of the star. Hybrid classes are described using a mix of symbols, in order of dominance, so a hydrogen dominated white dwarf with secondary metal absorption features, a debris disc and a $T_{\mathrm{eff}}$ of 16,800 K is a DAZd3.0 star.

**Table 1.1.** The white dwarf classification system.

| Class | $T_{\mathrm{eff}}$ Range (K) | Spectral Characteristics |
|---|---|---|
| H-rich | | |
| DA | 6,000–100,000 | Balmer lines only, no He or metal features |
| DAO | >45,000 | Balmer lines and weak He II features |
| He-rich | | |
| DO | 45,000–100,000 | Strong He II lines, some He I present |
| DB | 12,000–30,000 | HeI lines, no H or metals[*] |
| DBA | 12,000–30,000 | He I lines and weak Balmer lines present |
| Cool WDs | | |
| DQ | 6,000–12,000; 18,000–24,000[**] | C features (atomic or molecular) |
| DZ | <6,000[†]; 10,000[‡] | Metal lines only, no H or He |
| DC | <6,000[†]; 10,000[‡] | Featureless continuum (no lines deeper than 5%) |
| Additional | | Secondary Feature |
| P | | Magnetic with polarisation |
| H | | Magnetic with no detectable polarisation |
| E | | Emission lines present |
| V | | Variable |
| d | | Debris Disc |

[*]note that some DB stars with 30,000 K < $T_{\mathrm{eff}}$ < 45,000 K have been found in the 'DB gap' (Kleinman et al. 2004); [**]these stars correspond to the 'hot DQ' stars (Liebert et al. 2003; Dufour et al. 2008); [†]for a hydrogen atmosphere; [‡]for a helium atmosphere.

## 1.5. White dwarf formation.

All stars with masses ≤ 8 $M_{\odot}$ (e.g. Weidemann & Koester 1983; Weidemann 1987; Casewell et al. 2009), will end their lives as white dwarfs. Whilst on the stellar main sequence (see Figure 1.2), stars produce their energy from the fusion of hydrogen in their cores. The pressure from the radiation produced in this process counteracts the downward gravitational force acting on the star, holding the star up at a roughly constant radius; the star is in hydrostatic equilibrium. The amount of time a star spends in this phase depends strongly on its mass; stars like the Sun will spend of the order of ten billion years on the main sequence while stars ten times the mass of the Sun will spend only 30 million years in this state. Eventually, all stars will run out of useable hydrogen, leaving a helium core. The stellar core then contracts (Schönberg & Chandrasekhar 1942). This contraction increases the temperature of the core and the material surrounding it, since the temperature of a virialised mass is proportional to the reciprocal of its radius.

The cores of white dwarfs come in three distinct varieties, depending upon the initial mass of the progenitor star and its evolution. Stars with an initial mass < 0.5 $M_\odot$ never become hot enough to fuse helium into metallic elements, and are thus expected to evolve into stars with helium cores. The time for this process takes longer than a Hubble time (Laughlin et al. 1997) for single star evolution, leading to the assertion that the observed helium core white dwarf population must be the result of binary mass transfer (e.g. Liebert et al. 2004), where much of the mass of the progenitor star was stripped by the binary companion before the helium core can ignite. This leaves a naked He stellar core with a thin atmosphere; a He core white dwarf.

Stars with masses < 2 $M_\odot$ will burn hydrogen in a shell around the helium core, delivering ~100 times the luminosity of the previous burning stage (the evolution of a solar mass star is illustrated in Figure 1.3). This causes the star to expand to between 100 and 1000 times its previous radius, becoming cooler and less dense, forming a red giant. The atmospheres of these stars are highly convective, transporting hydrogen up into the burning hydrogen shell. This burning process increases the amount of helium in the stellar core. The core contraction halts when a temperature sufficient to ignite helium burning (producing carbon and oxygen) is reached, halting any further core contraction. The core then expands, allowing the atmosphere to shrink, moving the star onto the horizontal branch.

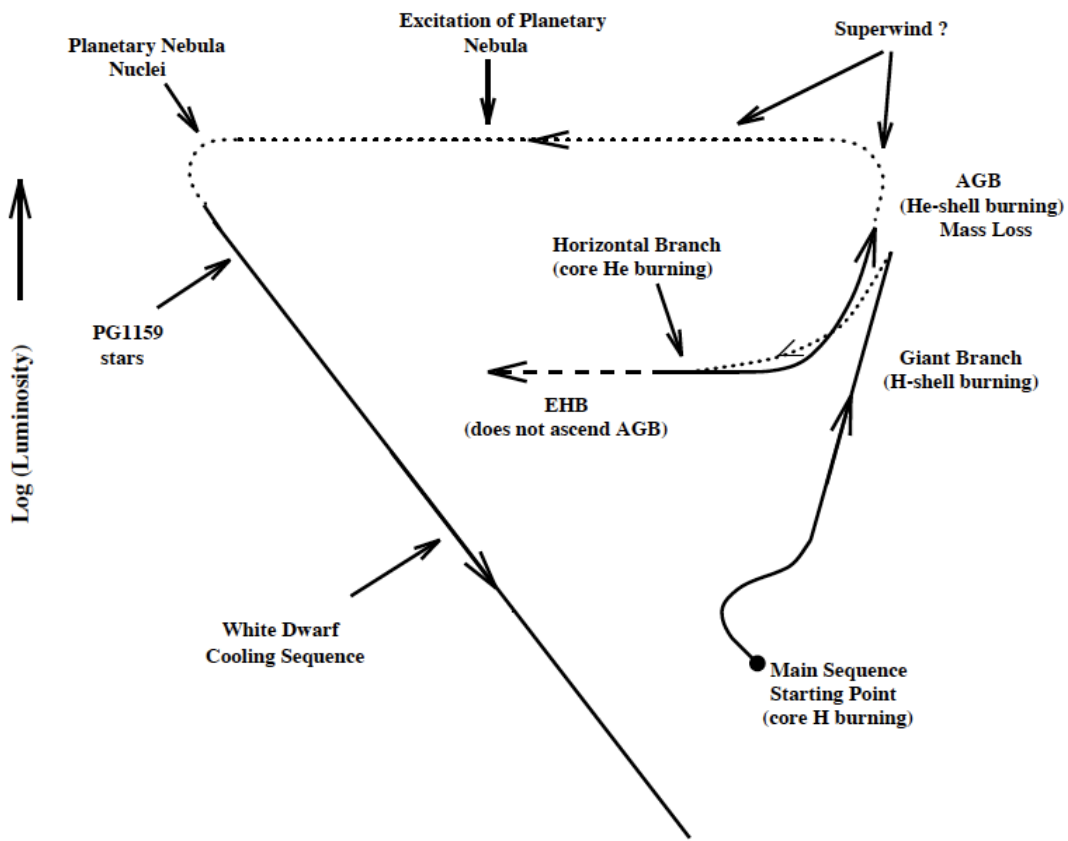

**Figure 1.3.** The Hertzsprung-Russell diagram, illustrating the evolution of a solar type star (from Marsh 1995).

The burning of helium can begin before the core contraction of stars with higher masses (2–8 $M_\odot$). In this case, since no contraction has occurred, the stellar core will not be degenerate. In lower mass stars, the core may be under high enough pressure to be partially degenerate when the burning of helium commences. Since degeneracy pressure is weakly dependent on temperature, the core does not expand. Degenerate material is a very good conductor of heat, allowing a runaway burning to take place over only a few seconds, during which the energy production of the star can increase by a factor of up to $10^{11}$; the helium flash. This occurs until the thermal pressure is sufficient to drive the rising temperature, causing the core to expand and the degeneracy to be lifted.

The helium in the core will eventually be exhausted, at which point a helium shell will burn around a carbon/oxygen core (produced via the alpha burning process). Again, the atmosphere expands and the star moves up the Asymptotic Giant Branch (AGB) and onto the Red Giant Branch (RGB). The most massive stars can contain cores of oxygen and neon (nothing heavier can be produced in stars with initial masses < 8 $M_\odot$), surrounded by shells of progressively lighter elements, all burning to produce progressively heavier elements. Eventually, though, this stage of the star's life will end. Thermal pulsations begin, ejecting the outer atmosphere of the star; for reasons not yet clear, a superwind phase may occur toward the end of this mass loss phase. The mass lost forms an expanding planetary nebula (PN) around the star, with the stellar core at the nucleus. Renzini & Voli (1981) estimate that Red Giant mass loss rates of at least $10^{-5}$ $M_\odot$ yr$^{-1}$ are required to account for PN observations. The exact details of post-AGB mass loss remain poorly understood.

The stellar core at the centre of the expanding PN continues to contract, until again this contraction is halted by degeneracy pressure. Over the course of this process, the star's $T_{eff}$ will increase to over 100,000 K and the log $g$ will rise by over four orders of magnitude. At a $T_{eff} \sim 30,000$ K the UV photons emitted by the stellar remnant will ionise the particles in the nebula. As electrons fall back to lower energy states visible photons are emitted, giving rise to the glowing rings often seen at PN (e.g. Figure 1.4). Eventually, the remnant PN dissipates into the ISM, the hydrogen and helium shells stop burning, and the star becomes a young, hot white dwarf.

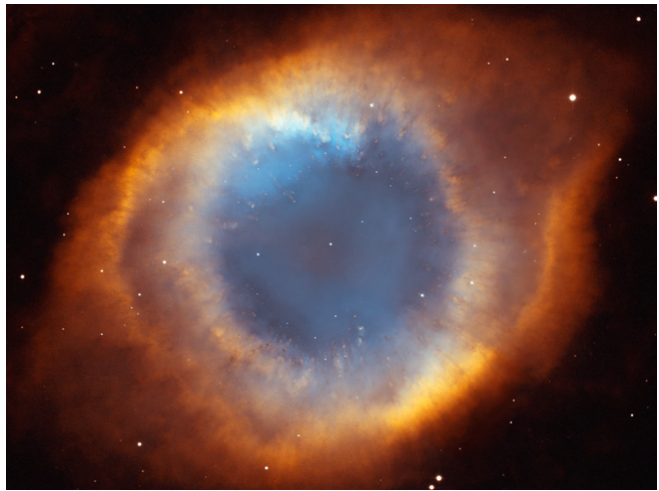

**Figure 1.4.** The Helix Nebula (NGC 7293), with the white dwarf WD 2226–210 at its centre. Image credit: NASA, WIYN, NOAO, ESA, Hubble Helix Nebula Team, M. Meixner (STScI) & T.A. Rector (NRAO).

## 1.6. White dwarf evolution.

As white dwarfs age, their evolution follows a relatively simple cooling sequence over a timescale of the order of $10^9$ years (this is, of course, ignoring any binary interaction). However, several complications arise, giving rise to the complex white dwarf classification system described in Table 1.1.

A carbon-oxygen core with a helium envelope, and in ~80% of white dwarfs a hydrogen layer, is predicted by stellar evolution theory. The evolution of hydrogen dominated DA stars is fairly straightforward. Above 50,000 K, DA white dwarfs have a significant radiation field, which acts to levitate the metals present in the star from either stellar nucleosynthesis or the nebula in which the star formed (if metals heavier than those that could have been produced earlier in the star's life are observed they must have been produced by an earlier generation of more massive stars) into the atmosphere, providing the photospheric metals observed at so many hot DAs (Chayer et al. 1994, 1995a,b). Below 50,000 K the effect of radiative levitation becomes greatly diminished, until at around 20,000 K gravitational diffusion becomes dominant (e.g. Koester & Wilken, 2006). At a $T_{\rm eff}$ ~14,000 K convection sets in (e.g. Bergeron et al. 1995, and references therein), though in most stars the convection zone is too shallow to dredge up significant enough amounts of helium and carbon to pollute the photosphere.

The evolution of helium rich objects is somewhat more complex. In a late helium flash at the end of the AGB phase, these objects are thought to lose the majority of their hydrogen. As the star cools and contracts, the carbon and oxygen sink out of the atmosphere, and a hot DO star is formed. Once this cools to ~45,000 K, some residual hydrogen is thought to emerge, turning the star into a DA with a thin hydrogen layer. This is used to explain the DB gap described in Section 1.4 (though, as noted in Table 1.1, objects are now being found inside the DB gap;

Kleinman et al. 2004). At the cooler end of this gap (~30,000 K), helium is dredged up, allowing the star to evolve into a DBA, then DB star.

At ~13,000 K a convection zone develops in DB stars, dredging carbon back up into the atmosphere to form a DQ white dwarf (Pelletier et al. 1986), with a carbon abundance peak predicted ~10,000 K (Fontaine & Brassard, 2005). Hot DQ white dwarfs (18,000 K < $T_{\rm eff}$ < 24,000 K) have been the subject of much recent study, though their origin and evolution is not yet clear (Dufour et al. 2007a,b; 2008).

Eventually all white dwarfs will cool to a point at which no spectral features can be seen at optical wavelengths. Thus, both hydrogen and helium rich white dwarfs end their lives as DC stars. When the star has eventually radiated away all of its residual heat, it will become a black dwarf (no black dwarfs have yet been observed, since the timescale required for this process to take place is longer than the current age of the Galaxy).

## 1.7. Metals in DA spectra.

Though white dwarfs are traditionally thought of as having pristine spectra, they often display signs of metals in their atmospheres. Two processes contribute to this metallic atmosphere content: radiative levitation and accretion (though convection can be influential at low $T_{\rm eff}$). This section will outline these processes, and discuss in more detail specific problems in our understanding of these mechanisms. Given the variety of hot white dwarfs and phenomena discussed in the proceeding sections of this introduction a table (Table 1.2) is given, detailing the stars explicitly discussed, the context in which they are mentioned, the introduction section in which that discussion takes place and the references used. Similarly, Table 1.3 details all of the metal absorption/emission lines discussed throughout the thesis.

### 1.7.1. The case of hot DA stars.

In DA white dwarfs with $T_{eff} > 50,000$ K, the upward radiation pressure from the residual heat left over from previous stages of the star's life counters the downward diffusion of heavy elements. This radiative pressure lends buoyancy to the metals in the star, an effect known as radiative levitation. Initial studies by Vauclair, Vauclair & Greenstein (1979) found that carbon, nitrogen and oxygen could levitate through bound-bound absorption. This theory was put on a more complete, formal footing by Chayer et al. (1994, 1995) and Chayer, Fontaine & Wesemael (1995), who computed predicted abundances and depth dependent distributions for the most commonly observed metals.

Observational evidence for such metal levitation in hot DA white dwarfs has been seen for some time, both through the presence of FUV absorption features due to metallic elements (e.g. Sion et al. 1992; Holberg, Barstow & Sion 1998; Barstow et al. 2003b) and the blanketing of the EUV/soft X-ray continuum by iron peak elements (e.g. Kahn et al. 1994; Koester 1989; Barstow et al. 1993).

Table 1.2. Individual white dwarfs discussed in the introduction to this thesis, their properties, location in the introduction and relevant references. 'CS' signifies 'circumstellar.'

| WD | Alt. name | Properties discussed | Section | References |
|---|---|---|---|---|
| 0050−335 | GD 659 | Stratified or homogeneous nitrogen (N V)? | 1.7.2 | Barstow et al. (2003b); Schuh, Barstow & Dreizler (2005); Chayer, Vennes & Dupuis (2005) |
| 0232+035 | Feige 24 | CS absorption lines (C IV) | 1.7.3 | Holberg, Barstow & Sion (1998); Bannister et al. (2003) |
|  |  | CS absorption lines (C IV) | 1.7.3 | Holberg, Barstow & Sion (1998); Bannister et al. (2003) |
|  |  | Strömgren spheres | 1.7.3 | Dupree & Raymond (1983) |
| 0455−282 | REJ 0457−281 | CS absorption lines (CIV, NV, SiIV) | 1.7.3 | Holberg, Barstow & Sion (1998); Bannister et al. (2003) |
| 0501+527 | G191-B2B | Stratified helium/iron | 1.7.2 | Barstow et al. (1997, 2005); Barstow & Hubeny (1998); Barstow, Hubeny & Holberg (1999); Dreizler (1999); Dreizler & Wolff (1999) |
|  |  | CS absorption lines (C IV, Si IV) | 1.7.3 | Bannister et al. (2003) |
|  |  | Strömgren spheres | 1.7.3 | Dupree & Raymond (1983) |
| 0556−375 | REJ 0558−373 | CS absorption lines (C IV) | 1.7.3 | Bannister et al. (2003) |
| 0939+262 | Ton 021 | CS absorption lines (C IV, Si IV) | 1.7.3 | Bannister et al. (2003) |
| 0948+534 | PG 0948+534 | Poor line profile fits; stratified metals? (C IV, NV, OV, SiIV) | 1.7.2 | Barstow et al. (2003b) |
| 1029+537 | REJ 1032+532 | Stratified or homogeneous nitrogen (N V)? | 1.7.2 | Barstow et al. (2003b); Holberg et al. (1999a); Schuh, Dreizler & Wolff (2002); Chayer, Vennes & Dupuis (2005) |
| 1337+705 | EG 102 | CS absorption lines (C II) | 1.7.3 | Holberg, Barstow & Sion (1998) |
| 1611−084 | REJ 1614−085 | Stratified or homogeneous nitrogen (N V)? | 1.7.2 | Barstow et al. (2003b); Holberg et al. (2000); Chayer, Vennes & Dupuis (2005) |
|  |  | CS absorption lines (C IV, Si IV) | 1.7.3 | Holberg, Barstow & Sion (1998); Bannister et al. (2003) |
| 1620−391 | CD −38°10980 | CS absorption lines (C II, Si II, Si III) | 1.7.3 | Holberg, Bruhweiler & Anderson (1995); Holberg, Barstow & Sion (1998) |
| 1729+371 | GD 362 | CS dust disc | 1.7.3 | Zuckerman et al. (2007); Jura et al. (2009) |
| 1738+669 | REJ 1738+665 | CS absorption lines (C IV, N V, O V, Si IV) | 1.7.3 | Bannister et al. (2003) |
| 2218+706 | WD 2218+706 | CS absorption lines (C IV, Si IV) | 1.7.3 | Bannister et al. (2003) |
| 2226−210 | Helix Nebula CSPN | Infrared excess at hot white dwarf | 1.7.3 | Chu et al. (2011) |
| 2326+049 | G 29−38 | CS dust disc | 1.7.3 | Jura et al. (2009) |
| − | SDSS 1228+1040 | Gas component to CS disc (Ca II emission) | 1.7.3 | Gänsicke et al. (2006); Brinkworth et al. (209); Hartmann et al. (2011); Melis et al. (2011) |
| − | SDSS 0845+2257 | Gas component to CS disc (Ca II emission) | 1.7.3 | Gänsicke et al. (2008); Melis et al. (2011) |
| − | SDSS 1043+0856 | Gas component to CS disc (Ca II emission) | 1.7.3 | Gänsicke et al. (2007); Melis et al. (2011) |

**Table 1.3.** The metal absorption/emission lines discussed in throughout this thesis, with their laboratory wavelengths (in Å). Included are the low ion absorption features that are used to characterise the ISM in Chapter 5. The features are grouped according to their origin. All wavelengths are taken from the Kurucz database[1].

| | |
|---|---|
| Photospheric high ion absorption lines | |
| C IV | 1548.187, 1550.72 |
| N V | 1238.821, 1242.804 |
| O V | 1371.296 |
| O VI | 1031.912, 1037.613 |
| Si IV | 1393.755, 1402.770 |
| Circumstellar high ion absorption lines | |
| C IV | 1548.187, 1550.72 |
| N V | 1238.821, 1242.804 |
| O V | 1371.296 |
| O VI | 1031.912, 1037.613 |
| Si IV | 1393.755, 1402.770 |
| Circumstellar metal emission lines | |
| Ca II | 8500 – 8660 triplet |
| Fe II | 5020, 5170 |
| ISM low ion absorption lines | |
| O I | 1302.168 |
| Si II | 1260.422, 1304.370, 1526.707 |
| S II | 1259.519 |
| Fe II | 1608.536 |

A common technique used to measure the $T_{\text{eff}}$ (and log $g$) of a white dwarf is to compare the observed Balmer line profiles to those predicted by model calculations (e.g. Holberg et al. 1985; Bergeron, Saffer & Liebert 1992). As first suggested by

---

[1]http://www.pmp.uni-hannover.de/cgi-bin/ssi/test/kurucz/sekur.html

Dreizler & Werner (1993), line blanketing by photospheric metals significantly affects $T_{eff}$ measurements for $T_{eff} > 55,000$K (Barstow et al. 1998). Barstow et al. (2001, 2003a) also found that a significant difference between the $T_{eff}$ values derived from Balmer and Lyman line analyses emerges in DAs with $T_{eff} > 50,000$K; a similar, more severe effect is seen in DAO stars (Good et al. 2004). Accurate measurements of parameters such as a white dwarf $T_{eff}$ are key to understanding stellar evolution; this gives how far along the white dwarf cooling sequence the star has travelled. Using these parameters as inputs to white dwarf evolutionary models (such as those of Wood 1995) can allow the age and mass of the star to be calculated. Reliable measurements of $T_{eff}$ are thus crucial to our understanding of white dwarf stars, and the application of this knowledge to stellar evolution and wider astronomy.

Further complications arise when examining the patterns in metal abundances. In their 2003 study, Barstow et al. (2003b) found that while the broad patterns predicted between abundance and $T_{eff}$ and log $g$ are reproduced, the precise abundance predictions are often not matched by the observed values. Furthermore, in some cases, stars with similar $T_{eff}$ and log $g$ values have quite different metal abundances. Given the importance of a proper understanding of metals in white dwarfs to our understanding of stellar evolution, a better understanding of the distribution of metals in hot white dwarf photospheres is therefore desirable.

### 1.7.2. The stratification of metals in hot DAs.

A simplifying assumption often made when modelling white dwarfs is that their atmospheres are homogeneous. However, for many hot white dwarfs this assumption conflicts strongly with observations. Indeed, radiative levitation calculations predict a varying metal abundance with depth (e.g. Chayer, Fontaine & Wesemael, 1995), and for the past three decades stratified atmospheres have been used to explain some white dwarf observations.

Vennes et al. (1988) showed that radiative levitation is not as efficient at lending buoyancy to helium when compared to carbon, nitrogen, oxygen, silicon, iron or nickel. This causes the helium to sink though the atmosphere to leave a polluted hydrogen layer above a helium envelope. Using similarly stratified models of G191-B2B (WD 0501+527) with homogeneously distributed metals, Barstow & Hubeny (1998) were able to reproduce the absence of the photospheric He II 1640 Å line and obtain a more typical interstellar helium ionisation fraction (since the helium observed along a given line of sight to a star is the sum of that observed in the photosphere and that in the ISM); when fit with a homogeneous model, the interstellar helium ionisation fraction was around 80 (±20) %, while a fraction of ~27 % was more representative of the local ISM (LISM; Barstow et al. 1997). The stratified models yielded a much lower ionisation fraction of 59 %, with the lower bound of possible ionisation fraction values at 37 % nearer values reported along other lines of sight (Barstow et al. 1997). However, a thicker hydrogen layer was required, heavily absorbing the EUV continuum in the model and giving a poor match to the data. The analysis of high resolution ($R = 4,000$) narrow band (226 Å – 246 Å) EUV data from the Joint Astrophysical Plasma-dynamic Experiment *(J-PEX)* showed a better fit was obtained with a homogeneous model (Barstow et al. 2005), in conflict with the conclusion of Barstow & Hubeny (1998). This was attributed to either a deficiency in the atomic data and/or a dual component, high ionisation fraction of He II along the line of sight, of which one component was consistent with measurements along other lines of sight.

The stratification of metals is also important in white dwarf atmosphere modelling. In another investigation of WD 0501+527, Barstow, Hubeny & Holberg (1999) found the star to have a stratified iron abundance. Here, the atmosphere was split into a series of horizontal, homogeneous slabs with an increasing iron abundance with depth. This model successfully explained the observed optical, FUV and EUV observations, with a high ISM ionisation fraction of He II (in keeping with the ISM later observed by Barstow et al. 2005). Radiatively driven mass loss was used to explain iron depletion in the upper atmosphere. Dreizler (1999) and Dreizler & Wolff (1999) also constructed stratified models to study the EUV spectrum of WD 0501+527, using depth dependent radiation intensity, and the chemical abundances at each depth point were those produced by the equilibrium of radiative levitation and downward diffusion. This method successfully modelled the EUV spectrum of WD 0501+527 without the interstellar He II column density detected, at odds with the interstellar measurements of Barstow & Hubeny (1998), Barstow, Hubeny & Holberg (1999) and Barstow et al. (2005).

Observations of metals other than iron have required stratified chemical configurations. Nitrogen stratification was used to explain the observed line profiles of the FUV N V doublet (1238.82 Å, 1242.80 Å) in the 44,350±715 K DA REJ 1032+532 (WD 1029+537; Barstow et al. 2003b, Holberg et al. 1999a). A homogeneous nitrogen distribution with a log(N/H) of –4.31 gave a line profile with a depth similar to that observed. However, such a high nitrogen abundance in the lower atmosphere caused the model absorption features to be heavily pressure broadened beyond the observed line profile. A thin nitrogen layer at the top of the atmosphere ($\Delta M/M = 3.1\times10^{-16}$) reproduced the observed line profiles well; the high nitrogen abundance provided the deep line profile, while not having nitrogen lower in the atmosphere avoided such heavy pressure broadening (Figure 1.5).

Holberg et al. (1999a) also considered the EUV spectrum of WD 1029+537 (Figure 1.6). The presence of nitrogen at high abundance in the lower atmospheric region of the homogeneous model causes significant EUV absorption, due to both nitrogen absorption edges at 180 and 260 Å and absorption lines at specific wavelength above 135 Å (Figure 1.6, lower curve). The stratified nitrogen removes this EUV absorber, better matching the data (Figure 1.6, upper curve). It was concluded that, given the stratified nitrogen distribution matched both the FUV N V line profiles and the EUV continuum, the star had a slab of high abundance nitrogen at the top of its atmosphere. This distribution was again put down to a radiatively driven mass loss process, which enigmatically affected only nitrogen (the carbon and silicon in this star were well modelled using a homogeneous distribution).

A similar nitrogen configuration has been suggested for GD 659 (WD 0050–332; $T_{eff}$ = 35,660±135 K, Barstow et al. 2003b), which displays a pure hydrogen EUV spectrum with FUV carbon, nitrogen and silicon absorption features (Barstow et al. 2003b). REJ 1614–085 (WD 1611–084; 38 840±480 K, Barstow et al. 2003b) has also been examined in this context, and was found to have strong N V FUV absorption lines, again indicative of a slab of nitrogen in the higher atmospheric region (Holberg et al. 2000; Barstow et al. 2003b).

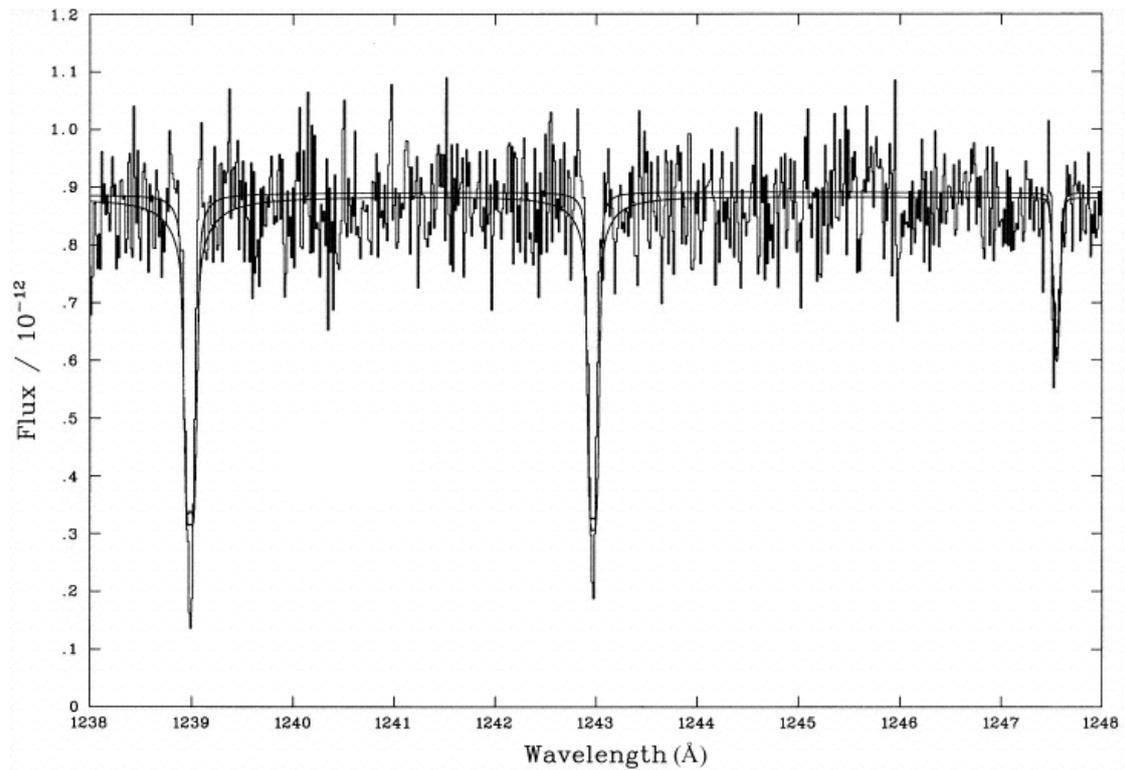

**Figure 1.5.** The N V resonance doublet of WD 1029+537 (Figure 5, Holberg et al. 1999a). A layer of nitrogen (log(N/H) = –4.31) in the topmost part of the atmosphere ($\Delta M/M = 3.1 \times 10^{-16}$) is illustrated with the upper curve. A homogeneous nitrogen distribution is shown with the lower curve.

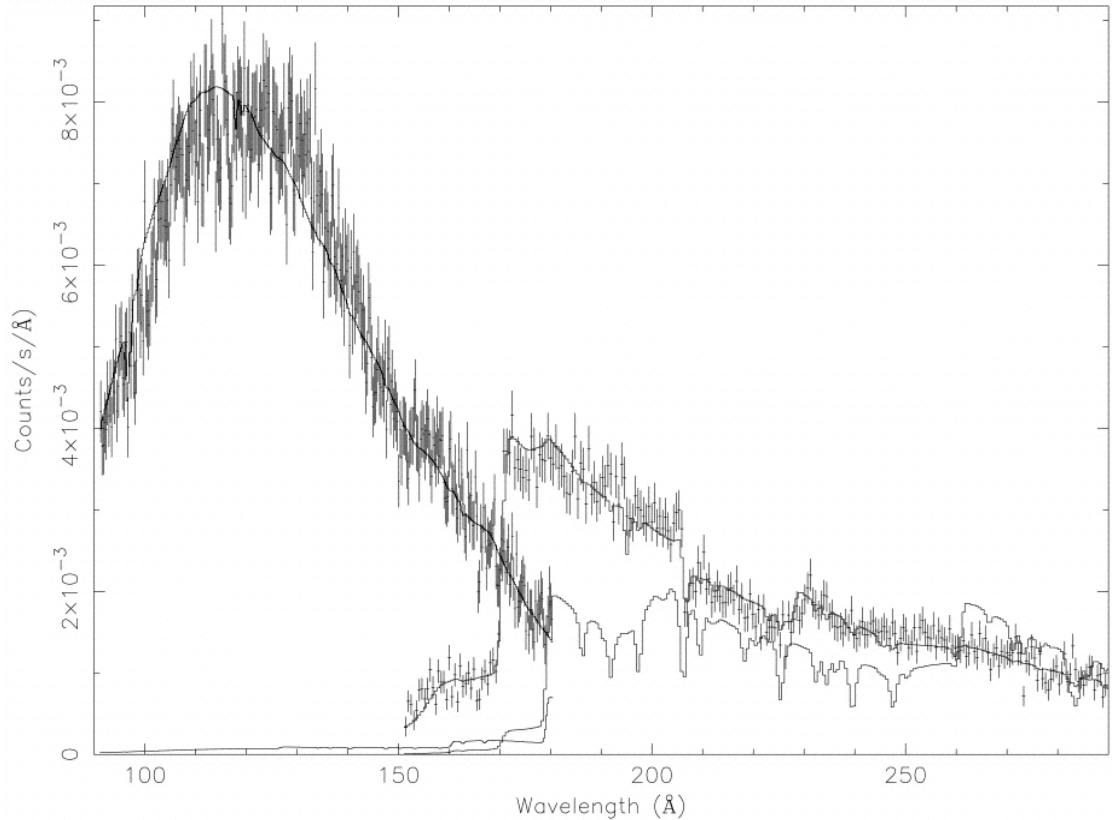

**Figure 1.6.** The *EUVE* spectrum of WD 1029+537 (Figure 6, Holberg et al. 1999a). The homogeneous model is again shown by the lower curve and the stratified nitrogen configuration is indicated with the upper curve.

When examining the patterns in the metal abundances of hot DA white dwarfs, WD 1029+537, WD 1611–084 and WD 0050–332 stand out as interesting objects. Barstow et al. (2003b) find no correlation between nitrogen abundance and $T_{eff}$, for stars with $T_{eff}$ > 50,000 K. Below 50,000 K, heavy elements should begin to sink out of the atmosphere due to the reduced dominance of radiative levitation. However, a dichotomy is observed. WD 1029+537, WD 1611–084 and WD 0050–332 show an increase in nitrogen abundance with decreasing $T_{eff}$, while all other white dwarfs show no nitrogen and only upper limits can be estimated (Figure 1.7), hinting that these three objects may in some way be special.

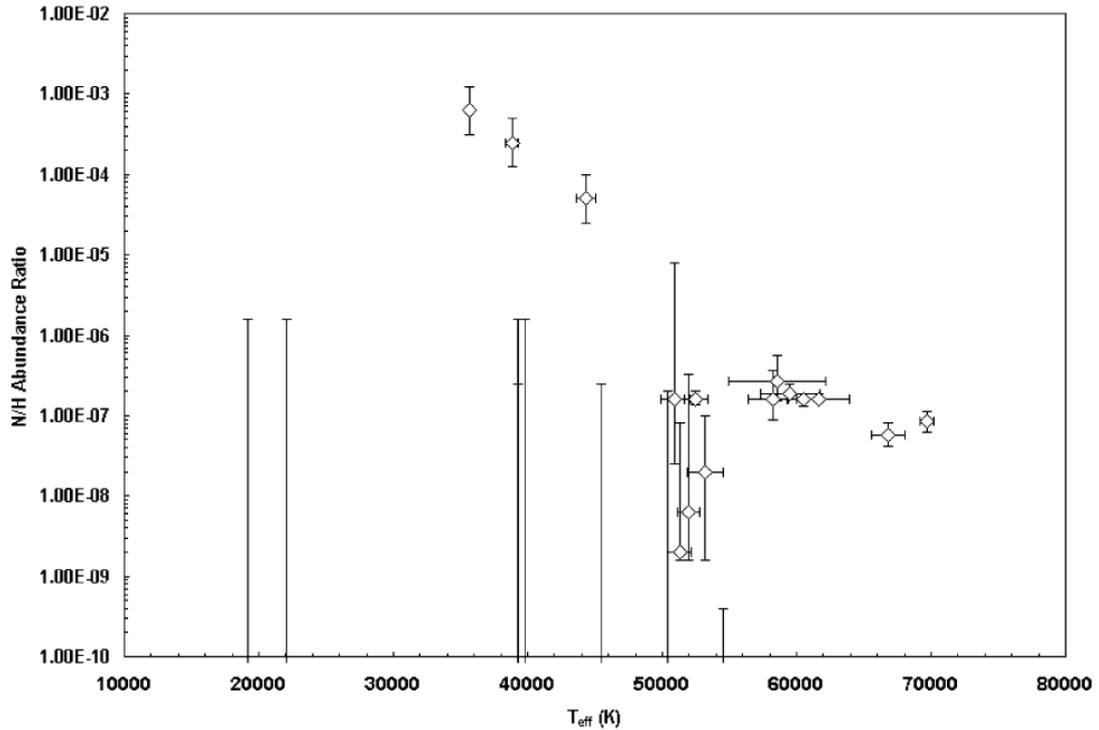

**Figure 1.7.** Measured nitrogen abundance as a function of $T_{\text{eff}}$ (Figure 10, Barstow et al. 2003b). WD 1029+537, WD 1611–084 and WD 0050–332 are the three objects with $T_{\text{eff}} <$ 50,000 K with nitrogen detections.

Schuh, Dreizler & Wolff (2002) used stratified model sets of the type of Dreizler (1999) and Dreizler & Wolff (1999) to model the *EUVE* spectra of a sample of DA white dwarfs, and explained the EUV properties of many of their objects well. However, four DAs (WD 1314+293, WD 1029+537, WD 2004–605 and WD 2152–548) were better represented by homogeneous models (the WD 1029+537 result being in direct conflict with that of Holberg et al. 1998a and Barstow et al. 2003b). Four other white dwarfs (WD 0027–636, WD 1056+516, WD 1234+481 and WD 2111+498) were not well fit; this was interpreted as accretion disturbing the radiative levitation/downward diffusion balance. Furthermore, a comparison by Schuh, Barstow & Dreizler (2005) of the abundance patterns measured using the stratified models of Schuh, Dreizler & Wolff (2002) to those measured by Barstow et al.

(2003b) found the nitrogen abundances of WD 0050–332, WD 1029+537 and WD 1611–084 were in keeping with the other white dwarfs of higher $T_{eff}$ (Figure 1.8), in stark conflict with the results of Barstow et al. (2003b). Oxygen abundances were roughly consistent. The stratified C III and C IV abundances were consistently over-predicted when compared to the homogneous models, while silicon was generally over predicted for $T_{eff}$ < 50,000 K and under predicted for $T_{eff}$ > 50,000 K. The iron and nickel abundances were within the systematic errors expected between both model sets, and the Fe:Ni ratio was ~20, consistent with the cosmic value and that of Barstow et al. (2003b).

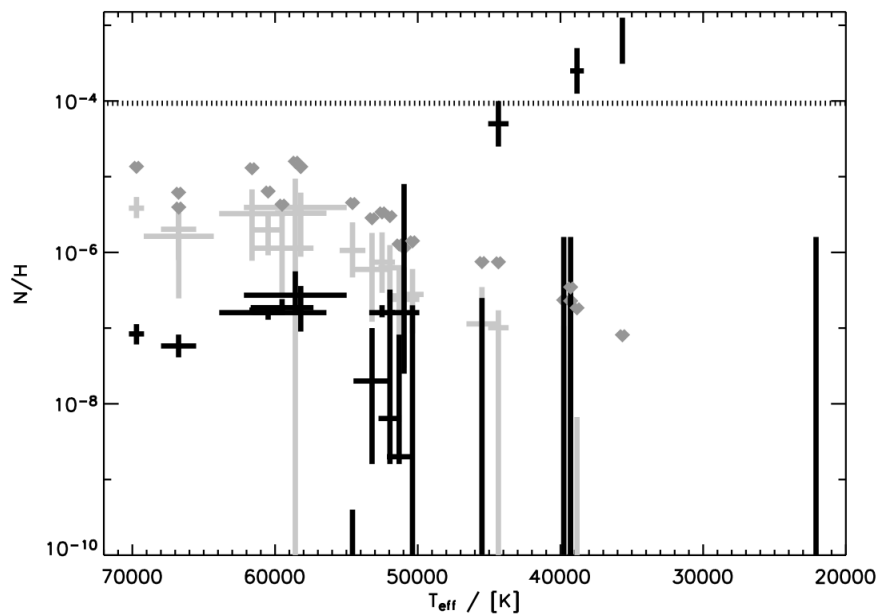

**Figure 1.8.** Figure 1 from Schuh, Barstow & Dreizler 2005. A comparison of the nitrogen abundances using the stratified models of Schuh, Dreizler & Wolff (2002; light grey data points) to the measurements of Barstow et al. (2003b; black data points). The radiative levitation predictions of Chayer et al. (1995) are denoted with dark grey symbols, while the cosmic abundance is shown with the dotted line.

A later study of WD 0050–332, WD 1029+537 and WD 1611–084 by Chayer, Vennes & Dupuis (2005) found that the nitrogen line profiles and EUV data could be explained with homogeneous nitrogen distributions, with much lower abundances (log(N/H) = –6.2, –5.2 and –5.9 for WD 1029+537, WD 1611–084 and WD 0050–332, respectively; Figure 1.9) than those found by Barstow et al. (2003b).

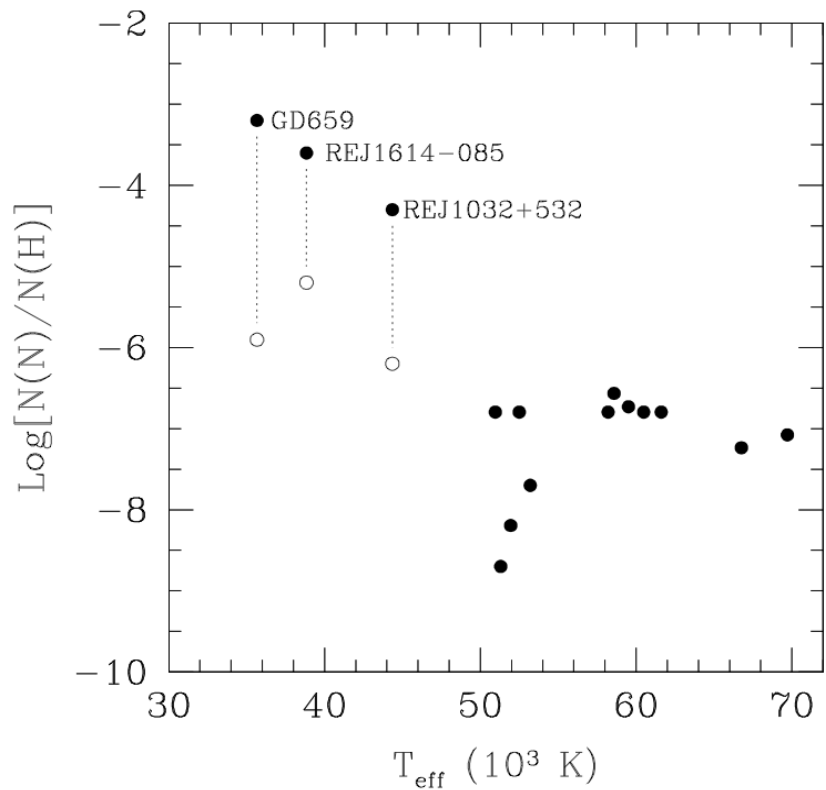

**Figure 1.9.** Figure 1 from Chayer, Vennes & Dupuis 2005. A comparison of the log(N/H) values found by Chayer, Vennes & Dupuis (open circles) to those found by Barstow et al. (2003b; filled circles).

Another star to show anomalous line profiles is the extremely hot DA PG 0948+534 (WD 0948+534, $T_{eff}$ = 110,000±2,500 K, Barstow et al. 2003b). The C IV, N V, O V and Si IV absorption features in the *STIS* data of this star are again extremely narrow and, in the case of C IV, almost completely saturated (Figure 1.10).

Given the similarity between this doublet and the N V doublets in WD 1029+537, WD 1611–084 and WD 0050–332, it was suggested by Barstow et al. (2003b) that the carbon might be similarly distributed in a slab in the upper atmosphere. Indeed, their preliminary calculations supported this hypothesis.

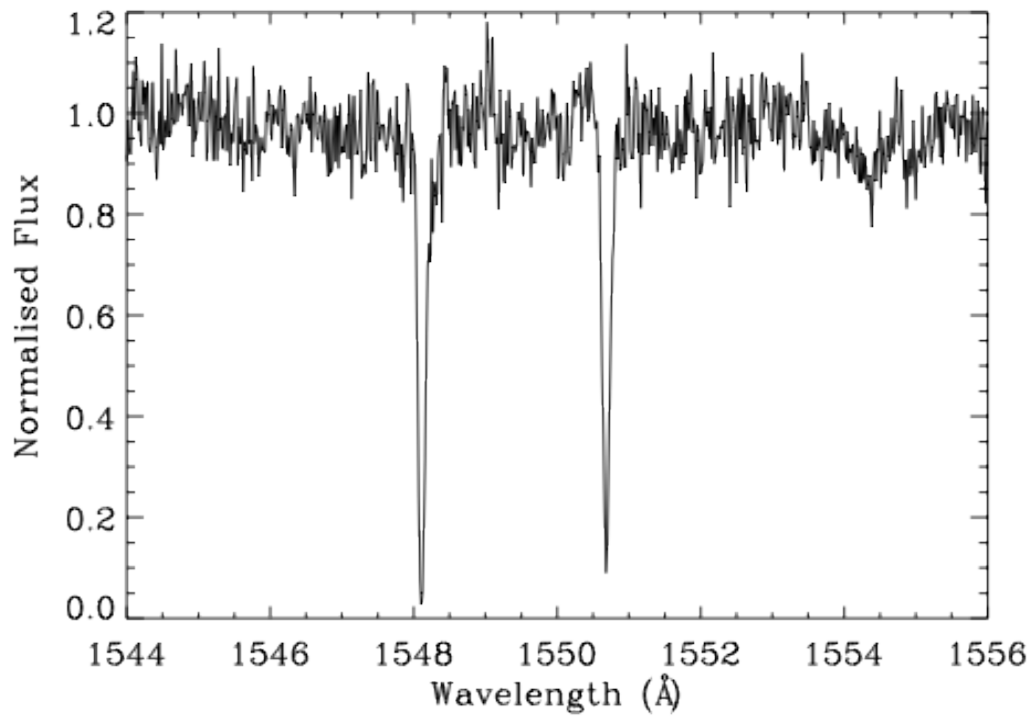

**Figure 1.10.** The C IV doublet in the *STIS* spectrum of WD 0948+534.

Given the differing conclusions of the analyses of the nitrogen abundance and distribution in WD 1029+537, WD 0050–332 and WD 1611–084, a detailed analysis of the nitrogen in these stars is desirable, and is described in Chapter 3. Since stratified, high abundance models have been used to explain the metal absorption features of WD 0948+534, the analysis is also extended to this object.

## 1.7.3. Circumstellar material at hot DA stars.

In addition to absorption features from highly ionised photospheric material, absorption features at non-photospheric velocities have been seen in the spectra of hot DAs during the past few decades of white dwarf research. The *IUE* spectrum of Feige 24 (WD 0232+035, a close binary system consisting of a hot DA white dwarf and an M dwarf) displays two sets of C IV absorption features (e.g. Dupree & Raymond 1982). This is interpreted as one set of photospheric features, where the changing velocity of the component reflects the orbital motion of the DA (Vennes et al. 1992), and a set of stationary features arising in a hot circumstellar gas. Again using *IUE* observations, Si II, Si III and C II absorption features are seen with a velocity far from the photospheric and interstellar velocities at CD $-38^{\circ}$ 10980 (WD 1620–391, Holberg, Bruhweiler & Anderson 1995), indicating the presence of a photoionised circumstellar cloud around the star.

A survey of 55 white dwarf *IUE* spectra (Holberg, Barstow & Sion, 1998) found 11 stars with evidence of circumstellar material, of which five were DAs (WD 0050–332, WD 0232+035, WD 0455–282, WD 1337+705, WD 1611–084 and WD 1620–391). These circumstellar features were all blueshifted, occupied a narrow velocity range (40 – 60 km s$^{-1}$), and were attributed to mass loss from the white dwarfs. Lines of sight close to the white dwarfs showed no ISM absorption features consistent with the observed circumstellar lines. Also, the ISM features seen along such sight lines had both red and blueshifted velocities with respect to the photosphere. In a more recent survey of 23 hot DA white dwarfs, using data from *IUE* and *HST STIS*/*GHRS*, eight white dwarfs were found with circumstellar material, with two further possible detections (Bannister et al. 2003). Potential sources of this material were put forward, and included the ionisation of nearby ISM in the 'Strömgren sphere' of the white dwarf, material inside the gravitational well of the star, mass loss in a stellar wind and ancient planetary nebulae (PNe). Indeed, WD 2218+706 (one of the objects exhibiting circumstellar absorption in the sample of Bannister et al. 2003) is located within the old planetary nebula (PN) DeHt5 (e.g. Napiwotzki & Schöberner, 1995).

In recent years, research into the circumstellar environments of white dwarfs has yielded many interesting results. The diffusion timescales of metals in the photospheres of cooler white dwarf stars ($T_{eff}$ < 25,000 K) is extremely short (e.g. Koester & Wilken, 2006), requiring an external source of metals near the white dwarf to maintain the observed abundances (radiative levitation has recently been found to have some effect in these cooler white dwarfs, although accretion is still required to explain the observed metal abundances; Chayer & Dupuis 2010; Dupuis, Chayer & Hénault-Brunet, 2010; Dupuis et al. 2010).

Dupuis et al. (1992, 1993) and Dupuis, Fontaine & Wesmael (1993) proposed a 'two phase accretion model' to explain the atmospheric pollution, where a white dwarf would encounter an ISM cloudlet and accrete metals. After passing through the cloudlet, the photosphere of the white dwarf would still contain observable metals until the diffusion timescale had passed. However, studies looking at white dwarf kinematics reported that cool, metal rich white dwarfs were sufficiently far from ISM cloudlets for all the metals in their photospheres to have fully diffused downwards, given their velocities and positions (e.g. Aannestad et al. 1993). Indeed, some white dwarfs displaying metallic photospheres were about to move into a cloudlet, and should not have had any metals in their photospheres. Subsequent studies demonstrated that metals may be accreted from more local sources, such as comets, mass lost from a binary companion (e.g. Zuckerman & Reid, 1998) or disrupted asteroids (e.g. Zuckerman et al. 2003; Jura, 2003, 2006, 2008).

Infrared studies show evidence of circumstellar dust discs at some of the cool DAZ stars, and the accretion of this dust introduces polluting metals to the white dwarf photospheres (e.g. Kilic et al. 2005, 2006; Kilic & Redfield, 2007; von Hippel et al. 2007; Farihi, Zuckerman & Becklin, 2008). A study with the Spitzer *IRAC* and *MIPS* instruments shows that, when combined with previous work, no more than 20% of all single white dwarfs with an implied metal accretion rate > $3 \times 10^8$ g s$^{-1}$ display infrared emission from a dust disc (Farihi, Jura & Zuckerman, 2009). It is reasoned that, since this accretion rate is between the dust production rates in the Solar System zodiacal cloud ($10^6$ g s$^{-1}$) and the debris discs around main-sequence A type stars ($10^{10}$ g s$^{-1}$), these discs are produced by the tidal disruption of extrasolar minor planets and/or asteroids, an idea first put forward by Graham et al. (1990) and developed by Debes & Sigurdsson (2002). Silicate emission from circumstellar dust at six externally polluted white dwarfs adds weight to this model (Jura, Farihi & Zuckerman, 2009). Farihi et al. (2010) report no relation between the accreted calcium abundances and the presence of clouds in the ISM for 146 DZ Sloan Digital Sky Survey (SDSS) white dwarf spectra. It was also found that for $T_{eff}$ < 12,000 K, the DBZ and DC white dwarfs belonged to the same stellar populations, implying that the metal pollution in the DBZ stars must be from tidally disrupted rocky planets since ISM accretion would also be evident in the DC stars.

The relative metal abundances of the DB GD 362 (9,850±100 K) show that the white dwarf is likely to be accreting from a large, disrupted asteroid/asteroids with an Earth-Moon composition (Zuckerman et al. 2007). An infrared and X-ray analysis reports evidence for the accretion of either 100 Ceres-like asteroids or one large object by both GD 362 (given the anomalously large relative amount of hydrogen in the material accreted by this star) and G29–38 (Jura et al. 2009). An alternative scenario is that a single parent body with a mass between that of Callisto and Mars, containing internal water, has been disrupted and is being accreted. Further studies have found many other white dwarfs harbouring the remains of extrasolar planets (e.g. Dufour et al. 2010; Klein et al. 2011; Melis et al. 2011; Zuckerman et al. 2011).

Gaseous components have been found at some white dwarf circumstellar discs. The optical spectrum of SDSS J122859.93+104032.9 ($T_{\rm eff}$ = 22,020 K, hereafter SDSS J1228+1040) displays emission from the Ca II 8500–8660 Å triplet (Figure 1.11, right hand panel), as well as weaker emission from Fe II at 5020 and 5170 Å (Gänsicke et al. 2006). The star also exhibits photospheric Mg II absorption (4482 Å, Figure 1.11, left hand panel), with roughly a solar abundance, suggesting a metal rich disc. The lack of photospheric He I absorption at 4470 Å (providing an abundance upper limit of 0.1 times the solar abundance) or Balmer and helium emission from the disc, lends further weight to the metallic composition of the circumstellar disc. The asymmetry in the double peaked Ca II emission line is indicative of an asymmetric disc, with an estimated outer disc radius of 1.2 $R_{\odot}$, comparable to the tidal disruption radius for a rocky asteroid (Davidsson, 1999). Time resolved spectroscopy and photometry do not reveal any radial velocity variations, showing no detectable interacting binary companion is present from which material could be accreted.

Detailed modelling of this circumstellar gas disc at SDSS J1228+1040 shows that the Ca II triplet emission can arise from a metallic gas disc inside the tidal disruption radius of the star, with a disc $T_{\text{eff}}$ ~ 6,000 K and surface mass density of ~0.3 g cm$^{-2}$ (Hartmann et al. 2011). The asymmetry in the emission is found to be due to either a spiral arm structure or disc eccentricity. However, the models of Hartmann et al. (2011), which assume a chemical composition typical for Solar System asteroids (including hydrogen, carbon, nitrogen, oxygen, magnesium, silicon and calcium), predict C II, O II, Si I, Si II, Mg I and Mg II emission that is not seen, and should therefore be treated with caution.

As shown in Figure 1.11, Gänsicke et al. (2007, 2008) have also found metallic gas discs around the DAZ SDSS J104341.53+085558.2 (hereafter SDSS J1043+0856) and the DBZ SDSS J084539.17+225728.0 (hereafter SDSS J0845+2257). GD 362 and WD 1337+705 (which has an anomalously high metal abundance) were also examined by Gänsicke et al. (2007), although no circumstellar gas discs were detected. Brinkworth et al. (2009) found a dust disc was also present at SDSS J1228+1040, demonstrating for the first time a debris disc with dust and gas components. A further study by Melis et al. (2011) found that both SDSS J1043+0856 and SDSS J0845+2257 also harbour dust discs that are spatially coincident with the gas discs.

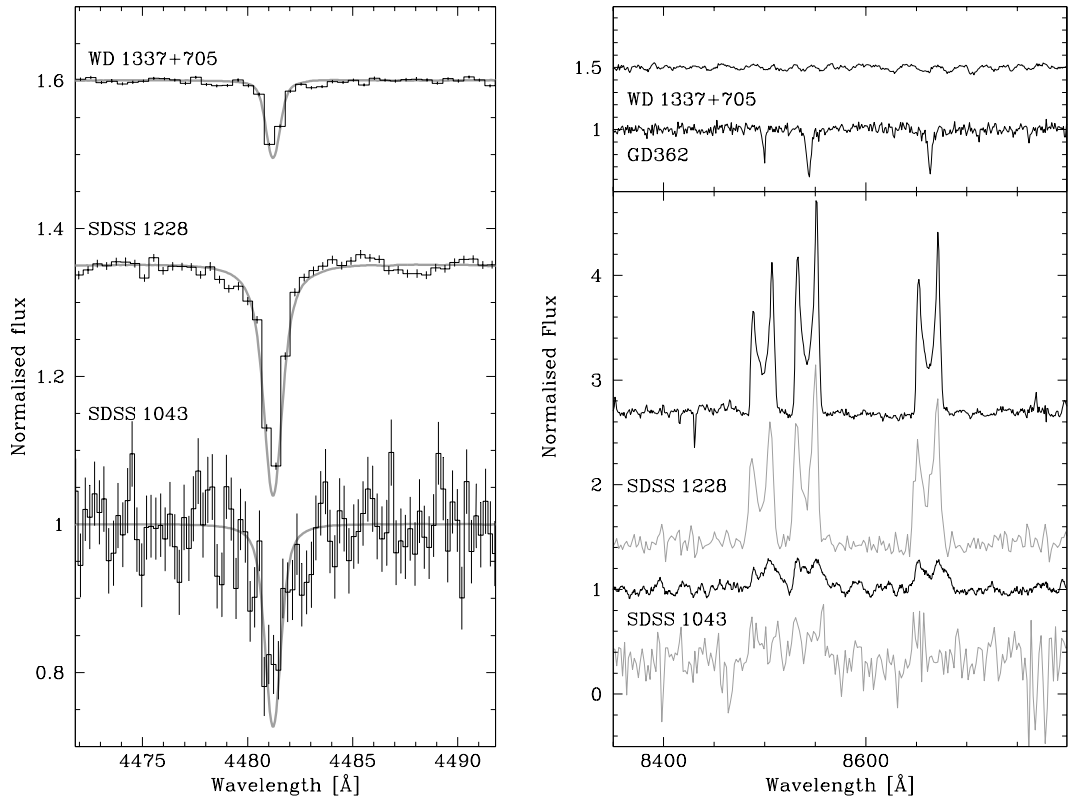

**Figure 1.11.** Figure 1 from Gänsicke et al., 2008. The left hand panel shows the photospheric Mg II (4481 Å) absorption lines in the William Herschel Telescope (WHT) spectra of WD1337+705, SDSS 1228 and SDSS 1043 (black lines), with the best fitting white dwarf models (grey). The right panel shows the 8350–8800 Å region of the spectra of WD 1337+705, GD 362, SDSS 1228 and SDSS 1043, normalised and offset for clarity. WHT spectra are shown in black, while the SDSS spectra are plotted in grey.

An infrared excess is seen at the central star of the helix nebula (Su et al. 2007), which may be due to a circumstellar disc. Unlike the circumstellar discs prevalent at cooler white dwarfs, these dust discs are around 30 – 150 AU from the central star of the PN (CSPN). In a recently published survey of 71 hot white dwarfs, with 35 CSPNe, Chu et al. (2011) report that ~20 % of the surveyed stars have infrared excesses. Only 5–6 % of non-PN stars have similar infrared excesses. While a Kuiper belt type dust disc has been put forward as a source for the infrared emission, it must be stressed that the exact origin of these infrared excesses is not fully understood, and could also be due to PN material or binary companions. A recent publication on the WIRED survey suggests that four of the DAs in the survey with $T_{\text{eff}}$ > 30,000 K have infrared excesses that could be due to circumstellar discs, although the observations may suffer from contamination from background objects (Debes et al. 2011).

In addition to circumstellar discs being a possible source of the circumstellar high ions, the ionisation of ISM material near to the star may provide a source of the shifted absorption components. Previous studies have attempted to quantify this effect. Dupree & Raymond (1983) looked to white dwarf Strömgren spheres to explain the high ion absorption features seen in the spectra of WD 0232+035 (Feige 24) and WD 0501+527 (G191-B2B), although these absorption features have since been found to originate in the photospheres of the stars (except the circumstellar C IV lines seen at these DAs). In an examination of the UV ionisation of the local ISM (LISM) within 20 pc of the Sun, Tat & Terzian (1999) concluded that material inside the Strömgren spheres of nearby white dwarfs could be responsible for the observed LISM ionisation. Of the 121 stars studied, 24 had Strömgren radii ($r_S$) > 0.5 pc. Since only stars within 20 pc were considered, the significant effect of B stars within 100 pc (Vallegra, 1999) was ignored, as was the presence of the rarefied "tunnel" towards $\beta$ CMa (Welsh, 1991) and the local chimney (Welsh et al. 1999). Welsh et al. (2010a) observed a series of cell-like cavities in their 3-D maps of the LISM (out to 300 pc), attributed to the ionisation of the ISM near B stars and some hot white dwarfs. The evaporation and photoionisation of circumstellar material by the radiation field of hot white dwarfs was used to explain the circumstellar high ion absorption features seen in the spectra of the three stars studied by Lallement et al. (2011).

A proper understanding of the origin of the circumstellar absorption seen at hot white dwarfs is crucial. Should, as highlighted by Lallement et al. (2011), the non-photospheric O VI attributed to the ISM by Savage & Lehner (2006) and Barstow et al. (2010) in fact be photoionised material local to the white dwarfs, then erroneous interpretations of the physical state of the LISM may have been made. This could explain the confused view of local hot interstellar gas. Indeed, Lallement et al. (2011) found evidence that suggests hot/cold gas interfaces cannot give rise to the observed non-photospheric features. This echoes the finding of Welsh et al. (2010a,b), that the Doppler widths of the observed non-photospheric high ion absorption features are inconsistent with hot/cold gas interfaces, and better suit photoionisation by B stars and hot white dwarfs. This shows a clear need to understand the true nature of the non-photospheric high ions seen in the spectra of hot DA white dwarfs.

## 1.8. Modelling white dwarf stars.

The white dwarf atmosphere models used in this work were constructed using the TLUSTY program (e.g. Hubeny & Lanz 1992, 1995, 2003; Hubeny, Hummer & Lanz, 1994; Lanz & Hubeny 1995, 2003a,b). Spectra were synthesised using the accompanying SYNSPEC code. Although the precise computational techniques used to construct stellar atmospheres vary, and an exhaustively detailed account of the programs used is not given here, the basic physical principles are discussed.

The model atmosphere is assumed to be plane parallel and horizontally homogeneous. The incoming flux from the stellar core is fixed by the model $T_{eff}$, and TLUSTY solves the hydrostatic, ionisation and excitation equilibrium in each of these atmospheric layers in the presence of specified chemical abundances. The results of these calculations provide the density, temperature, mass, electron density, log $g$, Rossland optical depth and flux at each depth point, along with details of the ionisation and excitation balance, which are then used to calculate the opacity in each layer due to each transition detailed during spectrum synthesis with SYNSPEC. The model ions used here are those constructed for the TLUSTY code[2], which contain the individual energy levels/super levels, oscillator strengths and photo-ionisation cross sections of each ion. These ionic data come from the Opacity Project TOPBASE[3] database. Significant variety can be found in the values of such data, depending on the source of the data (e.g. whether the data is taken from the TOPBASE database, the NIST Atomic Spectra Database[4], etc), and uncertainties in quantities such as the oscillator strengths (i.e. the probability of a give transition occurring) may have a significant impact on the atmosphere calculation and synthesised spectrum. Similarly, grouping energy levels into either into a "superlevel" (where a closely spaced of discrete energy states are approximated to one level to reduce the complexity of the model ion) may impact upon the calculated model in the high $T_{\text{eff}}$ regime where high ions of the complex iron peak elements are present. For iron peak elements, each model ion requires four files, one containing the model atom (with the superlevels, line transitions and energy limits), one with the photoionisation cross sections, a file with the Kurucz level data and another with the Kurucz line data, to allow the proper inclusion of line blanketing effects.

---
[2] http://nova.astro.umd.edu/Tlusty2002/tlusty-frames-data.html
[3] http://cdsweb.u-strasbg.fr/topbase/topbase.html

Many atmosphere calculations assume local thermodynamic equilibrium (LTE), where the gas is collisionally ionised, the thermodynamic variables of the system obey their classical relations, and the system is in ionisation equilibrium. Such an approximation considerably lightens the computational burden of the model calculation, and is an appropriate approximation for cooler stars. However, in hot white dwarfs, radiative ionisation is the dominant process. A series of studies in the mid-1990s (e.g. Lanz & Hubeny 1995; Lanz et al. 1996; Napiwotzki 1997) found that inclusion of non-LTE (NLTE) effects have the most significant effect on $T_{\text{eff}}$ measurements, more so than the inclusion of photospheric metals. These NLTE effects can easily be incorporated into model calculations using the TLUSTY program.

---

[4] http://physics.nist.gov/PhysRefData/ASD/lines_form.html

As already alluded to in Section 1.7.1, a white dwarf's $T_{eff}$ and log $g$ can be measured using models of the Balmer/Lyman line profiles. The strength of the absorption lines decrease as $T_{eff}$ increases, since more hydrogen atoms become ionised, leaving progressively less electrons in the n=2 state. Analogously, the line profile broadens as log $g$ increases, since density increases. This allows the energy level of each atom to be affected by its neighbouring ions and electrons, creating perturbations of the energy level structure that broaden the absorption feature, and the higher the density the more severe this effect. Grids of models are produced, with varying $T_{eff}$ and log $g$ values, which are compared to the observed data. The parameters of the model that best matches the observed data are adopted as those of the observed star.

Analogously, to measure metal abundances, grids of models are calculated with varying metal abundances. The metal abundance of the model that best reproduces the observed metal absorption line profiles (found at FUV and EUV wavelengths) is adopted as the metal abundance present in the star. The model fitting in this work was achieved using the XSPEC package (Arnaud, 1996), which utilised a $\chi_\nu^2$ minimisation technique based on a modified Levenberg-Marquardt algorithm.

## 1.9. Structure of thesis.

This thesis is split into six chapters. The proceeding chapter details the instruments used in this work. Chapter 3 examines the stratification of metals (particularly nitrogen) in some anomalous DA stars. Chapter 4 describes a search for circumstellar gas discs at a sample of hot DAs, while Chapter 5 presents a study of the origin of circumstellar absorption features in hot DA white dwarfs. Chapter 6 contains some concluding remarks and suggests the direction of future investigations in this field.

## 1.10. Summary.

- White dwarfs are the evolutionary end point of the vast majority of stars. They are compact objects (7 < log $g$ < 9), with degenerate cores, mainly hydrogen or helium dominated atmospheres, and are classified by their spectral characteristics.

- Metals are present in the photospheres of hot white dwarfs due to radiative levitation (and in some cases accretion). However, the metal abundances observed in hot white dwarfs are often not consistent with those predicted with radiative levitation theory. Some stars of similar $T_{eff}$ exhibit quite varied metal abundances, and some uncertainty exists in the distribution of metals in some stars. A thorough understanding of the abundance and distribution of metals in DAs with anomalous abundance measurements is desirable.

- Metal pollution from circumstellar discs is responsible for the photospheric metals seen in cooler white dwarfs where radiative levitation does not dominate the DA atmosphere. It is becoming widely accepted that these discs form via the tidal disruption of asteroids and minor planetary bodies in Solar System analogues. Gaseous components have also been detected at some discs, thought to be from vigorous collisions between dust particles. Searching for gas discs at hot DAs is an important task, since such discs could provide a polluting mechanism to account for the observed abundance anomalies.

- Non-photospheric metal absorption features are seen at some hot DA white dwarfs. Sources put forward for these features have been material in the potential wells of the stars, the ionisation of the ISM local to the stars, stellar mass loss and ancient PNe. A proper understanding of the origin of these non-photospheric absorption features is required to better understand the hot white dwarf circumstellar environment and the interaction of hot white dwarfs with the ISM in their locality.

# *Chapter 2.*

# Instruments used and white dwarf studied.

## 2.1. Introduction.

The instruments used to obtain the data studied in this thesis are outlined here. These include the Space Telescope Imaging Spectrograph (*STIS*) and Goddard High Resolution Spectrometer (*GHRS*) aboard the Hubble Space Telescope (*HST*), the International Ultraviolet Explorer (*IUE*), the Extreme Ultraviolet Explorer (*EUVE*), the Far Ultraviolet Explorer (*FUSE*) and the Intermediate dispersion Spectrograph and Imaging System (*ISIS*), at the William Herschel Telescope (*WHT*). Following these is a subsection detailing all of the stars studied in this thesis.

## 2.2. The Space Telescope Imaging Spectrograph (*STIS*).

*STIS* replaced the Faint Object Spectrograph (*FOS*), and was installed on the *HST* in 1997 during the second servicing mission by the astronauts Mark Lee and Steven Smith. In 2004 a fault developed in a circuit board, ending *STIS*'s use for some five years. The fourth servicing mission saw Michael Good and Mike Massimino replace the faulty part on 17$^{th}$ May 2009, after which the instrument performed almost as expected, when CCD radiation damage and changes in throughput during *STIS*'s downtime were taken into account.

Three detectors are present on *STIS*. The CCD has ~0.05 arcsecond square pixels, with a field of view (FOV) 52x52 arcsecond squared, and covers 2,000 – 10,300 Å. The near ultraviolet (NUV) $Cs_2Te$ Multi-Anode Microchannel Array (*MAMA*) covers 1,600 – 3,100 Å, and has ~0.024 arcsecond square pixels with a 25x25 arcsecond square FOV. The third detector, the far ultraviolet (FUV) solar-blind CsI *MAMA* (*FUV-MAMA*), has the same FOV and pixel size as the *NUV-MAMA*, but operates in the 1150 – 1700 Å range. The spectra used in this work were taken with *FUV-MAMA* detector using the E140M and E140H gratings, providing resolving powers of 40,000 and 110,000. The design and on-orbit performance of *STIS* are detailed by Woodgate et al. (1998) and Kimble et al. (1998).

## 2.3. The Goddard High Resolution Spectrograph (*GHRS*).

Replaced by the Near Infrared Camera and Multi-Object Spectrometer (*NICMOS*) during the second servicing mission in 1997, the Goddard High Resolution Spectrograph (*GHRS*) was one of *HST*'s original instruments. *GHRS* observed from 1150 – 3200 Å. The data used here were taken in the G160M mode, with a resolving power of 22,000.

## 2.4. Intermediate dispersion Spectrograph and Imaging System (*ISIS*).

Mounted on the 4.2m William Herschel Telescope (*WHT*), at the Isaac Newtown Group (*ING*) of Telescopes on La Palma, the Intermediate dispersion Spectrograph and Imaging System (*ISIS*) offers single slit spectroscopy over the 3,000 – 100,000 Å wavelength range. The R1200 grating was used in the observations carried out for the work in this thesis. With a central wavelength of 4899 Å, the blue arm observations (with the EEV12 chip) have a specific central wavelength resolution of 0.224 Å. The observations making use of the red arm of the spectrograph (the Red+ chip) have a specific central wavelength resolution of 2.41 Å for a central wavelength of 8600 Å.

## 2.5 The Far Ultraviolet Spectroscopic Explorer (*FUSE*).

The Far Ultraviolet Spectroscopic Explorer (*FUSE*), launched on 24$^{th}$ June 1999, operated until 18$^{th}$ October 2007. A detailed overview of the mission can be found in Moos et al. (2000), and the on-orbit performance of the observatory was reported by Sahnow et al. (2000).

Four co-aligned, prime focus telescopes and Rowland spectrographs with microchannel plate (MCP) detectors provided wavelength coverage from 905 – 1187 Å. Two telescope channels were coated with SiC to optimise the 905 – 1105 Å throughput and the channels covering 1000 – 1187 Å were covered with Al:LiF to optimise reflectivity. According to Sahnow et al. (2000), the velocity resolutions were ~17 km s$^{-1}$ in the 905 – 1105 Å region and ~13 km s$^{-1}$ in the 1000 – 1187 Å region.

Point source spectra often displayed a depression in flux, most noticeably spanning up to 50 Å of the LiF channel 1, detector segment B (1BLiF) spectra taken with the low resolution aperture (LWRS), caused by the shadow of the quantum efficiency (QE) grid wires. Given the time variability of these features (even within individual space craft orbits), they were termed 'worms', with the affected spectral region called the 'region of the worm' (Figure 2.1). The relationship between the precise position and alignment of the detector's QE grid wires, the optical elements, telescope illumination, pointing stability, the photon's wavelength and the instrument's optical design was highly complex, and the models used to calibrate for the worm failed to fully account for the effect; the minor errors in the telescope pointing and alignment caused the observed target to drift on the detector, causing the shadow's position to move, causing the worm to vary in both flux and wavelength space.

*FUSE* spectra also suffered severely with airglow emission ('sky lines'), caused by the excitation of H I, N I, N II, and O I in the Earth's upper atmosphere by solar activity (Feldman et al. 2001). This effect was more significant on dayside observations, when the observatory was orientated so that the path length of the light observed through the Earth's atmosphere was longer, and when the Sun was active. Given the variability of this effect, each data set needed to be inspected to establish its severity. In the most severe cases use of the data was limited.

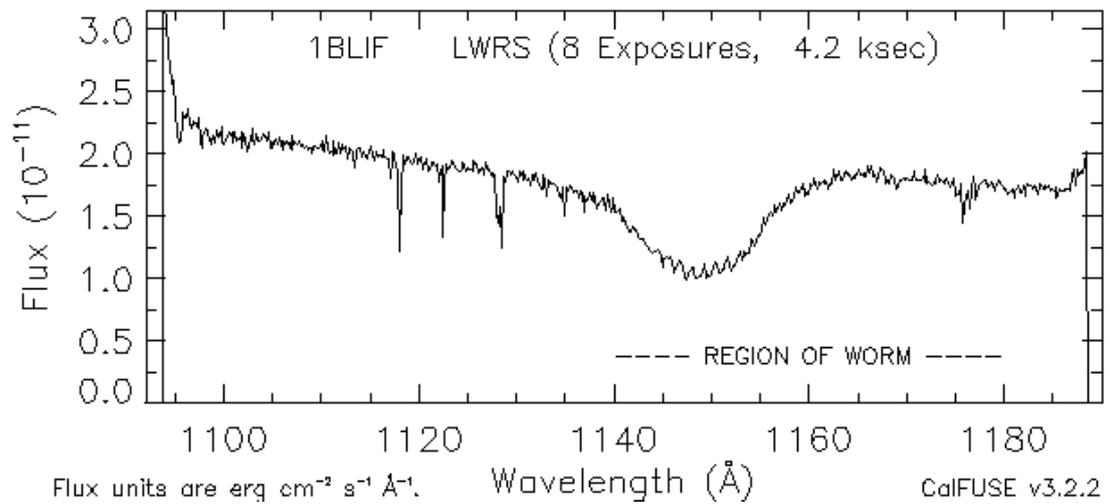

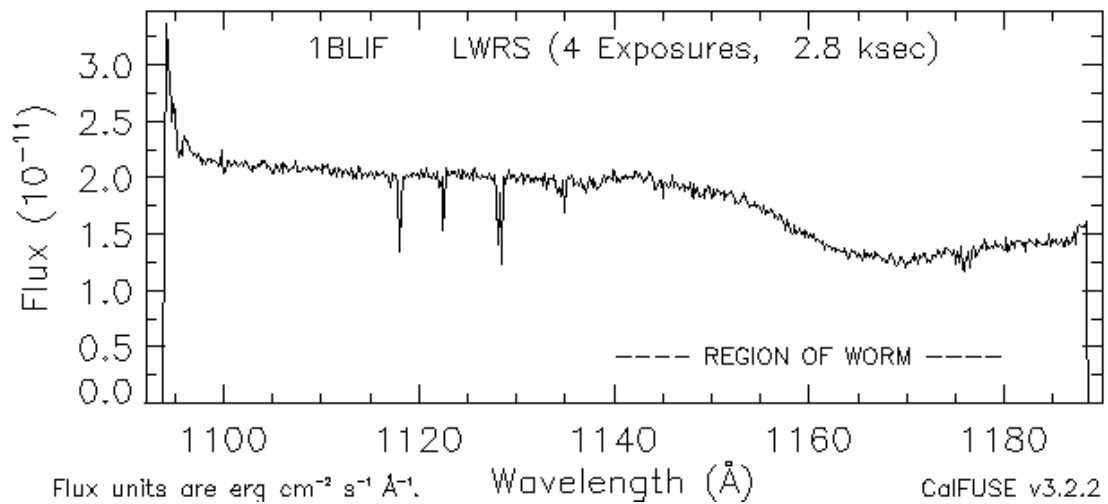

**Figure 2.1.** Two WD 0501+527 *FUSE* spectra, showing the 'region of the worm'. The upper panel shows the 1BLiF spectrum from observation M1010201000 (13[th] October 1999 01:25:31) and the lower panel is from observation M1030602000 (21[st] November 1999 11:39:56).

## 2.6. The Extreme Ultraviolet Explorer (*EUVE*).

The Extreme Ultraviolet Explorer (*EUVE*) was launched on 7$^{th}$ June 1992, stopped operation on 31$^{st}$ January 2001 and re-entered the Earth's atmosphere on 30$^{th}$ January 2002. *EUVE* observed in the 70 – 760 Å range, was the first dedicated EUV mission, and consisted of four photometric imaging systems and a three channel (short wave, SW, 70 – 190 Å; medium wave, MW, 140 – 380 Å; long wave, LW, 280 – 760 Å) spectrometer. The resolution of the three spectrometers was 0.5 Å, 1.0 Å and 2.0 Å for the SW, MW and LW channels, respectively. A complete sky survey was undertaken using the imaging systems, with the spectrometers being used for pointed spectroscopic observations of over 350 objects.

## 2.7. The International Ultraviolet Explorer (*IUE*).

Launched on 26$^{th}$ January 1978, the International Ultraviolet Explorer (*IUE*) is the oldest source of data used here. Greatly exceeding its expected operational lifetime of 3 years, *IUE* was finally shut down on 30$^{th}$ September 1996 after obtaining over 104,000 spectra. *IUE*'s spectrometers covered the 1,200 – 3,350 Å wavelength region at high (0.1 – 0.3 Å) and low (6 – 7 Å) resolutions. Two spectrographs were present on *IUE*, covering a short wavelength range (SW; 1,150 – 2,000 Å) and a long wavelength range (LW; 1,850 – 3,300 Å). Each spectrograph had two cameras, the prime (P) camera and a redundant (R) camera, carried in case of a failure of the prime camera. The data used here was taken in the SWP mode, with a resolving power of 20,000.

## 2.8. White dwarfs studied.

Table 2.1 details all of the white dwarfs studied in this thesis, their $T_{\text{eff}}$, the property of interest and the instruments from which the data analysed in each investigation was obtained. *Spitzer* infrared observations are included that, though not analysed here, prove useful in interpreting the results of the analyses.

**Table 2.1.** The white dwarfs studied here, their stellar parameters, interesting properties and the chapters in which they are investigated, and sources of data.

| WD | Alt. name | $T_{\text{eff}}$ (K) | Chapter | Properties of interest | Source(s) of data |
|---|---|---|---|---|---|
| 0004+330 | GD 2 | 47,219* | 4 | Circumstellar gas disc? | ISIS; Spitzer IRAC*‡ |
| 0050−335 | GD 659 | 35,660±135 | 3 | Stratified nitrogen? | HST STIS; EUVE |
| 0209+085 | HS 0209+083 | 36,000† | 5 | Circumstellar high ions? | IUE; HST STIS; Spitzer IRAC** |
| 0232+035 | Feige 24 | 60,487±1,100 | 4 | Circumstellar gas disc? | WHT ISIS |
| 0455−282 | REJ 0457−281 | 50,960±1,070 | 5 | Circumstellar C IV | HST STIS |
| 0501+527 | G191-B2B | 52,500±900 | 5 | Circumstellar C IV, NV, Si IV | IUE; Spitzer IRAC*‡ |
| 0556−375 | REJ 0558−373 | 59,508±2,200 | 5 | Circumstellar C IV, Si IV | HST STIS; Spitzer IRAC*‡ |
| 0621−376 | REJ 0623−371 | 58,200±1,800 | 5 | Circumstellar C IV | HST STIS |
| 0939+262 | Ton 021 | 69,711±530 | 5 | Circumstellar high ions? | IUE; Spitzer IRAC*‡ |
| 0948+534 | PG 0948+534 | 110,000±2500 | 5 | Circumstellar C IV, Si IV | HST STIS; Spitzer MIPS*** |
| 1029+537 | REJ 1032+532 | 44,350±715 | 3 | Stratified metals? | HST STIS |
| 1057+719 | PG 1057+719 | 39,770±615 | 5 | Stratified nitrogen? | HST STIS; EUVE |
| 1123+189 | PG 1123+189 | 54,574±900 | 5 | Circumstellar high ions? | HST STIS |
| 1254+223 | GD 153 | 39,290±340 | 5 | Circumstellar high ions? | HST GHRS |
| 1314+293 | HZ 43 | 50,370±870 | 5 | Circumstellar high ions? | HST STIS |
| 1337+705 | EG 102 | 22,090±85 | 5 | Circumstellar high ions? | IUE; Spitzer IRAC*‡ |
| 1611−084 | REJ 1614−085 | 38,840±480 | 3 | Circumstellar high ions? | IUE |
| 1738+669 | REJ 1738+665 | 66,760±1,230 | 4 | Stratified nitrogen? | IUE; Spitzer IRAC*‡ |
| 1942+499 | REJ 1943+50 | 34,056* | 5 | Circumstellar gas disc? | HST GHRS |
| 2023+246 | Wolf 1346 | 19,150±30 | 5 | Circumstellar high ions? | WHT ISIS; Spitzer IRAC*‡ |
| 2111+498 | GD 394 | 39,290±360 | 4 | Circumstellar gas disc? | HST STIS; Spitzer MIPS*** |
| 2152−548 | REJ 2156−546 | 45,500±1,085 | 5 | Circumstellar gas disc? | WHT ISIS; Spitzer IRAC*‡ |
|  |  |  |  | Circumstellar gas disc? | WHT ISIS |
|  |  |  |  | Circumstellar high ions? | IUE |
|  |  |  |  | Circumstellar gas disc? | WHT ISIS; HST GHRS; Spitzer IRAC*‡ |
|  |  |  |  | Circumstellar high ions? | IUE; HST GHRS; Spitzer IRAC*‡ |
|  |  |  |  | Circumstellar high ions? | HST STIS |

Note: All $T_{\text{eff}}$ values are from Barstow et al. (2003b) unless stated otherwise. *Mullally et al. (2007) and references therein; †Jordan et al. (1993); ‡Observations discussed, not analysed, **Chu et al. (2011)

**Table 2.1**-*continued*

| WD | Alt. name | $T_{eff}$ (K) | Chapter | Properties of interest | Source(s) of data |
|---|---|---|---|---|---|
| 2206+250 | REJ 2207+25[*] | 26,954 | 4 | Circumstellar gas disc? | *WHT ISIS* |
| 2211−495 | REJ 2214−492 | 61,613±2,300 | 5 | Circumstellar high ions? | *IUE* |
| 2218+706 | WD 2218+706 | 58,582±3,600 | 5 | Circumstellar high ions? | *IUE* |
| 2309+105 | GD 246 | 51,308±850 | 4 | Circumstellar gas disc? | *WHT ISIS* |
| | | | 5 | Circumstellar high ions? | *IUE; HST GHRS* |
| 2331−475 | REJ 2334−471 | 53,205±1,300 | 5 | Circumstellar high ions? | *IUE* |

*Napiwotzki, Green & Saffer (1999).

# *Chapter 3*

# Stratified metals in hot white dwarf atmospheres?

## 3.1. Introduction.

Inconsistent results have been obtained from investigations into the stratification of metals in the DAs WD 1029+537, WD 1614–084 and WD 0050–332. Some evidence exists for similar metal stratification in WD 0948+534 (Section 1.7.2). This is of particular interest, as the extraordinary metal abundances of these stars have been linked to the presence of circumstellar material by some authors (e.g. Section 1.7.3). Therefore, a clear understanding of what the photospheric abundances and distributions of such metals are is needed, to see if novel metal distributions exist in the stars in question.

The observations and methods used to examine the metal distributions in the stars are detailed in Section 3.2. The results for the stars in which nitrogen was studied are presented in sections 3.3–3.5; an examination of the metals in WD 0948+534 follows in Section 3.6. The results are discussed in 3.7 with a summary in Section 3.8.

## 3.2. Observations and method.

The observations used here are listed in Table 3.1. Models of WD 1029+537, WD 1611–084 and WD 0050–332 were computed using TLUSTY, with the $T_{eff}$, log $g$ and carbon, oxygen and silicon abundances in Table 3.1. Model ions for C III, C IV, N III, N IV, N V, O IV, O V, O VI, Si III, and Si IV were explicitly included (i.e. full model ions with all transitions and oscillator strengths, etc, were used), while C V, N VI, O VI and Si V were treated as one level ions. One level ions are used for the highest ion of each element to provide a simpler method of ionisation balance calculation without having to consider detailed partition functions for the highest ions (which for massive ions can become rather complex and cumbersome); in stars with high $T_{eff}$ and massive, non-hydrogenic elements the complex transitions that occur in the high ions of such heavy elements are not included in the models, in an analogous fashion to the use of superlevels. In a similar situation to the use of superlevels, these approximations may lead to some inaccuracies in atmosphere calculations where highly ionised, massive ions are present. Model grids were constructed to cover a range of nitrogen abundances spanning those measure by both Barstow et al. (2003b) and Chayer, Vennes & Dupuis (2005), rather than focusing on just the high abundance regime as in Barstow et al. (2003b), to allow a full examination of the evolution of the N V doublet and EUV continuum with nitrogen abundance.

FUV spectra were synthesised using SYNSPEC, from 1235 – 1245 Å to cover the N V doublet. XSPEC was used to fit the model spectra to the data. When a best fitting nitrogen abundance was found, a set of stratified models were constructed, with the best fitting nitrogen abundance at the top of the atmosphere and an abundance of zero below. In each model, the depth of this layer was extended, allowing a model grid to be constructed to investigate whether a stratified metal configuration better represented the data than a homogeneous one (it must be emphasised that the stratification of the metals here was not self-consistently formulated like the depth dependent models of Dreizler 1999, Dreizler & Wolff 1999, Schuh, Dreizler & Wolff 2002 and Schuh, Barstow & Dreizler 2005, but were 'exploratory' stratified models such as those of Hubeny & Holberg 1998 and Holberg et al. 1999a). The EUV analysis was conducted in the *EUVE* SW range (80 – 180 Å), since the EUV continuum absorption is most severe in this region. Again, XSPEC was used to fit the data. Interstellar absorption was charaterised using the column densities quoted in Table 3.1.

The FUV absorption features (Table 3.2) in the spectrum of WD 0948+534 were fit using the model grids used by Barstow et al. (2003b). Once initial abundance estimates were made, models with stratified metals were produced for C, N, O and Si in the same way as the nitrogen model grids constructed for the other three stars. The iron and nickel abundances measured by Barstow et al. (2003b) were included in all models of this object. Fe IV, Fe V, Fe VI, Ni IV, Ni V and Ni VI were included explicitly, with Fe VII and Ni VII treated as one level ions. Spectra around the wavelengths stated in Table 3.2 were again synthesised using SYNSPEC.

**Table 3.1.** The observation information, stellar and ISM parameters for the white dwarfs studied here (unless stated otherwise, all data are from Barstow et al. 2003b). The absence of data signifies where a measurement was unobtainable, either due to lack of spectral coverage or an inability to model the absorption features.

| | WD | 0050–332 | 0948+534 | 1029+537 | 1611–084 |
|---|---|---|---|---|---|
| STIS data | Mode | E140H | E140M | E140M | G160M |
| | $\lambda\lambda$ (Å) | 1159, 1357 | 1150, 1729 | 1150, 1729 | 1386, 1423 |
| EUVE data | Mode | SW | | SW | |
| | $\lambda\lambda$ (Å) | 80, 180 | | 80, 180 | |
| $T_{\rm eff}$ (K) | | 35,660 | 110,000 | 44,350 | 38,840 |
| log $g$ | | 7.93 | 7.58 | 7.81 | 7.92 |
| C III/H | | 0.00 | | 3.00x10$^{-7}$ | |
| C IV/H | | 5.00x10$^{-8}$ | | 1.60x10$^{-7}$ | 7.00x10$^{-7}$ |
| N V/H | | 6.30x10$^{-4}$ | | 5.00x10$^{-5}$ | 2.50x10$^{-4}$ |
| N V/H[a] | | 6.31x10$^{-6}$ | | 1.26x10$^{-6}$ | 6.31x10$^{-7}$ |
| O V/H | | 0.00 | | 1.20x10$^{-7}$ | |
| Si IV/H | | 4.80x10$^{-9}$ | | 9.50x10$^{-7}$ | 9.50x10$^{-9}$ |
| Fe V/H | | 0.00 | 1.90x10$^{-6}$ | 0.00 | 0.00 |
| Ni V/H | | 0.00 | 1.20x10$^{-7}$ | 0.00 | 0.00 |
| log($N_{\rm H\,I}$) (cm$^{-2}$) | | 18.46[b] | | 18.62[c] | |
| log($N_{\rm He\,I}$) (cm$^{-2}$) | | 17.37[b] | | 17.75[c] | |
| log($N_{\rm He\,II}$) (cm$^{-2}$) | | 17.17[b] | | 17.28[c] | |

[a]Chayer et al. (2005), [b]Barstow et al. (1997), [c]Holberg et al. (1999b).

**Table 3.2.** The laboratory wavelengths of the FUV absorption features examined here[5].

| Ion | Lab. Wavelength (Å) |
|---|---|
| C IV | 1548.187, 1550.772 |
| N V | 1238.821, 1242.804 |
| O V | 1371.296 |
| Si IV | 1393.755, 1402.770 |

## 3.3. WD 1029+537.

The FUV N V doublet yields a best fitting nitrogen abundance of $3.39^{+1.29}_{-1.31} \times 10^{-7}$ with the homogeneous grid (Figure 3.1), with a $\chi^2_\nu$ of 0.58. A secondary minimum was also found at $5.00^{+2.55}_{-0.48} \times 10^{-5}$, with a $\chi^2_\nu$ of 0.77. Forcing the nitrogen to a stratified configuration does not produce a significant improvement over the lower abundance, homogeneous fit.

An analysis was also performed of the *EUVE* data of this object. To account for the interstellar absorption observed in the *EUVE* dataset, the hydrogen and helium columns from Table 3.1 were used in the modelling process. The EUV spectrum generated using a homogeneous N/H of $3.39 \times 10^{-7}$ is shown in Figure 3.2; the data are represented well when compared to the Holberg et al. (1999a) homogeneous model (Figure 1.6, lower curve).

---

[5] wavelengths taken from the Kurucz database (http://www.pmp.uni-hannover.de/cgi-bin/ssi/test/kurucz/sekur.html).

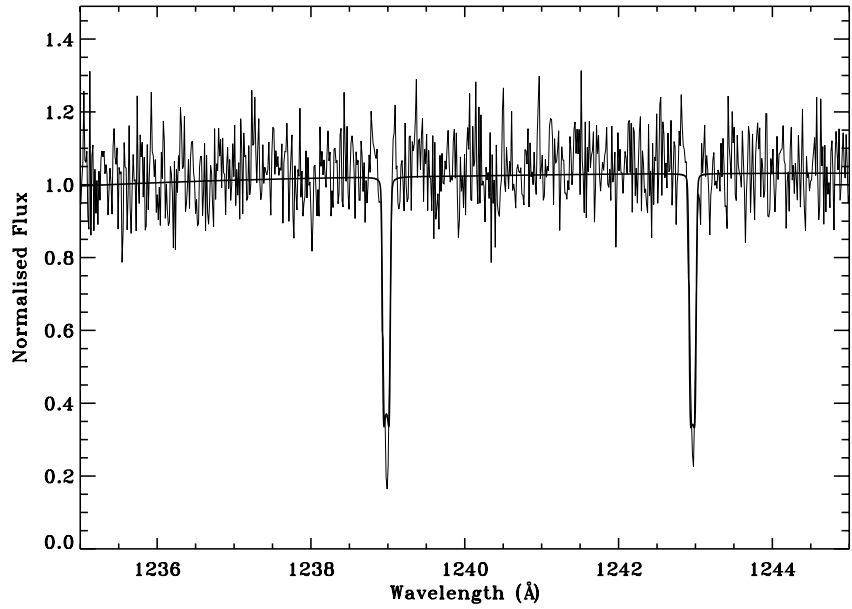

**Figure 3.1.** The best fitting model (N/H = $3.39 \times 10^{-7}$) of the N V doublet of WD 1029+537.

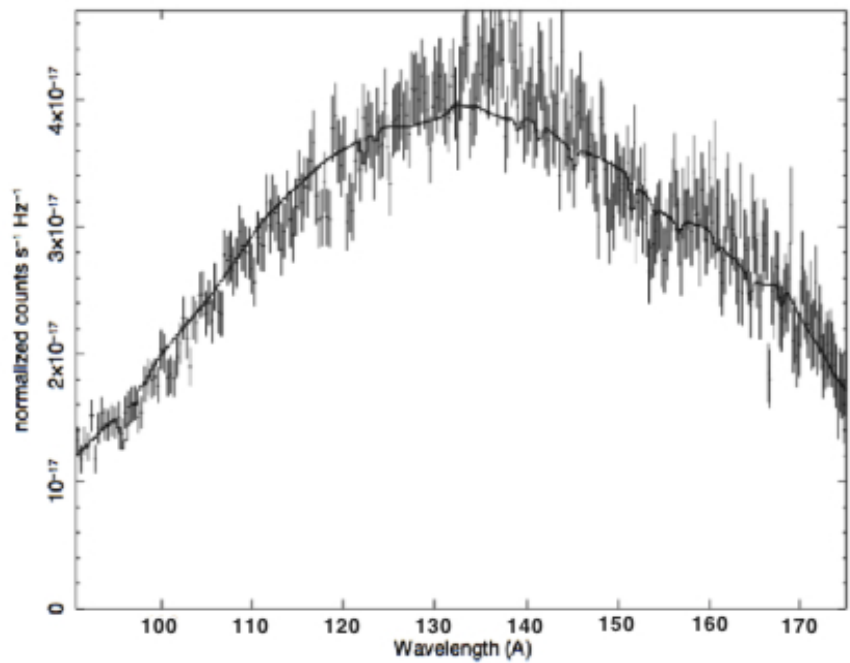

**Figure 3.2.** The *EUVE* SW data for WD 1029+537 fit with a homogeneous nitrogen distribution with N/H = $3.39 \times 10^{-7}$.

## 3.4. WD 1611–084.

Like WD 1029+537, two best fitting abundances are seen in the N V doublet fits of WD 1611–084 (Figure 3.3). The higher abundance model (N/H = $3.41^{+1.81}_{-1.50} \times 10^{-4}$) produces the global $\chi^2_v$ minimum ($\chi^2_v$ = 1.13; Figure 3.3, upper panel), while the secondary best fit was found at N/H = $1.76^{+1.65}_{-1.26} \times 10^{-6}$ ($\chi^2_v$ = 1.21; Figure 3.3, lower panel). The lower abundance model over predicts the depths of observed line profiles, while the N V absorption features of the higher abundance model do not quite extend into the observed line profiles. No *EUVE* data exists for this object.

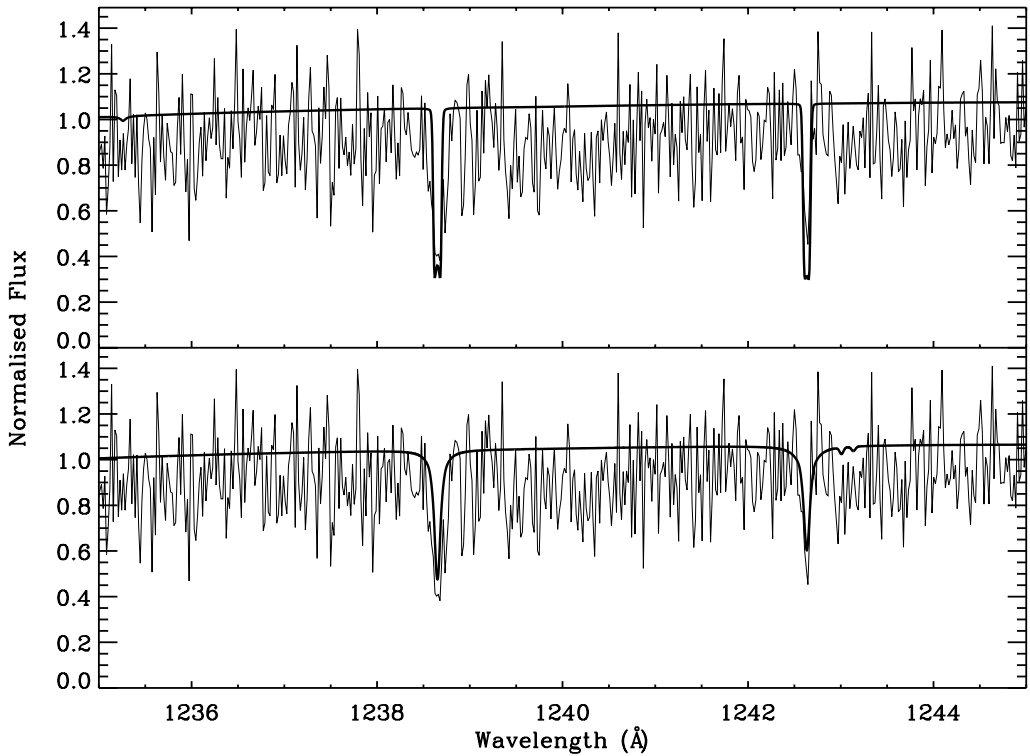

**Figure 3.3.** The lower nitrogen abundance model (N/H = $1.76 \times 10^{-6}$, $\chi^2_v$ = 1.21) of WD 1611–084 is shown in the upper panel. The lower panel shows the high abundance model (N/H = $3.41 \times 10^{-4}$; $\chi^2_v$ = 1.13).

## 3.5. WD 0050–332.

In keeping with the results of WD 1029+537, WD 0050–332 is better fit with a lower abundance, homogeneous model. The overall best fit is found at N/H = $6.05^{+0.64}_{-0.62} \times 10^{-7}$ ($\chi^2_\nu$ = 2.15; the best fitting model of the N V doublet is shown in Figure 3.4), with a secondary best fit at $5.70^{+0.41}_{-0.62} \times 10^{-5}$ ($\chi^2_\nu$ = 2.41). Again, the higher abundance, stratified model does not offer any improvement over the lower abundance, homogeneous model. The *EUVE* data are again well fit by the lower abundance, homogeneous model.

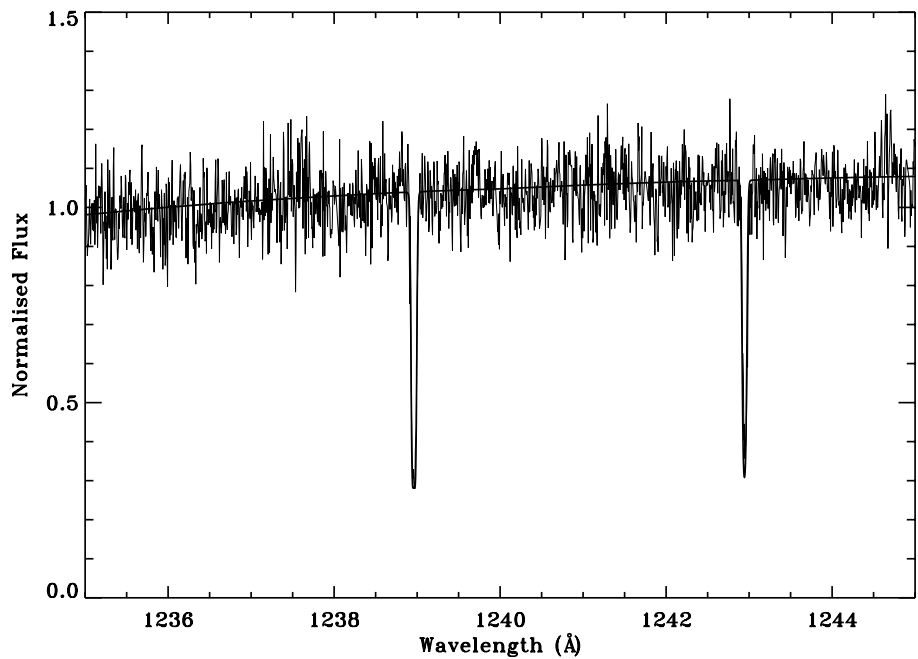

**Figure 3.4.** The best fitting model (N/H = $6.05 \times 10^{-7}$, $\chi^2_\nu$ = 2.15) of the N V doublet of WD 0050–332.

## 3.6. WD 0948+534.

The line profiles of the best-fit models do not reproduce the data well (e.g. Figure 3.5). The best fitting abundances are given in Table 3.3; above these abundances the model line profiles do not reproduce the narrow widths of the observed lines, and the NLTE core emission developing in the short wavelength component of the N V doublet in Figure 3.5 becomes more severe (due to the emission caused by the temperature inversion seen in NLTE atmospheres; Lanz & Hubeny, 1995). Stratification of the metals does not improve the model fits.

**Table 3.3.** The estimated metal abundances for WD 0948+534. Given the poor match to the data, errors were not computed for these models.

| C IV | N V | O V* | Si IV |
|---|---|---|---|
| $4.85 \times 10^{-6}$ | $1.6 \times 10^{-6}$ | $3.5 \times 10^{-5}$ | $3.15 \times 10^{-5}$ |

*this abundance is the grid upper limit

## 3.7. Discussion.

The modelling of nitrogen in the photospheres of WD 1029+537, WD 1611–084 and WD 0050–332 displays a degeneracy, where more than one minimum is present in the $\chi_v^2$ distribution of the fit (the abundances adopted in this study are given in Table 3.4; the reasons for those abundance choices are detailed in the following discussion). The absorption line profiles of WD 0948+534 are not well modelled here.

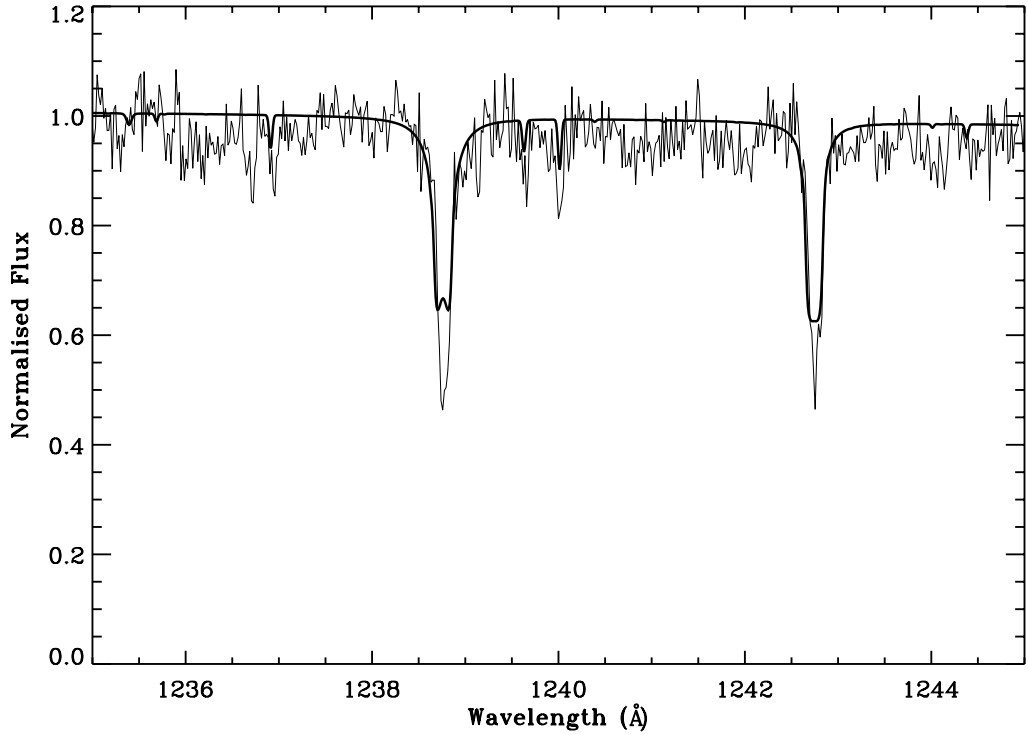

**Figure 3.5.** The N V doublet of WD 0948+534, fit with a model with N/H = $1.6 \times 10^{-6}$.

The $\chi_v^2$ distribution of WD 1029+537 is shown in Figure 3.6. The upper panel shows the full $\chi_v^2$ distribution, with a clear, global minimum at N/H = $3.39 \times 10^{-7}$ ($\chi_v^2$ = 0.58). A secondary minimum is seen at N/H = $5.00 \times 10^{-5}$ ($\chi_v^2$ = 0.77). The minima are shown in more detail in the lower panel of Figure 3.6. The global and secondary minima are clearly separated, with the secondary minimum well above the 3σ confidence limit (dotted line) of the global minimum (dashed line). The *EUVE* SW data are also well fit with the lower abundance model. A similar situation is seen at WD 0050–332, with a $\chi_v^2$ global minimum at N/H = $6.05 \times 10^{-7}$.

**Table 3.4.** The best fitting nitrogen abundances for WD 1029+537, WD 1611–084 and WD 0050–332 adopted in this study.

| Star | N/H |
|---|---|
| WD 1029+537 | $3.39^{+1.29}_{-1.31} \times 10^{-7}$ |
| WD 1611–084 | $1.76^{+1.65}_{-1.26} \times 10^{-6}$ |
| WD 0050–332 | $6.05^{+0.64}_{-0.62} \times 10^{-7}$ |

Two minima are also seen in the $\chi_v^2$ distribution of WD 1611–084 (Figure 3.7), although they were less easy to disentangle. The lower panel of Figure 3.7 shows that the difference between the two $\chi_v^2$ minima is not greater than 3σ confidence limit of the global minimum. Indeed, unlike WD 1029+537 and WD 0050–332, the global $\chi_v^2$ minimum for WD 1611–084 is found in the higher abundance regime (N/H = $3.41^{+1.81}_{-1.50} \times 10^{-4}$). Poorly defined minima can also be seen at N/H = $10^{-5}$ and $7 \times 10^{-3}$. Since WD 1029+537 and WD 0050–332 are well explained using a low abundance, homogeneous nitrogen model, the lower abundance model was also adopted for WD 1611–084. Data with a better signal to noise ratio may allow this degeneracy to be broken with more certainty.

An examination of the model NLTE population responsible for the N V doublet ($^2$S) and the N VI level shows that as the nitrogen abundance in WD 1029+537, WD 1611–084 and WD 0050–332 is increased, the value of log(N V($^2$S)/ N VI) begins to decrease above N/H = ~$10^{-6}$ (Figure 3.8). This 'over-ionisation' of nitrogen may provide a mechanism for the apparent degeneracy to arise;

as the nitrogen abundance increases the relative amount of N V ($^2$S) to N VI decreases. The trend seen in the log(N V($^2$S)/ N VI) of WD 1029+537 and WD 0050–332 is also seen in WD 1611–084, providing further support for a lower N/H in WD 1611–084.

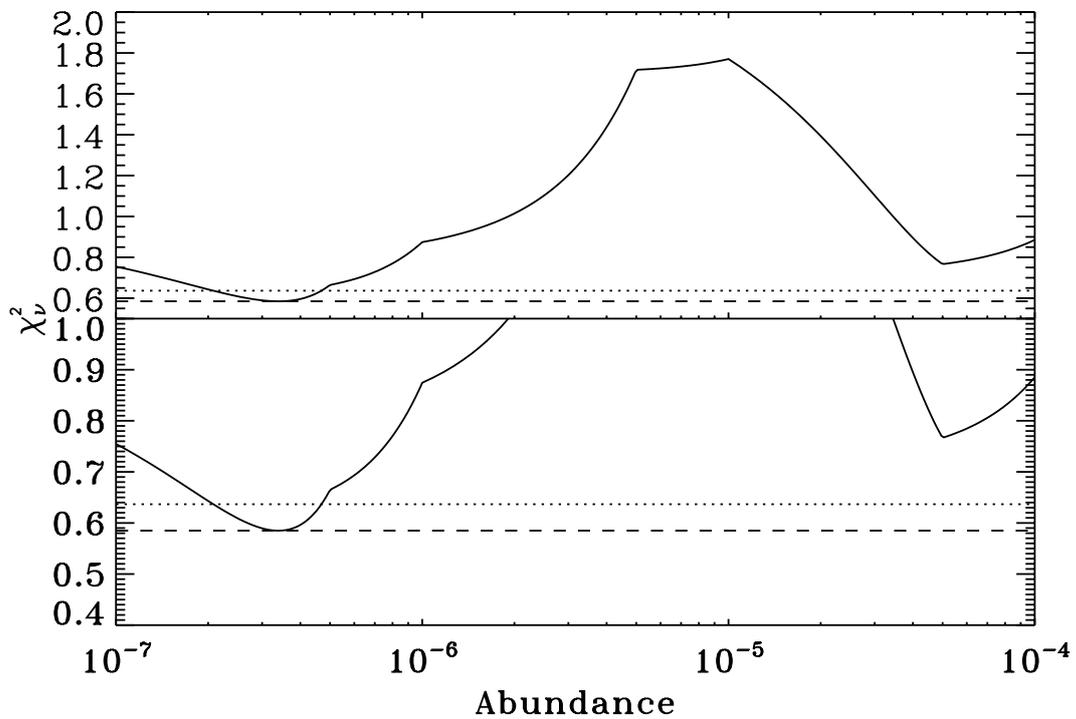

**Figure 3.6.** The $\chi^2_\nu$ distribution of WD 1029+537 as N/H is increased. The global minimum is represented with a dashed line, and its 3σ confidence limit is denoted with a dotted line.

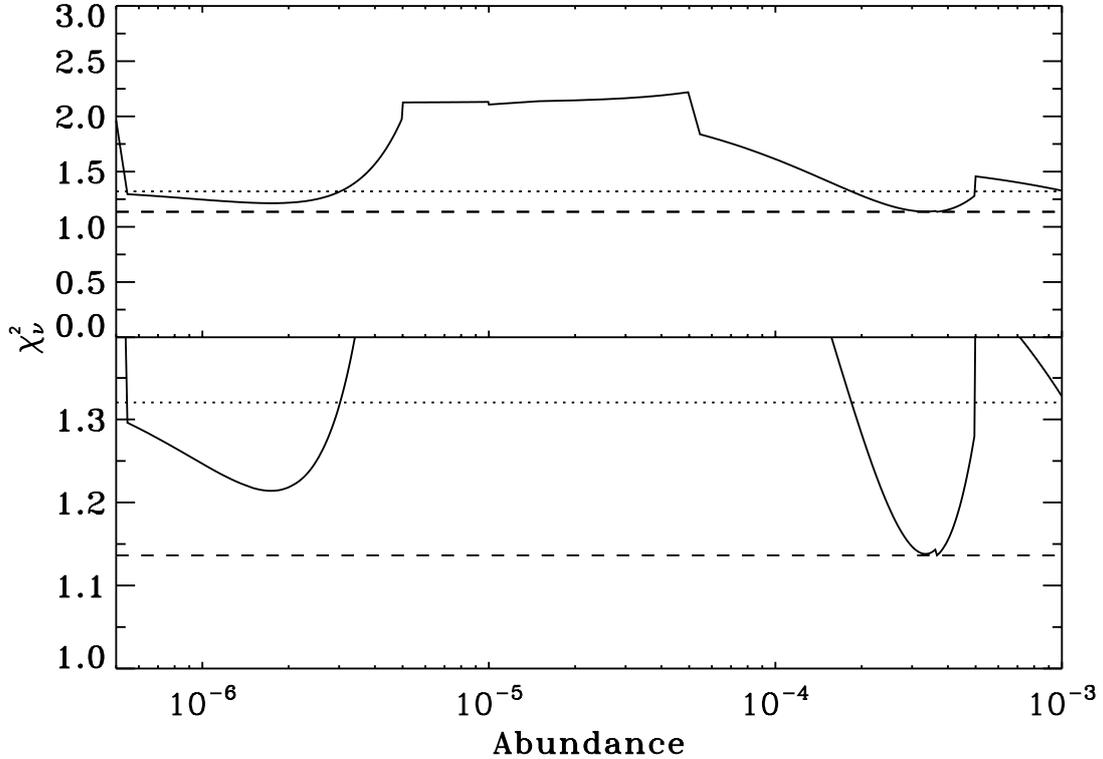

**Figure 3.7.** The $\chi_\nu^2$ distribution of WD 1611–084. The line representations are the same as in the previous figure.

Papers (e.g. Holberg et al. 1999a) that report high abundance, stratified nitrogen in the spectra of the stars cite radiatively driven mass loss acting on the photospheric nitrogen, enriching the upper atmosphere. Studies of mass loss in DA white dwarfs suggest that the radiative acceleration in white dwarfs in the $T_{\text{eff}}$ range studied here cannot overcome the log $g$ of the stars (Unglaub 2007, 2008). Given that no circumstellar nitrogen is seen at WD 0050–332, WD 1029+537 and WD 1611–084 (Bannister et al. 2003; Chapter 6 of this thesis), there is no mechanism for the enrichment of the upper atmospheres of these stars, further strengthening the argument that the stars have homogeneous nitrogen distributions. Another reason for the degeneracy seen here may be the use of the one level ions for the highest ionisation stages (though were to be an issues one may expect to see this effect in the absorption features in hotter stars). The development of more complex model ions is a subject of ongoing research; as this work progresses tests of more detailed model ions on stars such as these may provide the observed absorption line profiles.

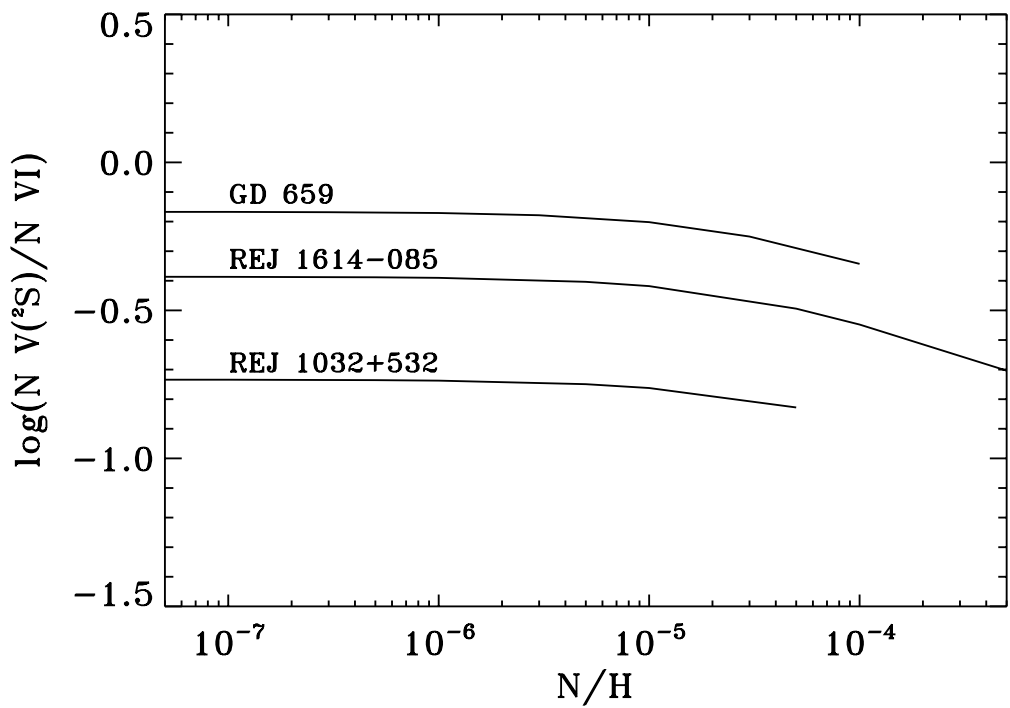

**Figure 3.8.** The change in the log of the ratio of the NLTE population responsible for the N V doublet ($^2$S) to the N VI level population, with nitrogen abundance. Note that not all of the model grids span the same abundance range; models were only computed over the range required to explain the observations.

A comparison of the nitrogen abundances found here to those measured by Barstow et al. (2003b) and Chayer, Vennes & Dupuis (2005) is shown in Figure 3.9.

The new results are in keeping with those found by Barstow et al. (2003b), for the stars with higher $T_{eff}$. The trend of increasing N/H with decreasing $T_{eff}$ seen by Barstow et al. (2003b) is not seen. The abundances measured by Chayer, Vennes & Dupuis (2005) are closer to those found here (although they are offset, probably due to systematic differences in the fitting procedure used by Chayer, Vennes & Dupuis, 2005). Figure 3.10 shows a comparison of the N/H values derived here to the radiative levitation predictions of Chayer et al. (1995); the abundances found here are closer to those predicted than those of Barstow et al. (2003b), where the stars had a stratified, high nitrogen abundance, with a trend of decreasing abundance with $T_{eff}$.

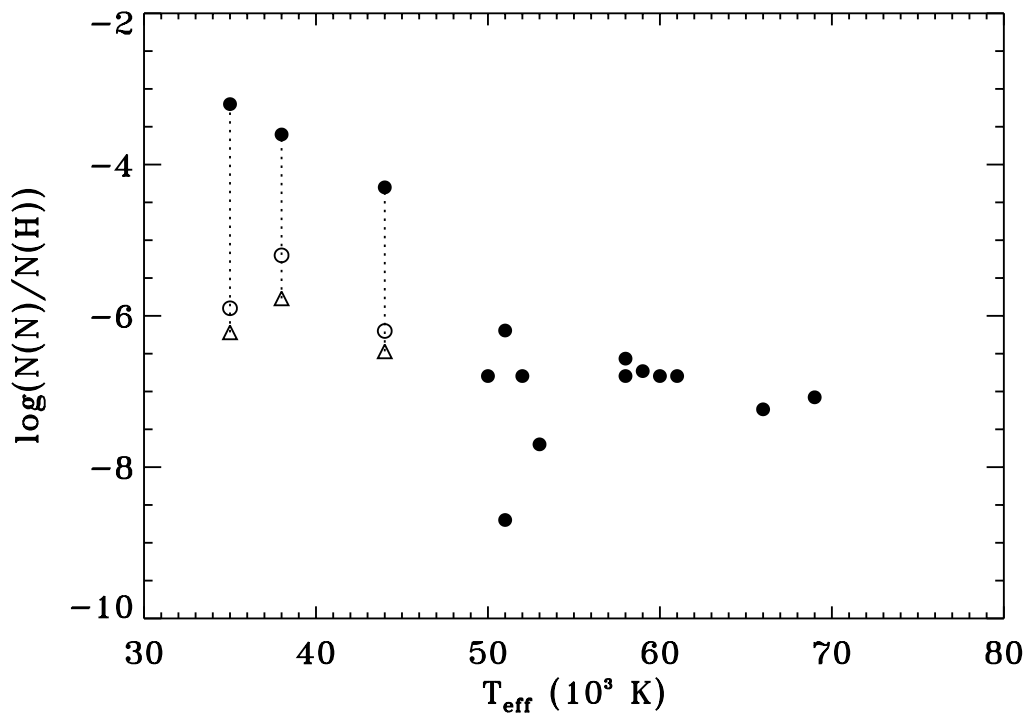

**Figure 3.9.** A comparison of the nitrogen abundances found here (triangles) to those found by Barstow et al. (2003b; filled circles) and Chayer et al. (2005; open circles). The dotted lines connect multiple measurements of individual stars for ease of comparison.

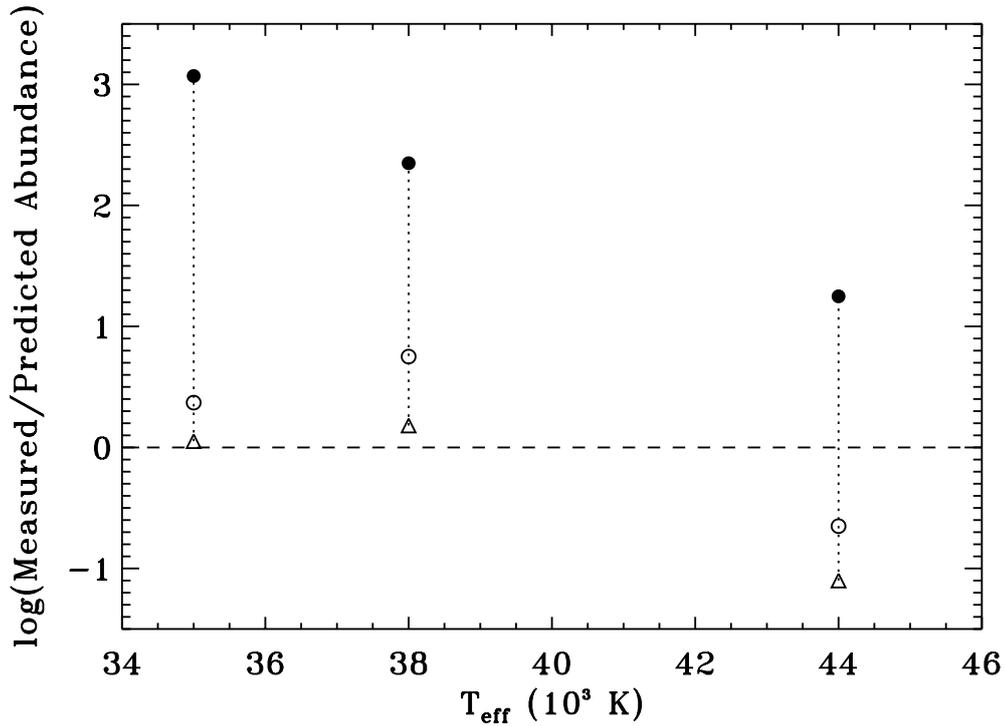

**Figure 3.10.** A comparison of the N/H values found here, by Barstow et al. (2003b) and Chayer et al. (2005), to the radiative levitation predictions of Chayer et al. (1995). The plot symbols are the same as in the previous figure.

Since homogeneous models with unremarkable abundances explained the observed N V doublets of WD 1029+537, WD 1611–084 and WD 0050–332 well, it is perhaps unsurprising that stratified models with high metal abundances failed to reproduce the line profiles of metallic high ions in WD 0948+534. Indeed, the $\chi_\nu^2$ distributions for the C IV, N V, O V and Si IV abundances of WD 0948+534 had single minima at the abundances detailed in Table 3.3. One possible explanation is that a physical effect that becomes significant at high $T_{eff}$ may be missing from the models. Given the much higher $T_{eff}$ of WD 0948+534 when compared to the other stars, this omission may cause the model-observation disagreement. The lack of detailed model ions for high ionisation stages may also cause a problem (particularly for Fe peak elements), as the lack of detailed ionisation and excitation equilibrium calculations may have a significant effect on the atmospheric structure. In addition to this, the lack of inclusion of small amounts of other heavier elements, such as Ar VII, Ge IV and possibly Sn IV that have recently been found in this DA (Werner, Rauch & Kruk, 2007) and others such as WD 0501+527 (Vennes, Chayer & Dupuis 2005), may also have an effect on the atmosphere model calculations, should the combined effect of all these ions on the structure of the atmosphere be significant. As more exotic metals are discovered in hot DA spectra and more detailed model ions become available they should be included in model calculations of this star to test this possibility. Another possibility is that some other material is present along the line of sight to the white dwarf, with a velocity unresolved from that of the photosphere (a hint of a secondary component can be seen in the C IV doublet near 1548.3 Å and 1550.9 Å; Figure 1.10). This is explored in Chapter 5.

## 3.8. Summary.

- Previous studies came to conflicting conclusions as to the abundance and distribution of photospheric nitrogen in WD 1029+537, WD 1611–084 and WD 0050–332.

- These stars were re-examined, and a degeneracy in nitrogen abundance is seen, with more than one minimum in the nitrogen abundance $\chi_v^2$ distribution. The N V doublets of WD 1029+537 and WD 0050–332 are better fit with lower abundance, homogeneous nitrogen. The degeneracy is difficult to break in the case of WD 1611–084. However, the lower abundance model is assumed.

- An examination of the model NLTE populations shows that as N/H increases, the values of log(N V($^2$S)/N VI) decrease in all three stars. This could explain the observed degeneracy.

- Mass loss is predicted to not occur at these stars, and circumstellar material is not present from which the upper atmospheres of the stars can be enriched, strengthening the argument for homogeneous, lower abundance nitrogen distributions.

- The abundances measured here follow the predictions of radiative levitation theory better than those measured by Barstow et al. (2003b).

- The high ion absorption features of WD 0948+534 cannot be satisfactorily matched with the model sets used here. This could be to an unaccounted for high $T_{eff}$ effect, or unresolved, non-photospheric C IV along the sight line to the star.

# *Chapter 4.*

# A search for circumstellar gas discs at hot white dwarfs.

## 4.1. Introduction.

In light of the unexplained, anomalous metal abundances at some white dwarfs with $T_{\text{eff}}$ > 20,000 K, it is desirable to study the circumstellar environment for evidence of polluting material. Indeed, this has provided a robust explanation for the presence of photospheric metals at cooler DA stars. Here, a search for circumstellar gas discs such as that seen at SDSS 1228+1040 (Section 1.7.3) at a sample of hot DA stars with anomalous metal abundances or UV circumstellar absorption features (Section 1.7.3) is presented.

Section 4.2 details the objects observed, the instruments used and the data reduction methods implemented. The resultant spectra are shown in Section 4.3. These results and their implications are discussed in Section 4.4, with a short summary of the work in 4.5.

## 4.2. Observations and data reduction.

The white dwarfs observed (selected due to their unusual metal abundances) are detailed in Table 4.1. The observations were made by M. Burleigh on the nights of the 6$^{th}$ and 7$^{th}$ of August 2007, at the William Herschel Telescope (WHT) on La Palma. Both the red and blue arms of the ISIS spectrometer were used, with the R1200 grating. The data were reduced in IRAF, using standard subroutines for the reduction of long slit spectroscopy.

**Table 4.1.** The white dwarfs observed in this study. The right ascension (RA) and declination (DEC) (J2000) of each star are from the McCook & Sion online catalog[6]. The $T_{eff}$ values are from the references detailed.

| WD number | Alt. name | $T_{eff}$ (K) | RA | DEC |
|---|---|---|---|---|
| WD 0004+330 | GD 2 | 47,219[a] | 00 07 32.3 | +33 17 27 |
| WD 0209+085 | HS 0209+083 | 36,000[b] | 02 12 04.8 | +08 46 51 |
| WD 1614–084 | REJ 1614–085 | 38,840±480[c] | 16 14 19.0 | –08 33 27 |
| WD 1942+499 | REJ 1943+50 | 34,056[a] | 19 43 42 | +50 04 41 |
| WD 2032+248 | Wolf 1346 | 19,150±30[c] | 20 34 22.3 | +25 03 58 |
| WD 2111+498 | GD 394 | 39,290±360[c] | 21 12 43.8 | +50 06 17 |
| WD 2205+250 | REJ 2207+25 | 26,964[d] | 22 07 44.8 | +25 20 22 |
| WD 2309+105 | GD 246 | 51,308±850[c] | 23 12 21.5 | +10 47 04 |

[a]Mullally et al. (2007) and references therein, [b]Jordan et al. (1993), [c]Barstow et al. (2003b), [d]Napiwotzki, Green & Saffer (1999).

---

[6] http://astronomy.villanova.edu/WDcatalog/index.html

## 4.3. Results.

The spectral region containing the Ca II triplet (8350–8800 Å) is shown in Figure 4.1, for all stars. Similarly, the spectral region containing the Mg II absorption line (4470–4490 Å) is shown in Figure 4.2. No emission from Si III or Fe II is seen, although absorption from Fe II (5020 Å) and Si III (4552 Å, 4568 Å) is detected in the spectra of WD 0209+085 and WD 2111+498, respectively (Figure 4.3).

## 4.4. Discussion.

No Ca II emission is seen in the spectra of the stars in this sample (Figure 4.1). However, the spectrum of WD 1942+499 is contaminated by a cosmic ray around 4482 Å (Figure 4.2), which might prevent the detection of Mg II absorption. Although Si II and Fe II absorption is seen at WD 2111+498 and WD 0209+085 (Figure 4.3), respectively, no emission lines are detected from these ions. Mullally et al. (2007) found no evidence for infrared excesses in the spectra of these stars. However, such excesses are present at no more than 20% of the cooler DAZ stars with metal abundances indicative of ongoing accretion (Farihi, Jura & Zuckerman, 2009), suggesting that the small sample size here may be responsible for a lack of such detections.

While no firm detections of circumstellar discs have been achieved here, many questions remain as to origin and abundance of the metals in these objects. WD 2111+498 (which has no binary companion from which metals can be accreted; Saffer et al. 1998) has an anomalously high silicon abundance (~100 times that of other DAs with a similar $T_{\text{eff}}$; Holberg et al. 1997), and its iron abundance is higher than that of WD 0232+035 and WD 0501+527 (Vennes et al. 2006). Given that these DAs have $T_{eff}$ values both significantly above that of WD 2111+498 and above 50,000 K (at which point radiative levitation becomes dominant), they are expected to have much higher metal abundances than WD 2111+498 (Chayer et al. 1995). WD 2111+498 cannot be modelled using the self-consistent, stratified models calculated by Schuh, Dreizler & Wolff (2002); this has been attributed to the accretion of circumstellar material disturbing the diffusion/levitation balance. This body of evidence strongly suggests the photosphere of WD 2111+498 is being enriched by the accretion of circumstellar metals. A periodic EUV variability ($p = 1.150 \pm 0.003$ days; Dupuis et al. 2000), suggests this accretion is inhomogeneous (possibly polar, although no significant magnetic field has been detected). However, since both this study and previous studies (e.g. Bannister et al. 2003) have not detected circumstellar metals, the source of accreted material at this star remains, thus far, a mystery. Similarly, the photospheric aluminium and silicon abundances of WD 2023+246 are in excess of those predicted by radiative levitation theory (Chayer & Dupuis, 2010; Dupuis, Chayer & Hénault-Brunet, 2010),

However, although no gas disc is detected here, non-photospheric absorption features inconsistent with the ISM along the sight line to WD 1611–084 are present (Holberg, Barstow & Green, 1997; Bannister et al. 2003). These features have, in keeping with WD 1620–391 (CD –38 10980; Holberg, Bruhweiler & Andersen 1995), been attributed to circumstellar material.

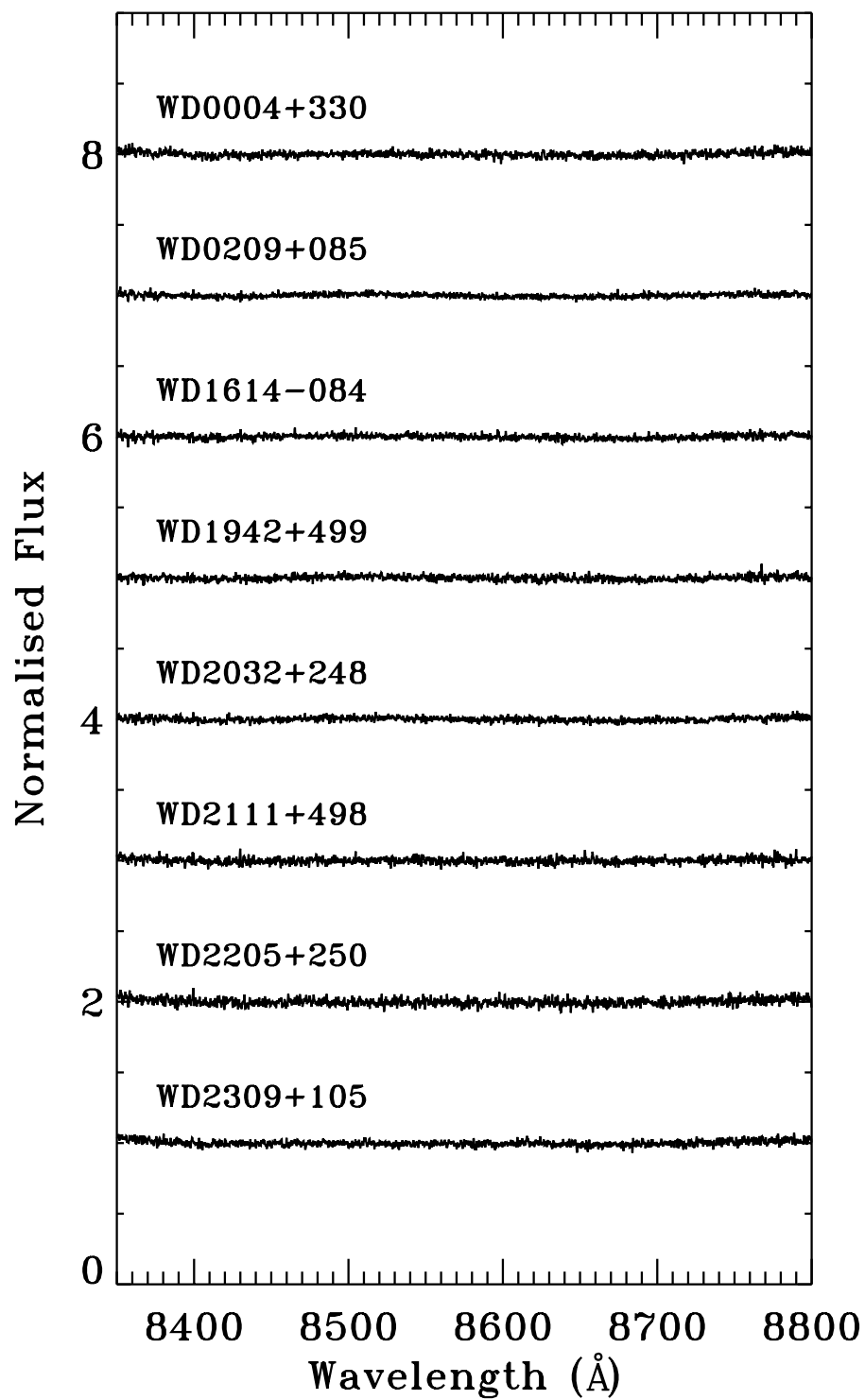

**Figure 4.1.** The spectral region containing the Ca II triplet. Each data set is normalised and offset for clarity.

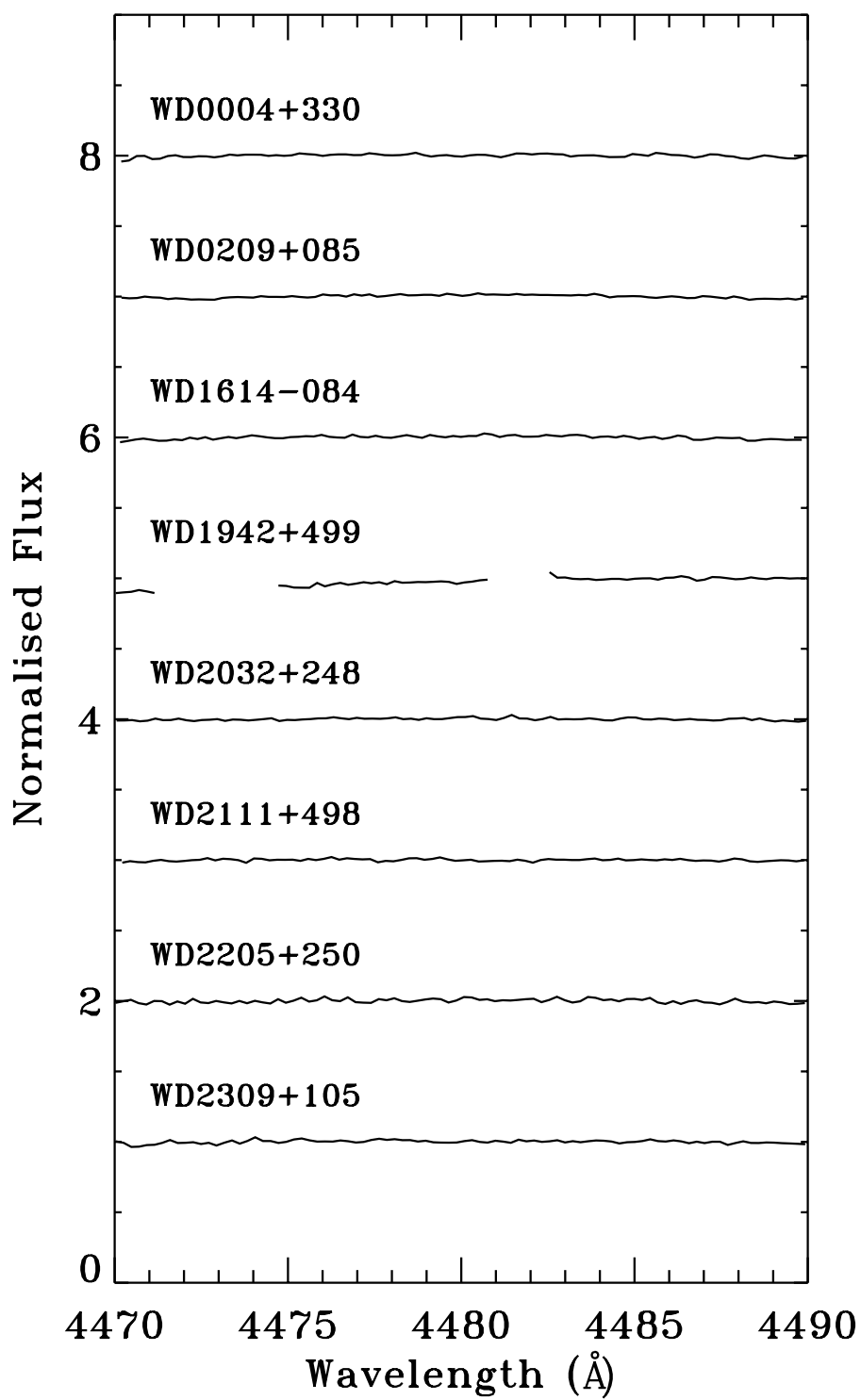

**Figure 4.2.** The spectral region containing the Mg II 4482 Å absorption line. The gaps in the spectrum of WD 1942+499 are due to the removal of cosmic ray contamination.

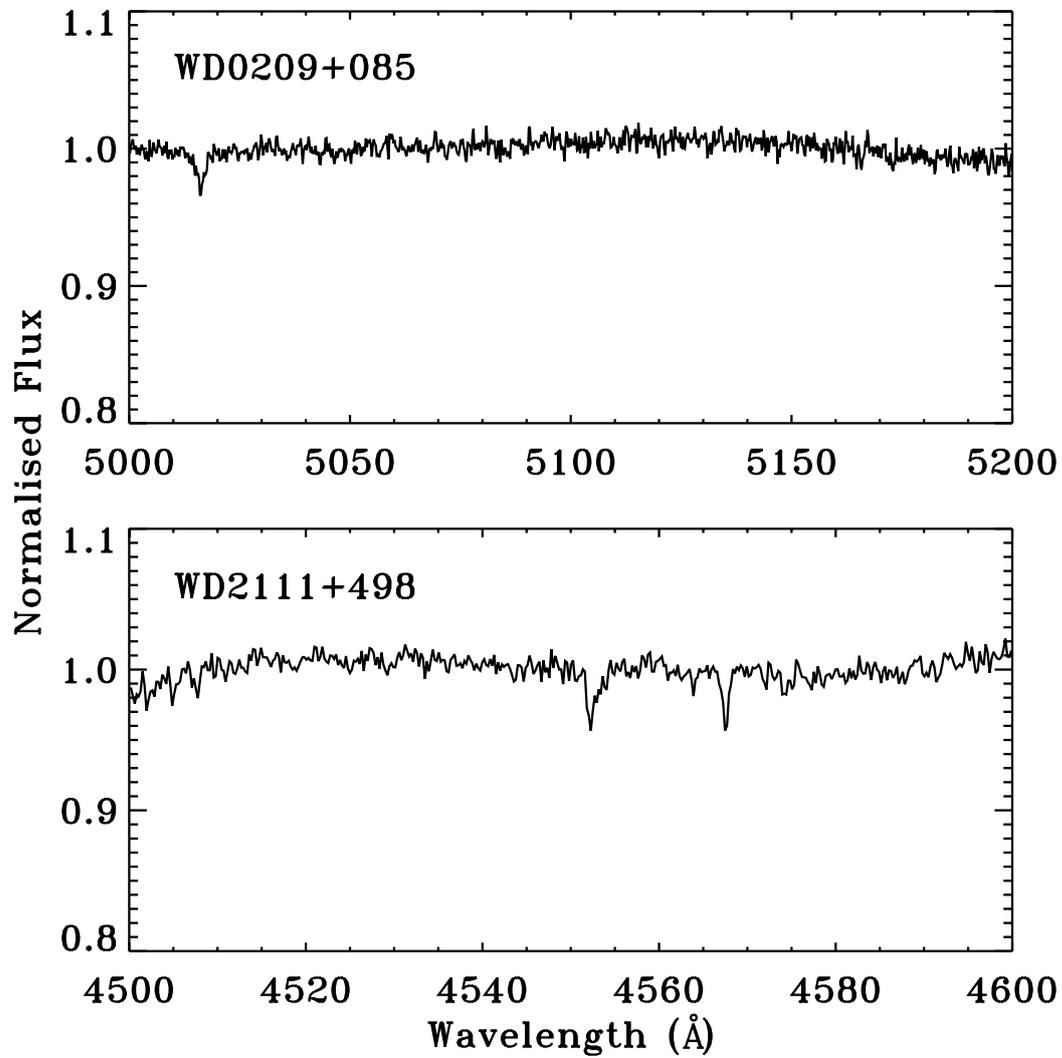

**Figure 4.3.** The upper panel shows the Fe II absorption (5020 Å) in the spectrum of WD 0209+085, while the bottom panel shows the Si III absorption features (4552 Å, 4568 Å) in the spectrum of WD 2111+498.

Further work is needed to understand the circumstellar environment of hot white dwarf atmospheres. An examination of the 'circumstellar' features seen by Bannister et al. (2003) is presented in the next Chapter, in an attempt to better understand the origin of these features and how they may relate to the observed white dwarf metal abundances.

## 4.5. Summary.

- Gaseous components to circumstellar dust discs have been detected at some white dwarfs. Accretion from circumstellar discs is used to explain the metal pollution of cooler DA stars ($T_{eff}$ < 20,000 K).

- A search for gas discs at hotter stars with metal abundances in excess of those predicted by radiative levitation is undertaken here, where one would expect any circumstellar discs present to have sublimated, in an attempt to identify a source of accretion.

- No circumstellar gas emission is detected.

- The stars still have metal abundances that cannot be explained by radiative levitation alone, and in some cases (such as WD 2111+498) the metal abundances are far higher than those found for other, hotter white dwarfs, Since circumstellar disc detection (via infrared excess emission) has only been achieved for approximately 20% of the cooler stars where accretion of metal rich material is required to explain the presence of photospheric metals, it may be the case that the lack of circumstellar gas detection here is related to the small sample size.

# *Chapter 5.*

# The origin of 'circumstellar' features in hot DA white dwarf spectra.

## 5.1. Introduction.

As discussed at length in Section 1.7.3, circumstellar absorption features have been observed in the spectra of hot DA white dwarfs. This Chapter re-examines the sample of Bannister et al. (2003), using a more advanced absorption line profile modelling technique. The results of this examination are compared to different physical scenarios in which circumstellar absorption may occur, including material in a circumstellar disc, the ionisation of ISM in the locality of the star, stellar mass loss and ancient PNe.

Section 5.2 details the sample and the method used to analyse it, while the results are presented in 5.3. The results are discussed in 5.4; a sub-section details each physical scenario (circumstellar discs, 5.4.1; ISM ionisation, 5.4.2; mass loss, 5.4.3; ancient PNe 5.4.4) in which such absorption may arise. A summary of the results and conclusions is presented in 5.5.

## 5.2. Observations and modelling circumstellar absorption components.

The white dwarfs studied were those included in the Bannister et al. (2003) sample; stellar parameters and observation details can be found in Table 5.1. The high ionisation stage absorption lines examined were the same as those studied in Chapter 3, and were detailed in Table 3.2. Where circumstellar absorption was observed, the O VI doublet (1031.912 Å, 1037.613 Å) in the *FUSE* spectrum ($R = 20,000$) of each object was examined, where data were present and the features exist.

The method used here to measure the absorption line components has been described extensively in studies of the ISM (e.g. Welsh & Lallement 2005, 2010; Welsh et al. 2010a,b). The absorption line profiles were fit with Gaussian absorption components. Non-photospheric absorption components were added when the fit was improved in a statistically significant way, i.e. if the probability of similar improvement in the fit upon inclusion of the additional component ($\Delta\chi^2 = 11.1$) being random was less than 1 % (Vallerga et al. 1993). The software used to make the measurements was written by, and is proprietary to, B. Welsh and R. Lallement. B. Welsh carried out the fitting, under guidance (the target list, lines of interest and objects at which previous circumstellar detections were made was provided).

Table 5.1. The stellar parameters and observation information for the DAs studied here.

| WD | Alt. name | $T_{\text{eff}}$ (K)[a] | $\log g$[a] | $L/L$[b] | $D$ (pc)[b] | Data source [Mode] | Resolving Power |
|---|---|---|---|---|---|---|---|
| 0050−335 | GD 659 | 35,660±135 | 7.93±0.03 | 0.24 | 53 | *STIS* [E140H]; *IUE* [SWP] | 100,000; 20,000 |
| 0232+035 | Feige 24 | 60,487±1,100 | 7.50±0.06 | 5.86 | 78 | *STIS* [E140M] | 40,000 |
| 0455−282 | REJ 0457−281 | 50,960±1,070 | 7.93±0.08 | 1.85 | 108 | *IUE* [SWP] | 20,000 |
| 0501+527 | G191-B2B | 52,500±900 | 7.53±0.09 | 3.16 | 50 | *STIS* [E140M] | 40,000 |
| 0556−375 | REJ 0558−373 | 59,508±2,200 | 7.70±0.09 | 4.61 | 295 | *STIS* [E140M] | 40,000 |
| 0621−376 | REJ 0623−371 | 58,200±1,800 | 7.14±0.11 | 11.69 | 97 | *IUE* [SWP] | 20,000 |
| 0939+262 | Ton 021 | 69,711±530 | 7.47±0.05 | 10.19 | 217 | *STIS* [E140M] | 40,000 |
| 0948+534 | PG 0948+534 | 110,000±2,500 | 7.58±0.06 |  | 193.8 | *STIS* [E140M] | 40,000 |
| 1029+537 | REJ 1032+532 | 44,350±715 | 7.81±0.08 | 1.03 | 127 | *STIS* [E140M] | 40,000 |
| 1057+719 | PG 1057+719 | 39,770±615 | 7.90±0.10 | 0.64 | 411 | *GHRS* [G160M] | 22,000 |
| 1123+189 | PG 1123+189 | 54,574±900 | 7.48±0.08 | 2.75 | 147 | *STIS* [E140H] | 100,000 |
| 1254+223 | GD 153 | 39,290±340 | 7.77±0.05 | 0.56 | 73 | *IUE* [SWP] | 20,000 |
| 1314+293 | HZ 43 | 50,370±870 | 7.85±0.07 | 1.49 | 71 | *IUE* [SWP] | 20,000 |
| 1337+705 | EG 102 | 22,090±85 | 8.05±0.01 | 0.03 | 25 | *IUE* [SWP] | 20,000 |
| 1611−084 | REJ 1614−085 | 38,840±480 | 7.92±0.07 | 0.43 | 86 | *GHRS* [G160M] | 22,000 |
| 1738+669 | REJ 1738+665 | 66,760±1,230 | 7.77±0.10 | 12.59 | 243 | *STIS* [E140M] | 40,000 |
| 2023+246 | Wolf 1346 | 19,150±30 | 7.91±0.01 | 0.03 | 14 | *IUE* [SWP] | 20,000 |
| 2111+498 | GD 394 | 39,290±360 | 7.89±0.05 | 0.45 | 57 | *IUE* [SWP]; *GHRS* [G160M] | 20,000; 22,000 |
| 2152−548 | REJ 2156−546 | 45,500±1,085 | 7.86±0.10 | 1.06 | 129 | *STIS* [E140M] | 40,000 |
| 2211−495 | REJ 2214−492 | 61,613±2,300 | 7.29±0.11 | 9.38 | 69 | *IUE* [SWP] | 20,000 |
| 2218+706 | WD 2218+706 | 58,582±3,600 | 7.05±0.12 | 9.64 | 436 | *STIS* [E140M] | 40,000 |
| 2309+105 | GD 246 | 51,308±850 | 7.91±0.07 | 2.23 | 72 | *STIS* [E140M]; *IUE* [SWP] | 40,000; 20,000 |
| 2331−475 | REJ 2334−471 | 53,205±1,300 | 7.67±0.10 | 2.88 | 104 | *IUE* [SWP] | 20,000 |

[a] from Barstow et al. (2003b); [b] from Bannister et al. (2003) and references therein.

The details of the absorption line modelling process are outlined in several papers (e.g. Vallerga et al. 1993; Sfeir et al. 1999), and a summary is presented here. The local continuum was characterised with a multi-order polynomial fit (absorption features were neglected), with the root mean square of the fit adopted temporarily as the error of each data point. The $\chi_v^2$ of the continuum model was calculated, and the errors on the data points previously assigned were adjusted by a constant factor until the $\chi_v^2$ was equal to 1.0. Errors on the fit to absorption feature were derived in a similar way (where the model Gaussian absorption components were included), allowing both the line profiles and continuum placement to be taken into account in the error calculation. The equivalent width of the absorbing component was calculated by integrating the absorbed flux in the absorption feature, propagating the error on the model data points.

The heliocentric velocities of the absorption components were allowed to vary simultaneously (note that in the case of the data used here, no heliocentric correction was applied since data from *IUE*, *HST* and *FUSE* are expressed in the heliocentric frame) through a $\chi^2$ minimisation technique, as was the broadening of the absorption line component (*b*). The column densities (*N*) of the circumstellar and interstellar material were derived as part of the line fitting routine. The measured equivalent widths and the oscillator strengths of the absorption lines were used to generate theoretical line profiles, which were fit to the absorption features to provide column densities. Reliable errors were produced for unsaturated line profiles. Since the oscillator strengths were used to model the absorption lines, the technique used by Bannister et al. (2003) of coadding absorption features in velocity space to improve the signal to noise and reveal hidden circumstellar components cannot be utilised here.

Bannister et al. (2003) used the curve of growth method to measure circumstellar column densities. In this method, the change in equivalent width with column density is calculated for the absorbing material, for a series of *b* values. The observed equivalent widths are then plotted on the curve corresponding to the measured *b* value and the column density is read off. This does method does not explicitly model the observed line profiles using the line transition data, making the work here a significant improvement over Bannister et al.'s (2003) previous study.

The velocities of the photospheric components of each high ion absorption line were measured (where present), and the mean of these measurements was calculated to give the averaged photospheric velocity ($v_{\text{phot}}$) of the star. Likewise, when circumstellar components were seen in the absorption line profiles, the mean of the measured velocities was calculated to produce a characteristic circumstellar velocity ($v_{\text{CS}}$) for the absorbing material. To characterise the ISM velocity, the velocities of unsaturated absorption lines (such as Si II; 1260.422 Å, 1304.370 Å and S II; 1259.52 Å) were measured since more accurate line properties could be derived. When additional measurements were required (e.g. where Si II or S II were not present), other low ion absorption lines such as O I (1302.168 Å) and Fe II (1608.451 Å) were used. Given the variety in observed ISM composition along each line of sight, the exact lines used varied in each case. The component with the largest equivalent width was deemed to be the primary ISM component (with $v_{\text{ISM,pri}}$), the next largest the secondary (with $v_{\text{ISM,sec}}$). Where three ISM components were present, the component with the smallest equivalent width was denoted as the tertiary component (with $v_{\text{ISM,ter}}$). Mean values of $v_{\text{ISM,pri}}$, $v_{\text{ISM,sec}}$ and $v_{\text{ISM,ter}}$ were calculated to allow easy comparison of $v_{\text{CS}}$ to all interstellar components. All errors were combined quadratically.

## 5.3. Results.

### 5.3.1. Summary of results.

Circumstellar absorption is unambiguously detected at eight white dwarfs. The measured velocities, circumstellar velocity shifts with respect to the photosphere ($v_{\text{CSshift}}$) and gravitational redshifts ($v_{\text{grav}}$, calculated by J. Farihi using stellar masses and radii derived from the evolutionary models of Fontaine, Brassard & Bergeron 2001) are presented in Table 5.2. Using maps of LISM morphology (Redfield & Linsky, 2008), the LISM clouds along each white dwarf sight line are identified and their projected velocity ($v_{\text{LISM}}$) is included[7] in the sixth column of Table 5.2. This allows the detected circumstellar and ISM component velocities to be compared to $v_{\text{LISM}}$, to examine any possible links the detected absorption may have to LISM clouds. Table 5.3 details which ions display circumstellar components, and gives their column densities. Detailed notes on the results for each object follow Tables 5.2 and 5.3. The objects grouped into those with circumstellar detections (Section 5.3.2), and those without (Section 5.3.3).

---

[7]the cloud identification and velocity projections were performed using the online tool available at http://lism.wesleyan.edu/LISMdynamics.html

Table 5.2. All measured velocities, circumstellar velocity shift ($v_{CSshift}$), predicted LISM velocities and gravitation redshifts ($v_{grav}$). All velocities are expressed in km s$^{-1}$.

| WD | $v_{phot}$ | $v_{CS}$ | $v_{CSshift}$ | $v_{ISM(pri,sec,ter)}$ | $v_{LISM \text{ (cloud name)}}$[*] | $v_{grav}$ |
|---|---|---|---|---|---|---|
| 0050−335 | 34.34±0.38 | 7.4±0.34 | | 6.7±0.3 | 4.56±1.36 (LIC) | 28.21 |
| 0232+035 | 30.11±0.52[a], 128.23±0.31[b] | | −22.19±0.68[a], −120.63±0.40[b] | 2.85±0.34, 17±1.3 | 18.1±1.13 (LIC) | 15.7 |
| 0455−282 | 79.28±1.79 | 18.8±3.47 | −60.48±3.90 | 12.1±1.5 | 12.56±1.03 (Blue) | 28.96 |
| 0501+527 | 24.51±0.16 | 8.9±0.07 | −15.61±0.17 | 8.15±0.18, 19.3±0.03 | 19.1±1.1 (LIC), 9.35±1.32 (Hyades) | 16.07 |
| 0556−375 | 25.37±2.03 | 10.2±1.07 | −15.17±2.30 | 7.8±1 | 11.36±0.95 (Blue) | 21.07 |
| 0621−376 | 39.44±0.25 | | | 15.8±0.4 | 11.09±0.93 (Blue) | 9.36 |
| 0939+262 | 36.5±0.47 | 9.38±6.6 | −27.12±6.61 | −2.1±0.2 | 10.81±1.29 (LIC) | 15.40 |
| 0948+534 | −17.09±1.73 | | | −18.45±0.42, −1.6±0.63, 22.6±0.8 | 10.07±1.31 (LIC) | 19.67 |
| 1029+537 | 37.98±0.21 | | | 0.95±0.79 | 7.72±1.33 (LIC) | 24.00 |
| 1057+719 | | | | −0.2±1 | 6.64±1.35 (LIC) | 27.18 |
| 1123+189 | | | | −4.75±3.18 | 3.03±0.79 (Leo) | 14.98 |
| 1254+223 | | | | −15.4±1.8 | −5.52±0.74 (NGP) | 22.36 |
| 1314+293 | | | | −6.6±0.1 | −6.15±0.74 (NGP) | 25.77 |
| 1337+705 | | | | −1.5±1.8 | 1.59±1.38 (LIC) | 32.84 |
| 1611−084 | −40.76±3.56 | −66.67±2.05 | −25.90±4.11 | −34.7±1.5 | −29.26±1.12 (G) | 27.95 |
| 1738+669 | 30.17±1.49 | −18.36±4.23 | −48.53±4.49 | −20.0±0.3 | −2.91±1.37 (LIC) | 23.65 |
| 2023+246 | | | | −16.3±1.7, 18.3±2.5 | | 26.41 |
| 2111+498 | 29.3±1.66 | | | −7.6±1.3 | −2.35±1.38 (LIC) | 26.77 |
| 2152−548 | −14.94±0.46 | | | −9.2±0.53 | −9.73±1.31 (LIC) | 25.9 |
| 2211−495 | 32.33±1.37 | | | −1.1±0.4 | −8.8±1.32 (LIC), 9.93±0.6 (Dor) | 11.43 |
| 2218+706 | −40.04±1.11 | −17.8±1.05 | 22.24±1.52 | −15.3±2.64, −1.2±4.01 | 4.47±1.37 (LIC) | 8.04 |
| 2309+105 | −13.45±0.13 | | | | −8.2±0.70 | 28.17 |
| 2331−475 | 38.88±0.72 | | | 14.3±0.7 | −3.41±1.37 (LIC) | 19.88 |

[*]from Redfield & Linsky (2008); [a]from binary phase 0.24; [b]from binary phase 0.74

**Table 5.3.** The stars with circumstellar detections, the identified species and measured column densities.

| WD | Species | Column density ($\times 10^{12}$ cm$^{-2}$) |
|---|---|---|
| 0232+035 | C IV | 28.0±1.3 |
| 0455–282 | C IV, N V, Si IV | 33.2±6.6, 4.17±0.83, 3.71±0.74 |
| 0501+527 | C IV | 104±10 |
| 0556–375 | C IV | 46.8±6.2 |
| 0939+262 | C IV, Si IV | 8.05±0.21, 1.81±0.36 |
| 1611–084 | C IV, Si IV | 9.77±1.9, 3.53±0.71 |
| 1738+669 | C IV, O V, O VI, Si IV | 55.5±6.1, 0.235±0.029, 7.04±1.41, 3.26±0.17 |
| 2218+706 | C IV, Si IV | 119±17, 7.34±0.06 |

## 5.3.2. Objects with circumstellar absorption.

### 5.3.2.1. WD 0232+035 (Feige 24).

WD 0232+035 is part of a binary system with a dM1.5–2 companion, with a binary period of 4.23160±0.00002 d (Vennes and Thorstensen, 1994). Data were obtained at two phases of the binary cycle, 0.73–0.75 (29 November 1997) and 0.23 – 0.25 (4 January 1998). The mean photospheric velocity changes between data sets ($v_{\text{phot}(0.24)}$ = 30.11±0.52 km s$^{-1}$; $v_{\text{phot}(0.74)}$ = 128.23±0.31 km s$^{-1}$), reflecting the orbital motion of the white dwarf. Two sets of absorption features are seen in the C IV doublet, one of which remains stationary at 7.4±0.34 km s$^{-1}$ ($v_{\text{CS}}$), and is thus the circumstellar component (Figures 5.1 and 5.2). The column density for this component is $(2.8\pm0.13)\times10^{13}$ cm$^{-2}$, with a *b* value of $6.4\pm0.5$ km s$^{-1}$. Barstow et al. (2010) found the single component O VI absorption line to arise in the photosphere.

Using the Si II 1304 Å and S II 1260 Å lines, a dual component ISM was found with the primary component at $2.85\pm0.34$ km s$^{-1}$ and the secondary at $17.1\pm1.3$ km s$^{-1}$. The predicted LIC velocity is $18.1\pm1.13$ km s$^{-1}$, in keeping with the secondary ISM component.

### 5.3.2.2. WD 0455–282 (REJ 0457–281).

Blue shifted features in the C IV and Si IV lines were first seen by Holberg, Barstow & Sion (1998), and some evidence for a second absorption component in the 1238.821 Å N V line was observed by Bannister et al. (2003). The mean photospheric velocity is detected at $79.28\pm1.79$ km s$^{-1}$, which compares well with the value of $76.91\pm0.83$ km s$^{-1}$ measured by Bannister et al. (2003). The mean velocity of the circumstellar features is $21.76\pm1.27$ km s$^{-1}$.

The mean C IV column density is $(3.32\pm0.66)\times10^{13}$ cm$^{-2}$, about one fifth of the value ($1.82\times10^{14}$ cm$^{-2}$) measured by Bannister et al. (2003). In addition to the C IV column density, the mean column densities for N V and Si IV are $(4.17\pm0.84)\times10^{11}$ cm$^{-2}$ and $(3.71\pm0.74)\times10^{12}$ cm$^{-2}$, respectively.

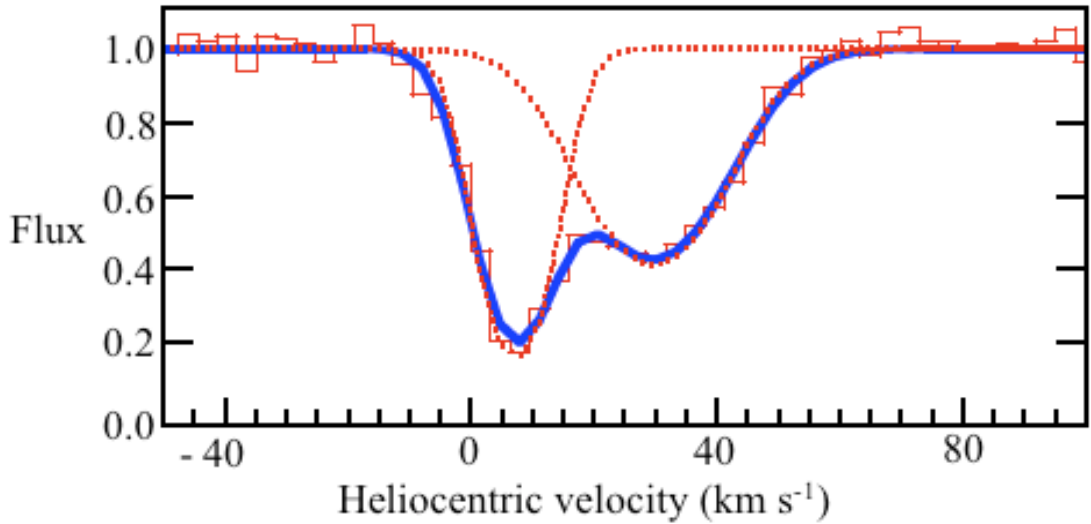

**Figure 5.1.** The 1548 Å C IV line of WD 0232+025, where the binary phase is 0.24. The circumstellar component (at 7.4 km s$^{-1}$) is blended with the photospheric components (at 30.11 km s$^{-1}$). The data are plotted with a solid red line, the model components are plotted with dotted red lines and the sum of the model components is plotted in blue; this plotting convention is also used in Figures 5.2–5.4.

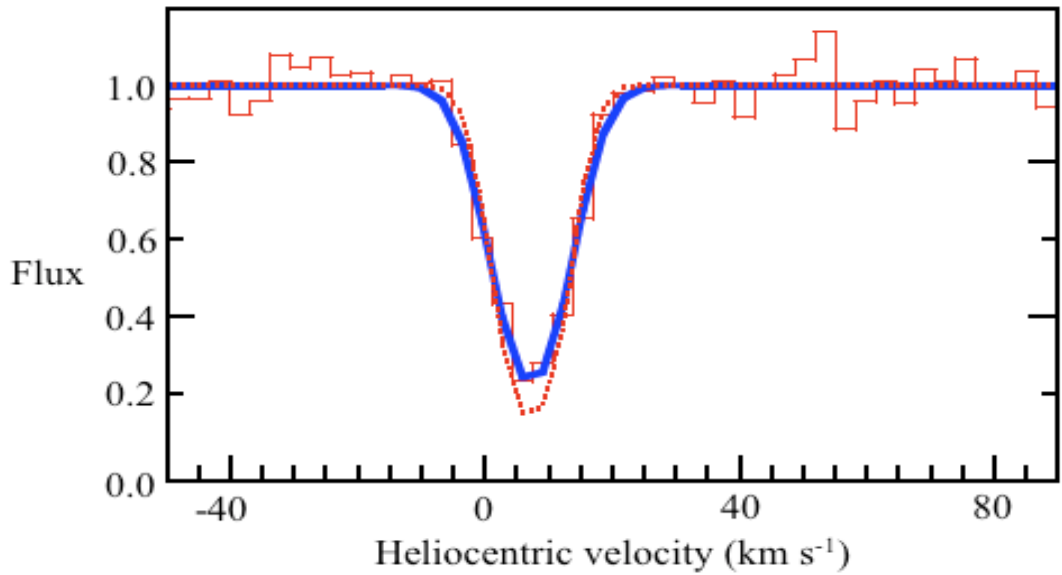

**Figure 5.2.** The circumstellar 1548 Å C IV line of WD 0232+025, when the binary phase is 0.74. The photospheric component (at 128.23 km s$^{-1}$) is not seen here.

The ISM is found at 12.1±1.5 km s$^{-1}$. The line of sight to WD 0455–282 traverses the Blue cloud, which has a projected velocity of 12.56±1.03 km s$^{-1}$, implying that the ISM observed along the sight line to the star resides in the Blue cloud. The O VI 1032 Å line displays two absorbing components, attributed to the photosphere and blue shifted ISM by (Barstow et al. 2010).

### 5.3.2.3. WD 0501+527 (G191-B2B).

WD 0501+527 is among the best-studied hot white dwarfs. In keeping with the results of Bannister et al. (2003), circumstellar C IV is observed in the *STIS* [E140M] spectrum, at a velocity of 8.9±0.07 km s$^{-1}$. The averaged photospheric velocity is 24.51±0.16 km s$^{-1}$. The circumstellar C IV column density is (1.04±0.1)x10$^{14}$ cm$^{-2}$, a factor of four higher than the column densities found by Bannister et al. (2003; 2.40x10$^{13}$ cm$^{-2}$) and Vennes & Lanz (2001; 6.31x10$^{13}$ cm$^{-2}$), probably due to the fact that a curve of growth was used in those analyses, instead of full absorption profile modelling as used here. The *b* value is 5.65±0.18 km s$^{-1}$, near the 5.2 km s$^{-1}$ of Vennes & Lanz (2001); the higher *b* = 10 km s$^{-1}$ reported by Bannister et al. (2003) is inconsistent with both of these values. A single component is detected in the O VI 1032 Å line at 19 km s$^{-1}$, agreeing with other photospheric lines in the *FUSE* spectrum, in keeping with both Barstow et al. (2010) and Savage & Lehner (2006).

The interstellar components (measured using the 1259 Å S II line) are found at 8.15±0.18 km s$^{-1}$ (similar to $v_{CS}$) and 18.0±0.03 km s$^{-1}$, agreeing with the values reported for the higher resolution *STIS* E140H data by Sahu et al. (1999) and Redfield & Linksy (2004). The projected LIC velocity is 19.1±1.1 km s$^{-1}$, suggesting that the LIC may account for the secondary component in the observed ISM features.

### 5.3.2.4. WD 0556–375 (REJ 0558–373).

Circumstellar counterparts to the photospheric C IV doublet are observed in this 59,508 K DA. The mean photospheric and circumstellar velocities are 25.37±2.03 km s$^{-1}$ and 10.2±1.07 km s$^{-1}$, respectively. The average C IV column density is (4.67±0.62) x 10$^{13}$ cm$^{-2}$.

The S II line is again used to characterise the ISM. Three components are seen at 7.8±1 km s$^{-1}$ ($v_{ISM, pri}$), 19.9±1.7 km s$^{-1}$ ($v_{ISM, sec}$) and 36.0±2.3 km s$^{-1}$ ($v_{ISM, ter}$). Comparing $v_{CSshift}$ (–15.17±2.30 km s$^{-1}$), the shift in the primary ISM component with respect to the photosphere (–17.57±2.26 km s$^{-1}$) shows an overlap in the error margins of these quantities, implying that they may be related. The sight line to WD 0556–375 traverses the Blue cloud, which has a projected velocity of 11.36±0.95 km s$^{-1}$, similar to $v_{CS}$ and near $v_{ISM, pri}$.

### 5.3.2.5. WD 0939+262 (Ton 021).

Circumstellar absorption is found in the C IV and Si IV doublets, giving a mean $v_{CS}$ of 9.38±6.59 km s$^{-1}$. When the N V doublet and the O V line are included, $v_{phot}$ averages to 36.5±0.47 km s$^{-1}$. The mean column density of the C IV doublet is (8.05±0.21)x10$^{12}$ cm$^{-2}$ with a *b* value of 8.3±1.75 km s$^{-1}$; with a Si IV column density of $(1.81\pm0.36)\times10^{12}$ cm$^{-2}$ and a $b$ = 11.65±5.28 km s$^{-1}$. These results are consistent with the findings of Bannister et al. (2003).

Measurements of the S II ISM line characterise the ISM velocity at –2.1±0.2 km s$^{-1}$. The predicted LIC velocity is far from this value, at 10.81±1.29 km s$^{-1}$.

### 5.3.2.6. WD 1611–084 (REJ 1614–085).

At 38,840 K, WD 1611–084 is the coolest white dwarf to show unambiguous signs of circumstellar absorption. Photospheric C IV, N V and Si IV is observed, yielding a mean $v_{phot}$ of –40.77±3.56 km s$^{-1}$. The *FUSE* spectrum of this object reveals a photospheric O VI 1032 Å absorption line. The column densities measured for the circumstellar C IV and Si IV are $(9.77\pm1.95)\times10^{12}$ and $(3.53\pm0.71)\times10^{12}$ cm$^{-2}$, respectively, with $b$ values are 4.55±2.6 and 2.3±0.46 km s$^{-1}$. The velocities of the absorption lines are inconsistent across the ions, as found by Bannister et al. (2003). Here, the circumstellar C IV is displaced by 23.55±3.96 km s$^{-1}$ with respect to the photospheric component, while the circumstellar silicon is displaced by 30.6±0.64 km s$^{-1}$, with an averaged $v_{CS}$ of –66.67±2.05 km s$^{-1}$.

Two components are observed in the ISM, at –34.7±1.5 ($v_{ISM,pri}$) and –13±3.2 km s$^{-1}$ ($v_{ISM,sec}$). The line of sight to WD 1611–084 traverses the G cloud, which has a projected velocity of –29.26±1.12 km s$^{-1}$, well separated from both $v_{CS}$ and both ISM components.

### 5.3.2.7. WD 1738+669 (REJ 1738+665).

The mean $v_{phot}$ for this star is 30.49 ± 0.28 km s$^{-1}$, with C IV, O V and Si IV exhibiting clear secondary absorbing components at a mean velocity of −18.36±4.23 km s$^{-1}$. The doublet components for N V cannot be coadded to identify a secondary absorbing component at $v_{CS}$ as in Bannister et al. (2003).

The column density of the C IV doublet is (5.55±0.61)x10$^{13}$ cm$^{-2}$ with a $b$ value of 6.8±0.43 km s$^{-1}$; the O V and Si IV column densities are (2.35±0.29)x10$^{11}$ and (3.26±0.17)x10$^{12}$ cm$^{-2}$, with $b$ values of 0.3±3.8 and 6.95±1.07 km s$^{-1}$. The C IV and Si IV column densities derived by Bannister et al. (2003) using a curve of growth were 1.70x10$^{13}$ cm$^{-2}$ and 1.66x10$^{13}$ cm$^{-2}$, both inconsistent with the values found here. However, the values measured by Dupuis et al. (2009) were (5.01±0.48)x10$^{13}$ cm$^{-2}$ and (5.50±0.13)x10$^{12}$ (for C IV and Si IV, respectively), and were obtained by the apparent optical depth technique (where the observed line profiles are converted to apparent optical depth profiles and apparent column densities per unit velocity, providing a form of absorption line data that can be directly interpreted, e.g. Savage & Sembach, 1991), in agreement with the column densities measured here. The ISM is again characterised using the S II line, which has a velocity of −20±0.3 km s$^{-1}$, far from the predicted LIC velocity of −2.91±1.37 km s$^{-1}$.

The *FUSE* spectrum shows two components in the 1032 Å O VI line, one at −32.7±3.4 km s$^{-1}$ and the other at 14.9±1.1 km s$^{-1}$. The −32.7 km s$^{-1}$ component is adopted as circumstellar (with a column density of $(7.04\pm1.41)\times10^{12}$ cm$^{-2}$ and a $b$ value of $12.6\pm5.1$ km s$^{-1}$) and the 14.9 km s$^{-1}$ is designated the photospheric component. This gives an O VI $v_{CSshift}$ of $-47.6\pm3.6$ km s$^{-1}$, in agreement with the value from the *STIS* measurements ($-48.53\pm4.49$ km s$^{-1}$).

### 5.3.2.8. WD 2218+706.

Circumstellar components have previously been seen in the C IV and Si IV doublets of this star (Bannister et al. 2003). The averaged $v_{phot}$ is $-40.04\pm1.11$ km s$^{-1}$, with the mean $v_{CS}$ at $-17.8\pm1.05$ km s$^{-1}$. No O VI is seen in the *FUSE* spectrum of this star. ISM components are found at $-15.3\pm2.64$ km s$^{-1}$ ($v_{ISM, pri}$) and $-1.2\pm4.01$ km s$^{-1}$ ($v_{ISM, sec}$). The line of sight to WD 2218+706 traverses the LIC, which has a projected velocity of $4.47\pm1.37$ km s$^{-1}$, nowhere near $v_{CS}$ or either of the ISM components.

The circumstellar C IV column density is $(1.19\pm0.17)\times10^{13}$ cm$^{-2}$ and the Si IV column is $(7.98\pm0.59)\times10^{12}$ cm$^{-2}$, with $b$ values of $6.75\pm1.67$ km s$^{-1}$ and $11.05\pm2.07$ km s$^{-1}$. These values are somewhat different to those derived by Bannister et al. (2003), who found C IV and Si IV column densities of $4.17\times10^{13}$ cm$^{-2}$ and $4.07\times10^{13}$ cm$^{-2}$, each with $b$ values of 6 km s$^{-1}$. However, the method used here models the circumstellar line fully, rather than performing a curve of growth on, in this case, two data points (Figure 6, Bannister et al. 2003).

### 5.3.3. Objects without circumstellar absorption.

### 5.3.3.1. WD 0050–335 (GD 659).

In this study, the mean photospheric and ISM velocities are 34.34±0.38 km s$^{-1}$ and 6.7±0.3 km s$^{-1}$ (using the S II 1259 Å line). A secondary component in the C IV doublet is not seen above the noise in the data. Indeed the circumstellar component seen by Bannister et al. (2003) was comparable to the noise in the data; the circumstellar component was only seen when adding the doublet components in velocity space, which cannot be achieved here. While the C IV, N V and Si IV resonance doublets are clearly visible, the O V resonance line is not. Photospheric O VI is detected by Barstow et al. (2010).

### 5.3.3.2. WD 0621–376 (REJ 0623–371).

The *IUE* spectrum of this 58,200 K white dwarf displays no clear circumstellar absorption. A mean $v_{phot}$ of 39.44±0.25 km s$^{-1}$ is found, along with a single component ISM at 15.8±0.4 km s$^{-1}$. Again, as was seen by Bannister et al. (2003), the centroids of the components of the C IV doublet are inconsistent with one another ($v_{1548\text{Å}}$ = 37.6±0.6 km s$^{-1}$, $v_{1551\text{Å}}$ = 48.5±0.7 km s$^{-1}$). Unusually (given the high oscillator strength of the short wavelength component), the equivalent width of the 1551 Å component (132.45 mÅ) is greater than that of the 1548 Å component (109.75mÅ). It may be the case that some unresolved material is present along the line of sight to this star; indeed given the poor resolving power of the *IUE* [SWP] data (20,000), details in the line profiles may be present that instruments of a higher resolution can resolve. A single component $v_{ISM}$ is found at 15.8±0.4 km s$^{-1}$.

### 5.3.3.3. WD 0948+534 (PG 0948+534).

In keeping with the findings of Bannister et al. (2003), the ISM observed along the line of sight to this star has three absorbing components. The 1260 Å S II line has components at −18.7±0.6 km s$^{-1}$, −1.6±0.6 km s$^{-1}$, and 22.6±0.8 km s$^{-1}$. When coupled with measurements of the absorbing components in the Si II 1304 Å line, $v_{ISM,pri}$ is found at −18.45±0.42 km s$^{-1}$, $v_{ISM,sec}$ at −1.6±0.63 km s$^{-1}$ and $v_{ISM,ter}$ (the velocity of the tertiary ISM component) at 22.6±0.8 km s$^{-1}$. In contrast, Bannister et al. (2003) found the −1.6 km s$^{-1}$ component had a larger equivalent width than the −18.45 km s$^{-1}$ component (and measured the components to be at −0.26±1.26 km s$^{-1}$ and −22.8±1.2 km s$^{-1}$). In the S II line, the model equivalent widths of the −18.7 km s$^{-1}$ and −0.6 km s$^{-1}$ components are comparable at 32.14 mÅ and 30.99 mÅ respectively; in the Si II line the equivalent widths of these components are 39.20 mÅ and 25.63 mÅ, showing clearly that the −18.7 km s$^{-1}$ is the primary absorbing component.

In keeping with previous studies, one absorbing component is statistically preferred to a two component model (i.e. the improvement in $\chi^2$ obtained upon the inclusion of a secondary component in less than 11.1; Section 5.2 of this thesis); in the C IV doublet this is found at −16±0.3 km s$^{-1}$ (Figure 5.3). All of the high ion lines give an average $v_{phot}$ of −13.06 km s$^{-1}$, although considerable spread is seen in the velocities ($v_{OV}$ = −10.03 km s$^{-1}$; $v_{CIV}$ = −16 km s$^{-1}$). Caution must be exercised here;

the single model fits used here do not include stellar atmosphere models and therefore do not reproduce physically accurate photospheric absorption line profiles. Gaussian models are fit, showing the statistically preferred number of absorbing components. Indeed, as detailed in Chapter 3 of this thesis, the narrow, almost saturated absorption line profiles of the high ion absorption features are yet to be explained using stellar models.

Although not statistically preferred, there is a hint of a second component in the C IV doublet, putting the mean C IV photospheric component at −17.6±0.39 km s$^{-1}$ and the circumstellar C IV photospheric component at 1.65±7.22 km s$^{-1}$ (Figure 5.4). Like WD 0621−375, the large inconsistencies in the centroid positions of these absorption features might be evidence of unresolved circumstellar absorption. Following this logic, the large spread in centroid positions in the absorption features of this object when fit with single absorption components may also be symptomatic of an unresolved secondary absorbing component. Applying this to the N V, O V and Si IV (Figure 5.5) absorption features, which have photospheric velocities of 15.4±1.35, −19.2±5.5 and 17.2±0.90 km s$^{-1}$, and circumstellar velocities of 2.0±1.39, −6.0±7.7 and 0.05±6.15 km s$^{-1}$, gives a mean $v_{\text{phot}}$ of −17.09±1.72 km s$^{-1}$ and a mean $v_{\text{CS}}$ of 0.2±5.40 km s$^{-1}$. The large error on the mean value of $v_{\text{circ}}$ is due to a combination of the large errors on the circumstellar velocities of each ion, and the large spread in circumstellar component velocities across the ions, possibly due to the inability to resolve the circumstellar components here, if present.

The O VI 1032 Å line displays components at −22.9±1.2 km s$^{-1}$ and 15.7±2.8 km s$^{-1}$ when fit with two components. Given that the *FUSE* velocity resolution (~15 km s$^{-1}$) is poorer than that of the *STIS* E140M data (~7 km s$^{-1}$), and that the absolute velocity calibration of each data set is not necessarily exactly the same, the secondary 15 km s$^{-1}$ component is deemed to be consistent with the circumstellar components of the other ions.

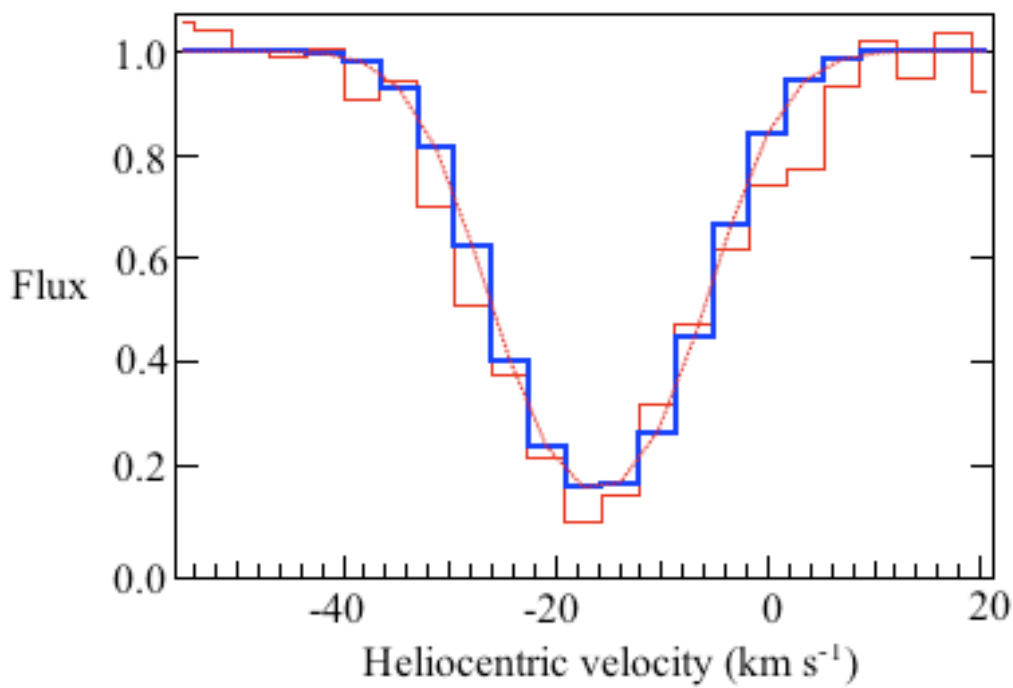

**Figure 5.3.** The 1548 Å component of the C IV doublet of WD 0948+534, fit with an absorbing component at −16 km s$^{-1}$.

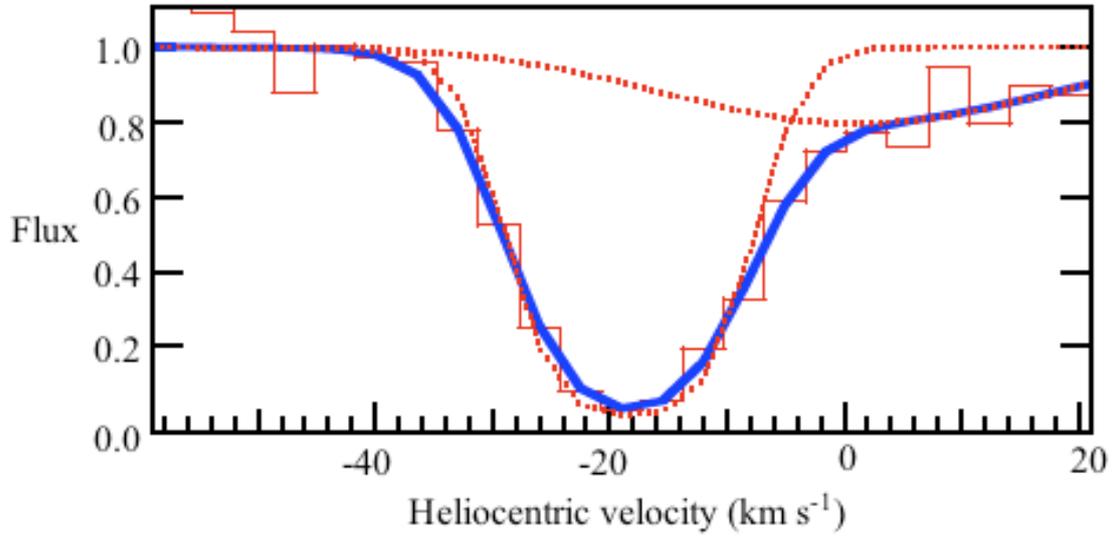

**Figure 5.4.** The 1548 Å component of the C IV doublet of WD 0948+534, fit with two absorbing components at −17.6 km s$^{-1}$ and 1.65 km s$^{-1}$.

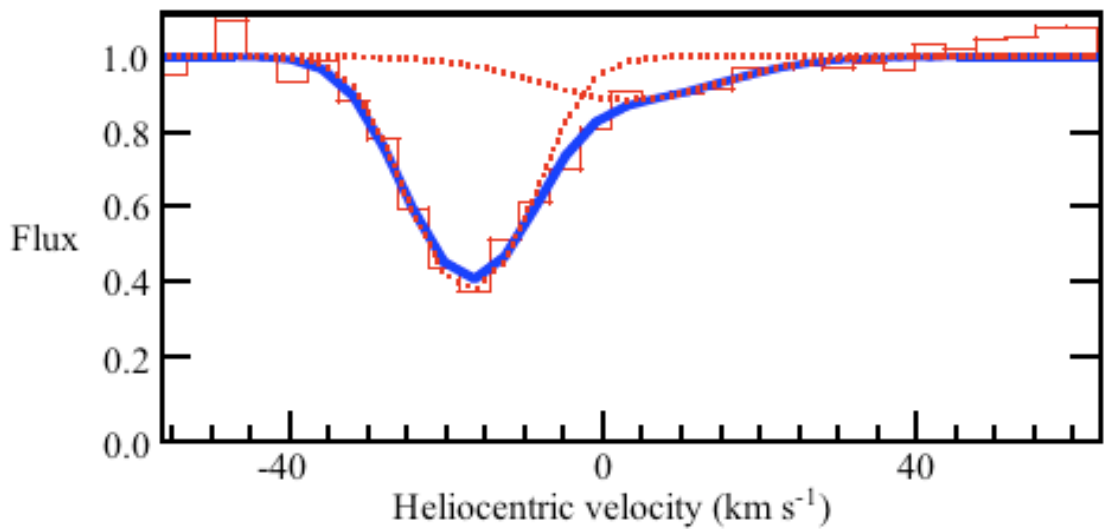

**Figure 5.5.** The Si IV 1393 Å line fit with two absorbing components at −16.9 km s$^{-1}$ and 3.5 km s$^{-1}$.

The mean column densities of the circumstellar C IV, N V, O V, O VI and Si IV are $(1.00\pm0.55)\times10^{13}$, $(9.95\pm0.96)\times10^{12}$, $(2.28\pm0.19)\times10^{13}$, $(2.05\pm0.64)\times10^{13}$ and $(2.41\pm0.31)\times10^{12}$ cm$^{-2}$, respectively. The $b$ values are 11.85±5.4, 2±1.39, 13.5±4.7, 8.4±4.2 and 10±4.73 km s$^{-1}$.

Using the *STIS* measurement, a $v_{\mathrm{CSshift}}$ value of 17.29±5.7 km s$^{-1}$ is computed, while the shift in the velocity of the secondary ISM component is 18.69±1.8 km s$^{-1}$, implying that the circumstellar component may be related to the secondary component of the ISM. The LIC velocity along the line of sight to WD 0948+534 is predicted to be at 10.07±1.31 km s$^{-1}$, far from any of the absorbing components measured here.

However, extreme caution must be exercised; while the two-component model of these high ion absorption features is an attractive explanation of the absorption line profiles, a single component is statistically preferred. Given the difficulty of explaining the absorption line profiles of this star with single component models, the hint of a secondary component in the absorption lines and the spread in velocities across the absorption features, obtaining higher resolution data is necessary, and may shed further light on this enigmatic object.

### 5.3.3.4. WD 1029+537 (REJ 1032+532).

The $v_{\mathrm{phot}}$ found here, 37.98±0.21 km s$^{-1}$, is in good agreement with the value of 38.16±0.40 km s$^{-1}$ found by Bannister et al. (2003). The mean $v_{\mathrm{ISM}}$ is at 0.95±0.79 km s$^{-1}$. No circumstellar absorption is detected in the spectrum of this star.

### 5.3.3.5. WD 1057+719 (PG 1057+719).

No evidence for circumstellar absorption is found at this star, nor has any been found previously (Bannister et al. 2003; Holberg et al. 1997). No definite absorption features are observed in the wavelength regions where C IV, N V, O V, or Si IV absorption is expected. Using the 1260 Å Si II line, a $v_{ISM}$ of $-0.2\pm1$ km s$^{-1}$ is obtained.

### 5.3.3.6. WD 1123+189 (PG 1123+189).

The ISM lines in the spectrum of this object display two components at $-4.75\pm3.18$ ($v_{ISM,\,pri}$) and $2.15\pm2.96$ km s$^{-1}$ ($v_{ISM,\,sec}$) in both the S II and O I lines. The high uncertainty in the measurements comes from the saturation of the O I line.

No definite photospheric detections are made. The *STIS* [E140H] data only covers the 1160 Å – 1360 Å wavelength range, and thus misses the C IV and Si IV doublets and the O V line. Bannister et al. (2003) estimated $v_{phot}$ (12.55±0.53 km s$^{-1}$), by coadding the N V doublet in velocity space.

### 5.3.3.7. WD 1254+223 (GD 153).

WD 1254+223 is another white dwarf in which no photospheric or circumstellar metals are seen. The interstellar Si II 1260 Å line is measured at $-15.4\pm1.6$ km s$^{-1}$. O VI is seen in the *FUSE* spectrum, and is attributed to the ISM (Barstow et al. 2010).

### 5.3.3.8. WD 1314+293 (HZ 43).

No photospheric metals or circumstellar absorption are observed in this star. The ISM is found at $-6.6\pm0.1$ km s$^{-1}$.

### 5.3.3.9. WD 1337+705 (EG 102).

With $T_{\text{eff}}$ = 22,090±85 K, well below the 50,000 K at which radiative levitation dominates, it is perhaps not surprising the high ion lines are not seen. No circumstellar absorption is found in the *IUE* spectrum of this star, although this remains an interesting object.

Possible evidence for the accretion of circumstellar absorption comes from several sources. As recalled by Bannister et al. (2003), Holberg, Barstow & Green (1997) found Mg II and Si II in the optical spectrum of this star, with Al II and Al III detected later in the *IUE* data (Holberg, Barstow & Sion, 1998). Holberg Barstow & Sion (1998) also categorise a strong C II line blueshifted with respect to the photosphere by 11 km s$^{-1}$ as circumstellar, as it also has a velocity inconsistent with the observed ISM lines (the mean $v_{\text{ISM}}$ found here is 1.5±1.8 km s$^{-1}$).

Using the 3922 Å Ca K line, Zuckerman & Reid (1998) found a significant amount of calcium in WD 1337+705 ($N$(Ca)/$N$(H) = 2.5x10$^{-7}$), which could be due a binary companion (Zuckerman & Reid, 1998) or the accretion of disrupted minor planets (Zuckerman et al. 2003). No infrared excess is found by Mullally et al. (2007), suggesting that this object does not in fact have a binary companion. Although the photospheric Mg II absorption symptomatic of a circumstellar gas disc is seen in the optical spectrum of this star, the Ca II and Fe I emission that accompanies it is not (Gänsicke et al., 2008). Therefore, one may expect circumstellar material to be present at WD 1337+705, although this is not evident in this analysis of high ion absorption features.

### 5.3.3.10. WD 2023+246 (Wolf 1346).

The coolest star in this sample ($T_{\rm eff}$ = 19,150±30 K), WD 2023+246 does not display circumstellar high ion absorption. In keeping with WD 1337+705, Bannister et al. (2003) found an Al III abundance of $N({\rm Al})/N({\rm H})$ = 2.2x10$^{-9}$, near the predicted value of ~10$^{-9}$ (Chayer et al. 1995), suggesting a photospheric origin. Bannister et al. (2003) also noted that the 1862 Å aluminium line requires a much higher abundance (6.0x10$^{-9}$) to yield a good fit (the unusual structure in the line profile, and thus the abundance, could also have been due to instrumental effects). Again, in spite of the inclusion of radiative levitation effects in the atmosphere of this DA, Dupuis, Chayer & Hénault-Brunet (2010) report that accretion must be occurring at this star to explain its metal photospheric abundances.

### 5.3.3.11. WD 2111+498 (GD 394).

Non-photospheric features are not seen at this star. Using the *GHRS* Si IV lines, the mean $v_{\rm phot}$ is found at 29.3±1.66 km s$^{-1}$, comparing well to the value of 28.75±0.91 km s$^{-1}$ derived by Bannister et al. (2003). $v_{\rm ISM}$ = −7.6±1.3 km s$^{-1}$. The *GHRS* data only covers the 1290 Å – 1325 Å and 1383 Å – 1419 Å ranges; the *IUE* data do not show any components of the C IV or N V doublets, or the O V line. Photospheric O VI has been observed in the *FUSE* spectrum of WD 2111+498. As discussed in some detail in the previous chapter, much evidence exists for the accretion of metal rich metal by this DA, although no source has yet been detected.

### 5.3.3.12. WD 2152–548 (REJ 2156–546).

The averaged $v_{phot}$ is calculated at −14.94±0.46 km s$^{-1}$, while the $v_{ISM}$ is −9.2±0.53 km s$^{-1}$. The projected LIC velocity is found at −9.73±1.31 km s$^{-1}$, suggesting that the observed ISM lines arise in the LIC. Akin to WD 0050–335, Bannister et al. (2003) suggested that WD 2152–548 might have circumstellar absorption in its C IV doublet co-add at −1.65±0.76 km s$^{-1}$, although this feature was on the edge of detectability. However, as such co-addition cannot be performed with the measuring technique utilised here, a circumstellar component is not recorded for this star.

### 5.3.3.13. WD 2211–495 (REJ 2214–492).

No clear evidence of circumstellar absorption is found at this star. Using the C IV, N V, O V and Si IV absorption features, a value of $v_{phot}$ of 32.33±1.37 km s$^{-1}$ is obtained, and the velocity of the ISM Si II line is found at −1.1 ± 0.4 km s$^{-1}$, consistent with the values of Bannister et al. (2003).

A large difference in centroid velocities across the C IV doublet seen by Bannister et al. (2003) is again noticed (the velocity of the 1548.187 Å component is 30.5±0.7 km s$^{-1}$ and the component 1550.772 Å is at 37.9±0.8 km s$^{-1}$), implying that a secondary component may be present. However, like WD 0948+534, fits of the slightly asymmetric 1548.187 Å component of the C IV doublet with two absorbers does not produce a substantially better fit. No other lines display any hint of a second component.

### 5.3.3.14. WD 2309+105 (GD 246).

No evidence of circumstellar absorption is present in the spectrum of this white dwarf. The *STIS* [E140M] $v_{phot}$ is found at −13.45±0.13 km s$^{-1}$, consistent with the value of −13.29±0.25 km s$^{-1}$ measured by Bannister et al. (2003). Given the lack of circumstellar features in the C IV, N V, O V or Si IV absorption lines, and that Barstow et al. (2010) report no detection of O VI in the *FUSE* data, the O VI line was not examined.

### 5.3.3.15. WD 2331−475 (REJ 2334−471).

Circumstellar absorption is not seen in the high ionisation absorption lines of this object. Bannister et al. (2003) found that double Gaussian fits were statistically preferred for the Si IV doublet (at 34.00 km s$^{-1}$ and 54.64 km s$^{-1}$) and in the N V 1243 Å component (at 19.7 km s$^{-1}$ and 43.51 km s$^{-1}$). However, since these dual components cannot be unambiguously identified, and the velocities of the dual absorbing components measured by Bannister et al. (2003) were not consistent between the ions, the absorption lines here are fit with single components. The $v_{phot}$ derived here is 38.88±0.72 km s$^{-1}$ and a single ISM absorbing component is found at 14.3±0.7 km s$^{-1}$.

## 5.4. Discussion.

Circumstellar detections are made at eight stars; WD 0232+035 (Feige 24), WD 0455–282 (REJ 0457–281), WD 0501+527 (G191-B2B), WD 0556–375 (REJ 0558–373), WD 0939+262 (Ton 021), WD 1611–084 (REJ 1614–085), WD 1738+669 (REJ 1738+665) and WD 2218+706. The results of this work are broadly consistent with those of Bannister et al. (2003), with a few exceptions.

Circumstellar detections are not made at some stars where they were made by Bannister et al. (2003), due to the modelling method used here. Similarly, the column densities measured here substantially differ in some cases, given that a curve of growth was used to derive column densities for the circumstellar material by Bannister et al. (2003), and a line profile modelling technique is utilised here. Since the method used here fully models the absorbing components rather than using a simplistic curve of growth, the values measured here are used in the discussion of this study. Indeed, in another study of WD 1738+669, Dupuis et al. (2009) measured C IV and Si IV column densities (5.01±0.48x10$^{13}$ cm$^{-2}$; 5.50±0.13x10$^{12}$ cm$^{-2}$) close to the values derived here (5.55±0.61x10$^{13}$ cm$^{-2}$; 3.26±0.17x10$^{12}$ cm$^{-2}$) using the apparent optical depth technique, lending weight to the assertion that the measurements made here are reliable. A summary of the differences between the work here and the findings of Bannister et al. (2003) is given in Table 5.4.

**Table 5.4.** The key differences between the findings of this study and that of Bannister et al. (2003).

| WD | Key differences. |
|---|---|
| 0050–335 | Bannister et al. (2003) found two components in the coadded C IV doublet; only absorbing one component is modelled here. |
| 0455–282 | The C IV column measured ($3.32\pm6.6\times10^{13}$ cm$^{-2}$) differs to that measured by Bannister et al. (2003; $1.82\times10^{14}$ cm$^{-2}$). Column density measurements are presented here for the N V and Si IV circumstellar components where they were not in the previous study. |
| 0501+527 | The C IV column measured ($1.04\pm0.1\times10^{14}$ cm$^{-2}$) is greater than the value stated by Bannister et al. (2003; $2.40\times10^{13}$ cm$^{-2}$). |
| 0556–375 | A C IV column density was measured here where it was not previously. |
| 0948+534 | A hint of circumstellar components to each of the high ion absorption features is seen here for the first time, though their velocities were inconsistent across absorption features and their addition to the line profile models is not statistically significant. |
| 1611–084 | Bannister et al. (2003) found a C IV column density of $3.16\times10^{13}$ cm$^{-2}$, whereas a column density of $9.77\pm1.9\times10^{12}$ cm$^{-2}$ is measured here. A Si IV column density measurement is made here where it was not by Bannister et al. (2003). |
| 1738+669 | No circumstellar N V is detected in this study. The column C IV and Si IV densities here differ ($5.55\pm0.61\times10^{13}$ cm$^{-2}$; $3.26\pm0.17\times10^{12}$ cm$^{-2}$) substantially to those measured by Bannister et al. (2003; $1.70\times10^{13}$ cm$^{-2}$; $1.66\times10^{13}$ cm$^{-2}$). |

**Table 5.4 -** *continued*

| WD | Key differences. |
|---|---|
| 2152–548 | Bannister et al. (2003) found two components in the coadded C IV doublet; only absorbing one component is modelled here. |
| 2218+706 | The C IV and Si IV column densities measured here ($1.19\pm0.17\times10^{13}$ cm$^{-2}$; $7.98\pm0.59\times10^{13}$ cm$^{-2}$) are somewhat different to those measured by Bannister et al. (2003; $4.17\times10^{13}$ cm$^{-2}$; $4.07\times10^{13}$ cm$^{-2}$). |

In this section, possible origins of the circumstellar absorption are explored, with a summary presented in Section 5.5. Debris discs are discussed in 5.4.1, while 5.4.2 details whether ionisation of the ISM by the white dwarfs can explain the observed circumstellar features. White dwarf mass loss is examined in 5.4.3 and PN material is explored in 5.4.4.

### 5.4.1. Circumstellar discs.

The circumstellar discs at cooler white dwarfs are located a few tens of stellar radii from their host stars. One would expect any solids in analogous circumstellar discs at hotter stars to sublimate, given the proximity of such discs to the parent star (e.g. von Hippel et al. 2007). A search for gaseous components to circumstellar discs at hot white dwarfs by Burleigh et al. (2010, 2011), which included the work from the previous chapter and further stars from the sample studied here, did not find any evidence of gaseous disc emission. Furthermore, no infrared excesses were seen at any of the stars included in this work in the surveys of Mullally et al. (2007) or Chu et al. (2011). WD 2218+706 does harbour an infrared excess (M. Burleigh, *priv. comm.*), although this may be due to the PN still present at the star. However, as discussed at length in the previous chapter, a lack of confirmed circumstellar disc detections at these stars does not completely rule out their presence.

Indeed, two white dwarfs (WD 0501+527 and WD 1611–084) have values of $v_{\text{CSshift}}$ comparable to $v_{\text{grav}}$, suggesting that the absorbing circumstellar material may be within a few tens of stellar radii of the star. Indeed, the $v_{\text{CS}}$ of WD 1611–084 is well separated from both the detected ISM and predicted LISM velocities, possibly an indication of circumstellar material. Furthermore, the $v_{\text{CS}}$ values at WD 0455–282, WD 0556–375 and WD 0939+262 are slightly separate from the detected ISM velocities along the sight lines to the stars. It may be the case, as suggested by Lallement et al. (2011), that these objects are ionising circumstellar planetesimals or material in a circumstellar disc. However, given how close $v_{\text{CS}}$ is to $v_{\text{LISM}}$ at these objects, it is not immediately clear that these stars are not ionising an ISM cloudlet loosely associated with the LISM; this is discussed in the next section. A key problem with interpreting the observed absorption features is, as discussed by Gänsicke et al. (2012), the debris accreted by cooler DZ stars is depleted of volatile elements, i.e. carbon poor and silicate rich; the converse is true here, with strong carbon detections and secondary silicon detections (and in some cases circumstellar nitrogen and oxygen detections). Indeed, the one circumstellar high ion detection made in Gänsicke et al.'s (2012) study was in the Si IV doublet only, lending weight to the argument that if the material observed here were in a circumstellar disc, its origin is not the tidal disruption of planets or asteroids.

The precise interaction of circumstellar discs with the intense radiation field to a hot white dwarf is poorly understood. A model that predicts the details of the star-disc interaction, and makes testable, quantifiable predictions is required.

## 5.4.2. Ionised ISM.

Given the inability of material inside the potential wells of the stars to explain the observed circumstellar lines, another explanation is the ionisation of the ISM. Indebetouw & Shull (2004) outlined possible reasons for the existence of high ions in the ISM, including evaporating ISM cloudlets, planar conduction fronts, cooling Galactic fountain material, the evaporation of hot gas in stellar wind and supernova bubbles, turbulent mixing layers and white dwarf Strömgren spheres. If the high ion absorption features observed in this work arise in ISM processes such as these, one would expect such features to be seen along the lines of sight to both the white dwarfs and nearby stars (except for the white dwarf Strömgren sphere model, where the ionised material is local to the star). A search[8] of *IUE* and *HST STIS, GHRS* and *COS* data show no evidence of similar non-photospheric absorption in the C IV, N V, O V or Si IV absorption lines in the spectra of objects along sight lines within five degrees of the white dwarfs. This indicates that if ionisation of the ISM is occuring, it is in the vicinity of the white dwarfs and not in an ISM process.

Since the study of Bannister et al. (2003), improvements have been made in the mapping the morphology of the LISM (Redfield & Linsky, 2008). Using these maps, the LISM cloud(s) traversing the line of sight to each white dwarf was identified and the projected velocity of each cloud calculated (Table 5.2, column six). This allowed a comparison between both the detected ISM components and the predicted LISM, and an examination of other (unresolved) ISM components related to the LISM along each the sight line to each objects. Comparing $v_{\text{CSshift}}$ to the shift in the predicted and observed ISM velocities shows the circumstellar and interstellar velocities line up or are very close in many cases (Figure 5.6).

---

[8] using the online tool at http://archive.stsci.edu/xcorr.php

The photoionisation of the Hyades cloud by WD 0501+527 (at 50 pc) has been used to explain both the similarity in high and low interstellar ion dynamical properties (Redfield & Linsky 2008) and the high electron density (~0.5 cm$^{-3}$) along the sight line to the star (Redfield & Falcon 2008). Indeed, Tat & Terzian (1999) found that WD 0501+527 has a Strömgren sphere that extends into the space within 20 pc of the Sun. Thus, the circumstellar absorption seen here is attributed to the ionisation of the Hyades cloud. However, caution must be exercised; the metal depletion along the sight line to the white dwarf is much higher than those along other sight lines through the Hyades cloud. This may be evidence of multiple ISM components along the sight line to this star, which are unresolved here. This has been seen in other studies of the ISM; Welsh et al. (2010b) found that using high-resolution observations, multiple component, complex absorption profiles can be seen where only one absorber was detected in lower resolution data. In such cases, the velocities all occupy a narrow range along the line of sight, around the LISM velocity. The LIC does not extended beyond 13 pc along this sight line (Redfield & Falcon 2008), constraining the distance to the detected secondary ISM component, which has a velocity in keeping with $v_{\text{LIC}}$.

At WD 1738+669 and WD 2218+706, $v_{CS}$ and $v_{ISM,pri}$ match up well, indicating that these stars may be ionising the ISM in their localities, causing the observed absorption. These stars are too distant (at 243 and 436 pc) to be able to ionise the LISM; indeed the projected LISM velocities (due to the LIC) are well separated from $v_{CS}$ and $v_{ISM}$ ($v_{ISM,pri}$ at WD 2218+706). The secondary ISM component detected in the spectrum of WD 2218+706 has a velocity in keeping with $v_{LIC}$, although the error on $v_{ISM,sec}$ is large (4.01 km s$^{-1}$).

At WD 0455–282, $v_{CS}$ (18.8±3.47 km s$^{-1}$) is separate from both $v_{ISM}$ (12.1±1.5 km s$^{-1}$) and $v_{LISM}$ (12.56±1.03 km s$^{-1}$, due to the blue cloud). Given that these velocities are, within errors, quite close, it may be that this star (at 108 pc) is ionising some ISM in its Strömgren sphere that is loosely associated with the LISM, and thus has a velocity similar to it. Similarly, the $v_{CS}$ at WD 0939+262 (9.38±6.6 km s$^{-1}$) is far from $v_{ISM}$ (–2.1±0.2 and –34.7±1.5 km s$^{-1}$) and yet close to $v_{LIC}$ (10.81±1.29 km s$^{-1}$), suggesting that a 'LIC like' cloudlet may be being ionised. The $v_{CS}$, $v_{ISM,pri}$ and $v_{LISM}$ values (10.2±1.07, 7.8±1.0 and 11.36±0.95 km s$^{-1}$) at WD 0556–375 indicate an analogous situation may be present along the line of sight to the DA. As well as a lack of high ion absorption features along nearby sight lines to these stars, Welsh et al. (2010b) found no Na I absorption along sight lines near these stars with velocities near $v_{CS}$ (although no data were present within ten degrees of WD 0939+262), indicating that whatever the source of the circumstellar high ions, they do not arise in the nearby ISM.

The $v_{CS}$ detected in the spectra of WD 0232+035 and WD 1614–084 are well separated from all detected and predicted ISM components, suggesting that the ionisation of the ISM by these stars does not explain the observed circumstellar ions. Other scenarios (such as the circumstellar discussion in Section 5.4.1, or the ionisation of mass lost from the binary companion of WD 0232+035) may explain the observations of these stars.

In addition to velocity data, column densities were measured for all circumstellar and ISM detections. Indebetouw & Shull (2004) collated a table of predicted C IV, N V, Si IV, and O VI column densities for the models considered in their study. The column density ranges of only two models coincide with the column densities seen here; the white dwarf Strömgren sphere model and the 4 $M_\odot$ cooling fountain (where hot, supernova heated gas rises to a 1 kpc scale height above the disc, cools and falls back to the galactic plane; Shapiro & Field, 1976). However, the range of column densities predicted by the 4 $M_\odot$ cooling fountain covers only a small part of the range of detected column densities, while the white dwarf Strömgren sphere covers the full range of observed column densities.

The predicted metal column densities from the Strömgren sphere model in Table 5.5 relate to a limited $n_H$ range. The full predicted column density ranges span $0.44 \times 10^{12}$ < Si IV < $4.4 \times 10^{12}$ cm$^{-2}$, $7.8 \times 10^{12}$ < C IV < $77 \times 10^{12}$ cm$^{-2}$, $1.2 \times 10^{12}$ < N V < $12 \times 10^{12}$ cm$^{-2}$ and $1.4 \times 10^{12}$ < O VI < $20 \times 10^{12}$ cm$^{-2}$ (Dupree and Raymond, 1983). On an object-to-object basis, the predicted metal column densities do not match those observed. However, when the limits of the range of expected metal column densities are applied to the observed column densities, good agreement is observed (Figure 5.7).

Several limitations exist in the model of Dupree & Raymond (1983). As noted by Bannister et al. (2003), a single DA Strömgren sphere model with $T_{eff}$ = 60,000 K was constructed; given that the range in $T_{eff}$ for the stars with circumstellar absorption here is 38,840 - 69,711 K, one can expect significant variety in Strömgren radii ($r_S$) values when compared to the model, causing a variation in the expected column density. Also, the model did not include photospheric metals, which act to reduce the radiated stellar flux, reducing $r_S$. This effect is more significant for $T_{eff}$ > 50,000 K, when iron peak elements blanket the observed continuum (although the redistributed flux may still contribute to the Strömgren sphere through lower energy transitions). Seven of the eight stars with circumstellar absorption have $T_{eff}$ > 50,000 K, making this omission a serious flaw in the physical model. The ISM was also assumed to be isothermal (at 40,000 K) and of uniform density. For these reasons, the column densities predicted should be treated as a rough estimate of scale, rather than precise predictions.

Coupling the distances to the white dwarfs (Table 5.1) with $r_S$ estimates can yield an estimate to the distance of the ionised material. Tat & Terzian (1999) provided $r_S$ estimates for a variety of hot white dwarfs, for $n_e$ = 0.01 cm$^{-3}$ and 0.03 cm$^{-3}$. Using the $r_S$ values for the model star with the $T_{eff}$ closest to each of the stars with circumstellar absorption, a series of $r_S$ estimates can be made for the DAs here (Table 5.6).

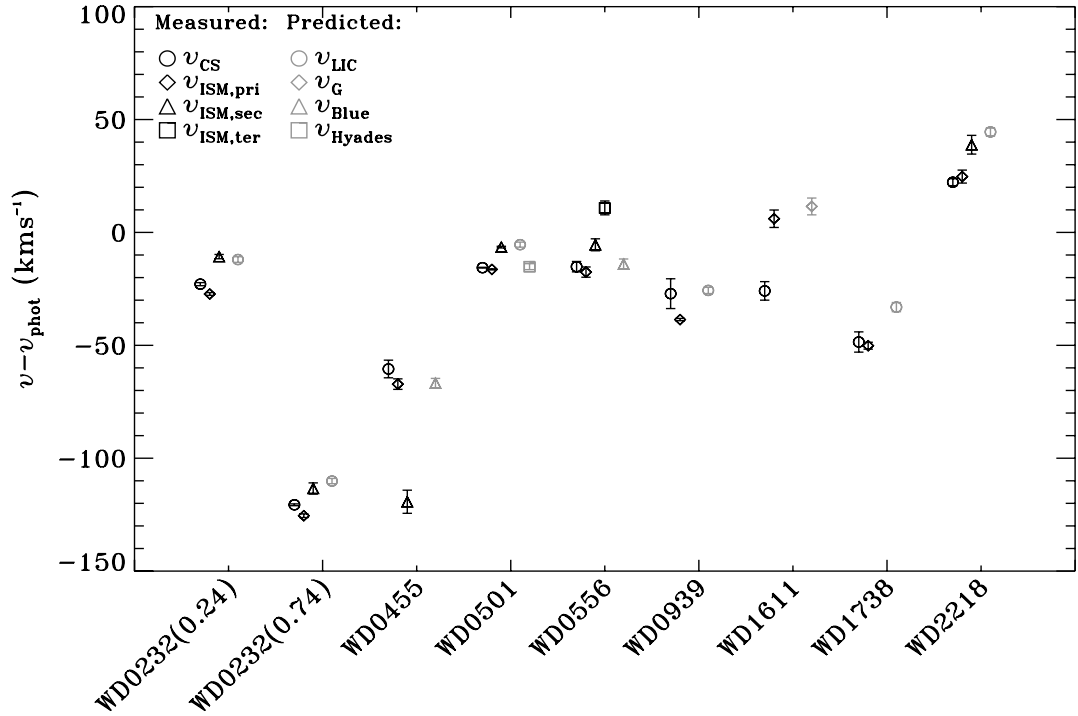

**Figure 5.6.** A plot of $v_{CSshift}$, with the shifts in the measured and predicted ISM components (from Table 5.2). The measured velocity shifts are plotted in black, while the predicted shifts in $v_{LISM}$ are plotted in grey. In some cases the error bars are smaller than the plot symbols; the symbols are open to allow the error bars to be seen.

**Table 5.5.** The metal column densities for a range of ISM models (Table 1, Indebetouw & Shull, 2004). All column densities are expressed in units of $10^{12}$ cm$^{-2}$.

| Model | Si IV | C IV | N V | O VI |
|---|---|---|---|---|
| Evaporating cloudlet[*] | . . . | 1.2 – 1.5 | 0.5 – 0.6 | 9 – 12 |
|  | 0.10 – 0.14 | 2.7 – 3.8 | 1.0 – 1.2 | 12 – 14 |
| Planar Conduction front[*] | 0.10 – 0.16 | 1.6 – 3.2 | 0.6 – 1.0 | 8 – 10 |
|  | 0.029 – 0.097 | 0.89 – 2.7 | 0.40 – 1.0 | 6.7 – 14 |
| Stellar wind bubble | 0.21 – 0.25 | 3.3 – 4.0 | 1.3 – 1.6 | 21 – 25 |
| SNR bubble[*] | 0.4 – 0.6 | 6.3 – 10 | 3.2 – 5.0 | 40 – 79 |
|  | ~ 0.52 | ~ 7.8 | ~3.6 | ~47 |
| Halo SNR bubble | . . . | 8 – 15 | 3.4 – 7.9 | 35 – 150 |
| 4M$_\odot$ cooling | 3.3 – 6.4 | 43 – 79 | 28 – 36 | 580 – 600 |
| 40 pc cooling cloud | ~ 25 | ~ 50 | ~ 13 | ~ 200 |
| Turbulent mixing layer | 0.0010 – 0.47 | 0.025 – 6.8 | 0.0022 – 0.32 | 0.017 – 0.81 |
| White dwarfs ($0.1<n\mathrm{H}<1$ cm$^{-2}$)[†] | 1.4 – 4.4 | 25 – 77 | 3.7 – 12 | 5.6 – 20 |

[*]the different rows correspond to different models references; [†]from Dupree & Raymond (1983).

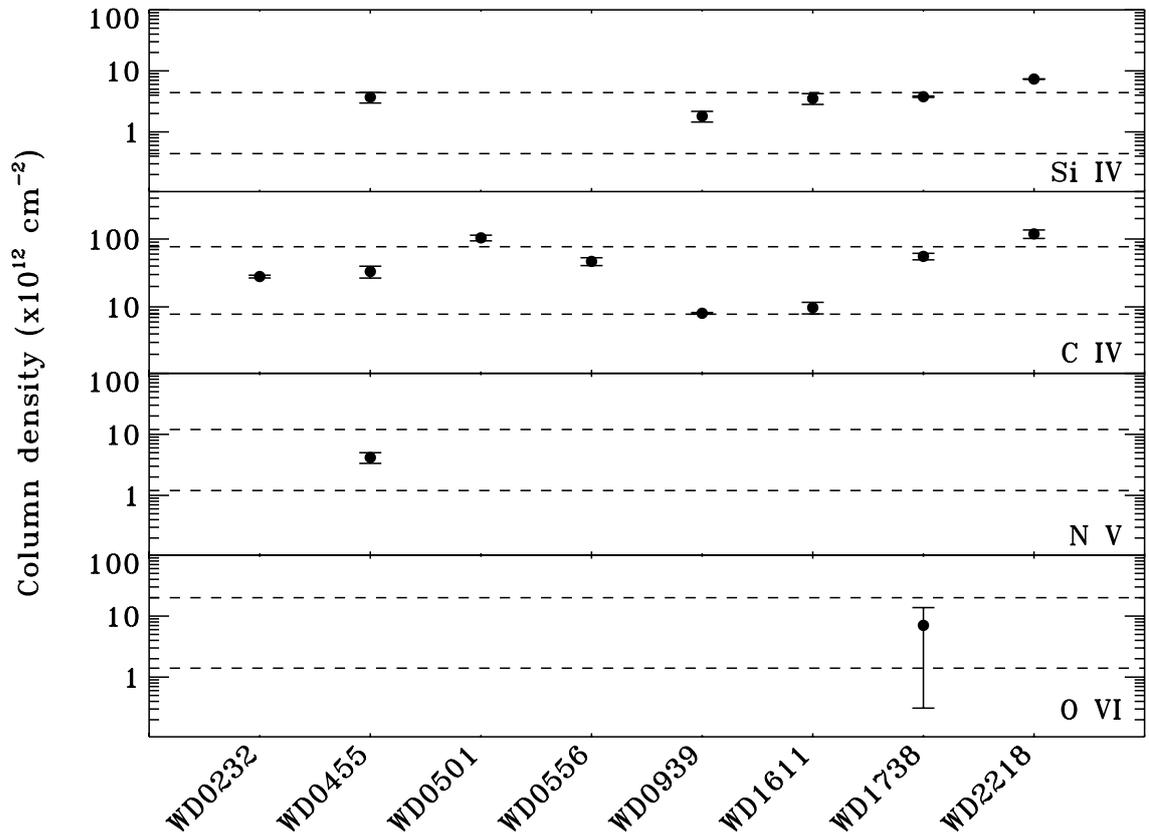

**Figure 5.7.** The column densities measured in this study, with the column density ranges predicted by Dupree & Raymond (1983) for a DA Strömgren sphere (dashed lines).

**Table 5.6.** The estimated $r_S$ (from Tat & Terzian, 1999) for each of the DAs with circumstellar absorption, and estimated distances ($D$) to the absorbing material (for $n_e = 0.01$ and $0.03$ cm$^{-3}$). $D$ was calculated by subtracting $r_S$ from the distance to the star (from Table 5.1). The ISM component with a velocity matching $v_{CS}$ is stated; a '?' denotes the tentative association with the Hyades cloud. All $r_S$ and $D$ values are expressed in pc.

| Star | $n_e = 0.01$ cm$^{-3}$ | | $n_e = 0.03$ cm$^{-3}$ | | ISM component |
|---|---|---|---|---|---|
| | $r_S$ | $D$ | $r_S$ | $D$ | |
| WD 0232+035[a] | 33.66 | 44.34 | 16.18 | 61.82 | |
| WD 0455−282[b] | 26.00 | 82 | 12.50 | 95.5 | |
| WD 0501+527[b] | 26.00 | 24 | 12.50 | 37.5 | ISM, pri (Hyades?) |
| WD 0556−375[a] | 33.66 | 261.34 | 16.18 | 278.82 | |
| WD 0939+262[c] | 39 | 178 | 18.86 | 198.14 | |
| WD 1611−084[d] | 18.43 | 67.57 | 8.86 | 77.14 | |
| WD 1738+665[c] | 39 | 204 | 18.86 | 224.14 | ISM, pri |
| WD 2218+706[b] | 33.66 | 402.34 | 16.18 | 419.82 | ISM, pri |

Note: the $r_S$ values adopted here were those calculated for the star with the nearest $T_{eff}$ by Tat & Terzian 1999 to the star in question here, and are signified as follows: [a]WD 0501+527 ($T_{eff} = 61\,160$ K); [b]WD 0232+035 ($T_{eff} = 50\,000$ K); [c]WD 1211+332 ($T_{eff} = 70\,000$ K); [d]WD 2111+498 ($T_{eff} = 39\,800$ K).

In the absence of consistent ISM components, the distances represent the distance to the boundary of the Strömgren sphere.

Again, these $r_S$ estimates were made in the presence of several oversimplifications similar to those already discussed; the basic formalism of Strömgren (1939) was used, with an isothermal ISM at 7,500 K and discreet $n_e$ values. A more thorough approach to estimating $r_S$ for the stars in this sample would be to fully model the interaction of the stellar flux (taking proper account of UV photospheric metal absorption) with the LISM (using measured ISM temperatures and electron densities).

The study of the relationship of $v_{CSshift}$ and $T_{eff}$ by Lallement et al. (2011) using the three DAs in their sample, as well as the samples of Holberg et al. (1998), Bannister et al. (2003) and Barstow et al. (2010), reported considerable spread around a trend of decreasing $v_{CSshift}$ with $T_{eff}$. This was put down to the inclusion of a variety of white dwarf environments, such as PN and binary systems. The variation in photospheric and ISM component velocities along the sight lines to the stars here offers another explanation for the observed dispersion in $v_{CSshift}$.

This clearly demonstrates the importance of the white dwarf circumstellar environment to our understanding of the ISM, and the need for a detailed analysis of the spectra of hot white dwarfs in which 'circumstellar' high ions are seen, to ascertain the true origin of these absorption features.

### 5.4.3. White dwarf mass loss.

A possible source for the observed circumstellar absorption put forward by previous authors (e.g. Holberg et al. 1998, 1999; Bannister et al. 2003) is hot white dwarf mass loss. Using the mass loss theory of Abbott (1982), Bannister et al. (2003) found that the stars without circumstellar absorption had a decreasing mass loss rate with decreasing $T_{\text{eff}}$, while the stars with circumstellar absorption did not follow this decay, concluding that mass loss played some part in the presence of the circumstellar absorption features.

Since the calculations of Bannister et al. (2003), the metal abundances of the stars in the sample have been revised (Barstow et al. 2012, *in preparation*; Chapter 3 of the thesis). In light of this, the mass loss calculations of Bannister et al. (2003) are repeated here, to check whether a relationship between mass loss rate and circumstellar features persists when the updated photospheric metal abundances are considered. The formalism of Abbott (1982) provides a mass loss rate ($\dot{M}$), given a stellar metallicity (relative to solar abundances; $Z$) and luminosity ($L$), using

$$\dot{M} \approx 2 \left(\frac{L}{L_\odot}\right)^2 \frac{Z}{0.02} \text{M}_\odot \text{yr}^{-1}.$$

The metallicity of each star is calculated using

$$Z = \left[\sum_{z>2} A_*(z)\right] \left[\sum_{z>2} A_\odot(z)\right]^{-1},$$

where $A_*(z)$ is the abundance of the element with the atomic number $z$ relative to hydrogen ($A_\odot(z)$ is the solar abundance of the metal). The metal abundances used in the calculations are detailed in Appendix A and the stellar luminosities are stated in Table 6.1.

When the mass loss rates computed by Bannister et al. (2003) are compared to the mass loss rate calculated here, a few differences emerge (Figure 5.8). The mass loss rates for WD 0050–335, WD 1029+537 and WD 1611–084 are revised down significantly, due to their greatly reduced N V abundances (see Chapter 3). WD 0050–335, having been moved to the non-circumstellar population, does not look out of place with its new mass loss rate. Given the lower mass loss rate of WD 1611–084, the pattern identified by Bannister et al. (2003) of the white dwarfs with circumstellar absorption having high mass loss rates is broken. Indeed, the strongest determinant of mass loss rate calculated by the formalism of Abbott (1982) is $T_{eff}$, and the star with the highest mass loss rate (WD 0621–376) does not display circumstellar features.

While the Abbott (1982) mass loss calculation method was used by Bannister et al. (2003), and is repeated here to allow a comparison to their results, several limitations exist in its application to hot white dwarfs. As discussed by Bannister et al. (2003), this formalism was developed for the stellar envelopes of O-G type stars, and is found to be most applicable to OB stars. Being concerned with main sequence stars, the theory does not explain the mass-loss of Wolf-Rayet stars, which have a different stellar structure to main sequence stars. Since white dwarfs are also structured somewhat differently to main sequence stars, this casts doubt on the applicability of this theory to white dwarfs.

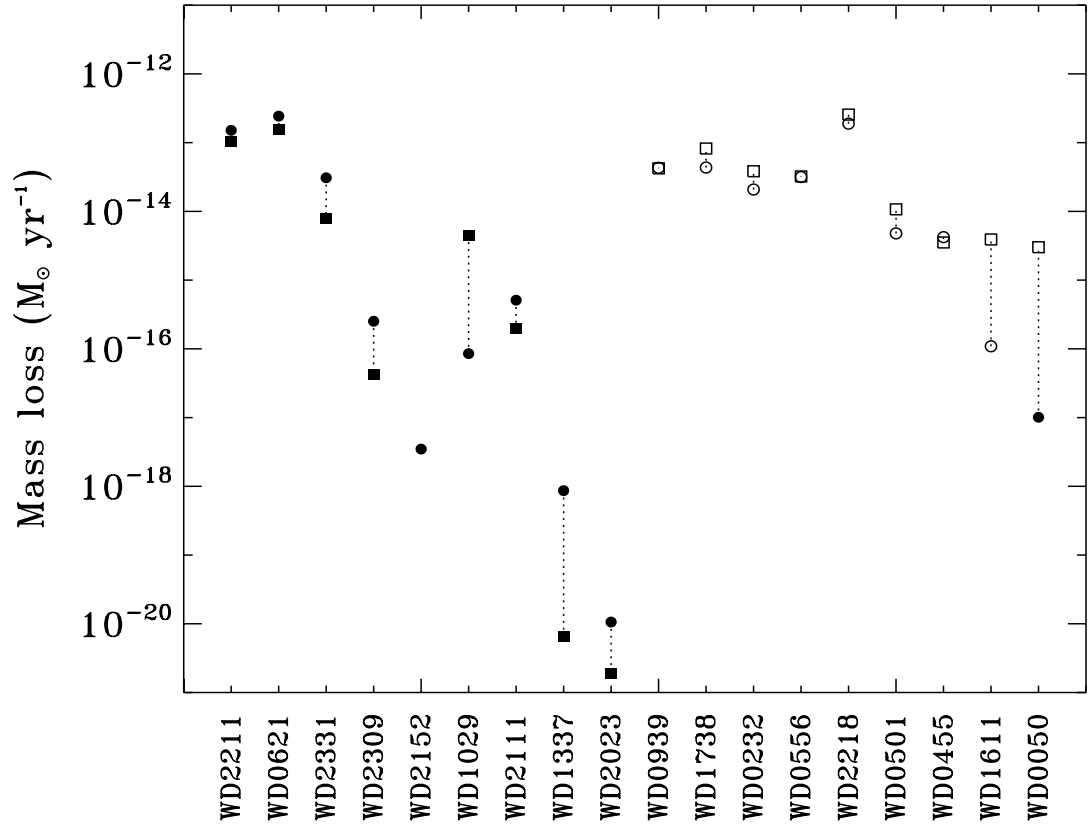

**Figure 5.8.** A comparison of the mass loss calculations performed here using the metal abundances in Appendix A (circles) to those of Bannister et al. (2003; squares). The stars without circumstellar absorption lines are the filled symbols, while the stars with circumstellar absorption are plotted with open symbols. The stars are plotted from left to right in order of decreasing $T_{eff}$ to show the trend seen by Bannister et al. (2003).

The formula of Abbott (1982) is based on the parameterisation of the dependence of the radiative force on the wind optical depth, which is only valid in a limited range, and predicts an increasing radiative force with decreasing wind density. However, this force will saturate (at a point depending on the metallicity of the material; Kudritzki, 2002); this is not included in the Abbott (1982) model. Therefore, the formalism always predicts a non-zero mass loss, even for compact stars such as white dwarfs where at some point the radiative force will not be able to overcome the gravity of the star.

More recent studies of hot white dwarf mass loss show that for 25,000 K ≤ $T_{eff}$ ≤ 50,000 K, no winds can exist for a star with log $g$ > 7.0 and a solar or sub-solar metallicity, since the radiative acceleration saturates below the gravitational acceleration of the star (Unglaub, 2008). Similarly, a steady mass loss cannot be sustained for a DA with $T_{eff}$ = 60,000 K and log $g$ > 7.0 (Unglaub, 2007). Given that the lowest log $g$ of any of the white dwarfs with circumstellar absorption is 7.05 (WD 2218+706), it can reasonably be concluded that mass loss does not account for the observed circumstellar lines in the spectra of WD 0455–282, WD 0501+527, WD 0556–375, WD 1611–084 and WD 2218+706, given that the $T_{eff}$ values of these stars are all lower than 60,000 K. The remaining white dwarfs (WD 0232+035, WD 0939+262 and WD 1738+669) do not have $T_{eff}$ values much greater than 60,000 K.

No rigorous mass loss calculations have been made for $T_{eff}$ > 60,000 K, although it seems unlikely mass loss can exist in these hotter DAs. In thin winds, the majority of the radiative force comes from the CNO elements (Vink et al. 2001). When more and more particles of these elements change into the helium like ionisation stage at higher $T_{\text{eff}}$, the radiative force is reduced, since the lines of this ionisation stage are at very short wavelengths away from the flux maximum. Thus, it is not immediately clear that winds are more likely to exist in DA stars with $T_{\text{eff}} > 60{,}000$ K, despite their higher luminosity.

If any winds were to be present at all, they would consist only of metals since hydrogen would be in hydrostatic equilibrium. Following the arguments of Seaton (1996), initial estimates for a DA with $T_{\text{eff}} = 66{,}000$ K and $\log g = 7.7$ suggest that the outward flow of CNO elements in an otherwise hydrostatic atmosphere cannot exceed a rate of the order $10^{-19}$ $M_\odot$ yr$^{-1}$ (Unglaub, *priv. comm.*). Indeed, a mass loss rate much higher would rapidly empty the white dwarf atmosphere of CNO elements. If mass loss were to occur, the winds must have a velocity of the order of a few thousand km s$^{-1}$, since that is roughly the escape velocity of a white dwarf. In the case of sdB stars, mass loss occurs at a rate of the order $10^{-11}$ $M_\odot$ yr$^{-1}$, and the metallic winds are accelerated up to several thousands of km s$^{-1}$, exceeding the escape velocity of the star (Votruba et al. 2010). No circumstellar absorption features like those studied here have been reported. Though no calculations have been performed detailing the mass loss rates required to produce such circumstellar absorption, it could be argued that even if such winds were to exist at the hottest white dwarfs with circumstellar absorption, the combination of a low mass loss rate and high wind velocity would cause a thin wind to be present, which is unlikely to account for the observed absorption features since they are not seen in sdB stars with higher mass loss rates and similar wind velocities.

## 5.4.4. Planetary nebula material.

Since one of the stars (WD 2218+706) with circumstellar absorption in its spectrum is a bona fide CSPN, the possible link between ancient, diffuse CSPN and circumstellar UV absorption is again examined. Following the methodology of Bannister et al. (2003), the $v_{phot}$–$v_{CS}$ values for the white dwarfs found here are plotted with the planetary nebula expansion velocities of Napiwotzki & Schönberner (1995; Figure 5.9).

Given the broad consistency between the velocities measured here with those measured by Bannister et al. (2003), it is perhaps not surprising that this line of analysis produced similar results. Again, the $v_{phot}$–$v_{CS}$ for WD 2218+706 is inconsistent with the expansion velocity of the PN at the star. The interpretation of Bannister et al. (2003) was that the non–photospheric absorption seen at this star is not located inside the expanding PN, but that the PN was a possible source of material. Bannister et al. (2003) also suggested that the suspected PN at WD 1738+669 might also account for the observed circumstellar absorption.

Since the study of Bannister et al. (2003), however, evidence has emerged that WD 1738+669 is not a CSPN, and the material associated with it is ionised ISM inside the Strömgren sphere of the star (Frew & Parker, 2006). At WD 1738+669, $v_{CS}$ matches $v_{ISM,pri}$; a similar situation exists at WD 2218+706. Given this, and that the material at WD 1738+669 is now thought not to be related to an ancient PN, further strength is given to the argument that the material seen at WD 2218+706 is not contained within the PN at the star and it is in fact ionised material inside the white dwarf Strömgren sphere. Whether the material was once resident in the PN and has now become part of the ISM, or whether the ionised ISM is from another source, cannot be determined from this study alone.

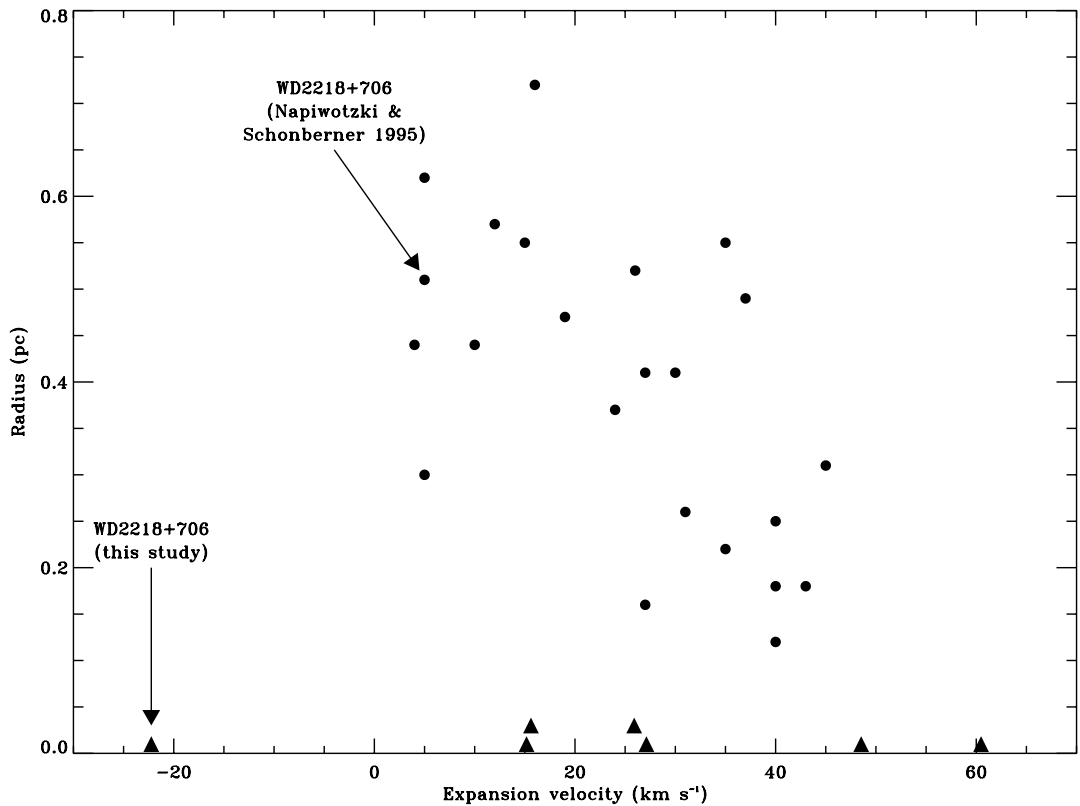

**Figure 5.9.** A comparison of the $v_{\text{phot}} - v_{\text{CS}}$ values here (triangles) to the $v_{\text{exp}}$ measurements from Napiwotzki & Schönberner (1995; circles). Since no nebula radius measurements exist for the stars in this sample, the $v_{\text{phot}} - v_{\text{CS}}$ values are plotted at zero pc. Overlapping values are offset for clarity, and the values for WD 2218+706 are labelled.

Given the highly complex, localised structures seen in high-resolution observations of PNe such as the Helix Nebula (Figure 1.4). It may be that highly localised, thus far undetected, PN remnant structures lie along the sight lines to these hot DAs, providing the observed absorption. A detailed investigation of this scenario will form the basis of future work.

## 5.5. Summary.

- Circumstellar features have previously been seen at eight hot DA white dwarfs, with two further possible cases reported. Given the recent advancements made in the study of circumstellar discs at cooler white dwarfs, the morphology of the ISM and hot white dwarf mass loss it is desirable to better understand the source of this material.

- The material was measured using a technique that, for the first time, provided column densities for all non-photospheric absorption features, rather than providing a curve of growth for a few discreet absorption line *b* values for most circumstellar features. Unambiguous signatures of circumstellar absorption were again detected at eight stars.

- Several physical scenarios from which the material could originate were discussed, including circumstellar discs, the ionisation of the ISM near the stars, stellar mass loss and ancient PNe.

- If circumstellar discs are present at a few tens of stellar radii, it is reasonable to assume they would not persist as solids. However, no signs of gas discs have been found at the stars, although accretion is required at some of the white dwarfs to explain their photospheric abundances. While such discs have not been detected, the results here do not rule out their presence. The differences between $v_{CS}$ and the measured and projected ISM velocities may be explained by the ionisation of circumstellar material that is not associated with the ISM. Indeed, the $v_{CS}$ detected at WD 1611–084 is far from all ISM components. However, it is unlikely that any circumstellar discs present would be due to the tidal disruption of planets or asteroids, given the ubiquitous C IV detections.

- The circumstellar absorption seen at WD 0501+527 has been attributed to the photoionisation of the Hyades cloud by the intense UV radiation of the star, in keeping with other studies. The ionisation of the ISM by WD 1738+669 and WD 2218+706 may explain the circumstellar absorption, given the agreement in $v_{CS}$ and $v_{ISM,pri}$ at these stars ($v_{ISM}$ at WD 1738+669). Some difficulty is met when linking the circumstellar absorption at WD 0455–282, WD 0556–375 and WD 0939+262, where $v_{CS}$ does not match up well with the detected ISM components. It may be the case, given the distances to the stars and the similarity in $v_{CS}$ and $v_{LISM}$, that currently unresolved ISM cloudlets with velocities near $v_{LISM}$ are being ionised.

- The range of measured circumstellar column densities are consistent with the range of column densities predicted by hot white dwarf Strömgren sphere models. Although the observed $v_{CS}$ of WD 0232+035 is inconsistent with both measured and predicted ISM velocities along the sight line to the star, the Strömgren radius is estimated to be of the order of 16–33 pc. The ionisation of mass lost from the dwarf companion to this star could account for the observed circumstellar features.

- The mass loss calculations made by Bannister et al. (2003) were repeated with more up to date stellar metal abundances, and the relationship previously seen between mass loss and circumstellar features no longer holds. Indeed, the method used by Bannister et al. (2003) has since been found to be inappropriate for hot DA stars. More recent studies show that mass loss cannot occur for DAs with $T_{eff}$ < 60,000 K, and is unlikely to occur in sufficient quantities to explain the observed circumstellar features in hotter stars, if at all.

- The circumstellar absorption at the CSPN WD 2218+706 is again found not to share the PN expansion velocity. Studies since that of Bannister et al. (2003) have found the ionised gas around WD 1738+669 is not from a PN but is instead ionised ISM inside the white dwarf Strömgren sphere. This suggests that the circumstellar features observed at theses stars are again ionised ISM inside the Strömgren spheres of the star, though a localised structure left from the PN phase may also be present.

- This study indicates that the origin of the observed circumstellar absorption may be the ionisation of material inside the white dwarf Strömgren spheres.

This highlights the great influence the radiation field of white dwarfs has over the nearby space, and contributes to our understanding of the ionised ISM.

- Alternatively, the circumstellar material observed may be due to the ionisation of circumstellar discs/planetesimals. A model that details testable predictions that can be compared to these observations will be insightful, and offer an understanding of the evolution of extrasolar planetary systems through the hot white dwarf phase.

- Higher resolution observations of the ISM along the sight lines to the stars will allow a better charaterisation of any as yet unresolved interstellar cloudlets. More realistic models of the Strömgren spheres of hot white dwarfs, that account properly for the properties of the ISM near the stars, is needed, as is a model of the interaction of circumstellar material with hot DA stars.

# *Chapter 6.*

# Conclusions and suggestions for further work.

## 6.1. Introduction.

This chapter, the final in this thesis, pulls together the results of the previous three science studies. A brief statement of the findings of each investigation is given, followed by a discussion of how these results help to further shape our understanding of white dwarf stars. Suggestions for future work that can extend the findings of this study and improve our understanding of these fascinating objects are made.

## 6.2. Concluding remarks and suggestions for further work.

The first science chapter of this work deals with the controversy over whether the nitrogen in the stars WD 0050–335, WD 1029+537 and WD 1611–084 can be modelled with homogeneous nitrogen distributions, with abundances not inconsistent with those of DAs with higher $T_{eff}$, or whether higher abundance, stratified nitrogen configurations are required. The observations of these stars are found to be well explained by homogeneous models. A degeneracy is observed, where two minima are present in the nitrogen model $\chi_v^2$ distribution of WD 0050–335 and WD 1029+537, near the two abundances previously reported for these stars. In each of these stars, the lower abundance model has a global $\chi_v^2$ minimum that was well separated (>3σ) from the secondary minimum. In WD 1611–084, the minima are less easy to disentangle. The model line profiles for the best fitting abundance do not reproduce the observed data well, and the lower abundance, homogeneous model is adopted as the best fit, since this is the case at the other stars. A comparison of the abundances derived here show that they are much closer to the predictions of Chayer et al. (1995) than previous measurements.

A search was carried out for signatures of circumstellar gas discs in a sample of eight white dwarfs, which have enhanced metal abundances when compared to theoretical predictions. At none of these stars is any metal line emission observed. In other studies, no infrared excesses were seen at any of the stars.

The final research chapter contains a re-analysis of a sample of 23 hot white dwarfs, of which eight had previously been found to exhibit non-photospheric high ion absorption features, with two further marginal detections. Circumstellar absorption is measured at the eight objects where it has been seen previously and, for the first time, proper modelling of the circumstellar absorption lines allowed column densities to be derived for all circumstellar absorption features.

Four possible scenarios for the origin of this material were examined. Mass loss is found not be significant at these stars, if it occurs at all, and thus is not thought to be the cause of the circumstellar absorption. No firm evidence that ancient planetary nebula are the origin of the material is found, as only one of the objects (WD 2218+706) is a CSPN, and the observed velocity of the circumstellar material at that star is inconsistent with measurements of the PN expansion velocity. This may, however, be an oversimplification, since the morphology of planetary nebulae can be highly inhomogeneous with extremely localised structures (such as those seen in the HST image of the Helix Nebula; Figure 1.4). It may be the case that the circumstellar absorption observed is due to some highly localised material left over from the PN that would have been around the white dwarfs earlier in their lives. A more detailed investigation of this will be the subject of future work.

While no evidence of circumstellar discs has been found at the stars, and the precise behaviour of a circumstellar disc at such hot white dwarfs is poorly understood, the observations do not allow their dismissal; the circumstellar velocity shifts in two of the stars are in keeping with the gravitational redshifts expected at the objects, while at the other DAs (except WD 1611–084) all of the circumstellar velocity shifts are blue shifted with respect to the photosphere, as may be expected were the material in a disc about the star. Furthermore, the $v_{CS}$ detected at three other stars (WD 0455–282, WD 0556–375 and WD 0939+262) do not align exactly with the measured ISM component velocities, suggesting that the circumstellar material may be due to the ionisation of planetesimals near the stars, or circumstellar discs.

The characteristics of the circumstellar material detected at WD 0501+527 are in keeping with what would be expected if this star were ionising the Hyades cloud. The $v_{CS}$ at WD 0455–282, WD 0556–375 and WD 0939+262 are near the projected $v_{LISM}$ values along the sight lines to the stars. This suggests, that instead of the ionisation of circumstellar material, these stars may be ionising material associated with the LISM. These stars, with distances ranging from 108 – 295 pc, are too distant to be able to ionise the LISM. However, it could be the case that ISM cloudlets along the sight lines to the stars with velocities near that of the LISM are being ionised. Indeed, Welsh et al. (2010b) found that when re-observed with high-resolution instruments, multiple absorbing components can be detected where single ISM components were previously seen in lower resolution data.

This suggests that the UV flux of the stars may be ionising the nearby ISM, forming a Strömgren sphere type structure. Estimates of the radii of these spheres shows that the binary companion of WD 0232+035 sits well inside this ionisation sphere, and the mass lost from the dwarf companion may be being ionised by the DA. The column densities of the observed circumstellar absorption features are consistent with the predictions of Dupree & Raymond (1983), for a DA Strömgren sphere. However, this model contains several limiting assumptions, and needs to be recalculated, taking proper account of the effect of photospheric metals, NLTE effects and local ISM temperatures and densities.

While the finding that the nitrogen in WD 0050–335, WD 1029+537 and WD 1614–084 can be modelled homogeneously closes a minor controversy, the application of this result to our understanding of circumstellar absorption at hot white dwarfs is interesting. A link between the previously measured, anomalously high metal abundances was previously made to the observed circumstellar absorption at WD 1611–084, and the possible circumstellar absorption at WD 0050–335 (e.g. Bannister et al. 2003). This work shows that such a link can no longer be claimed, since the abundances measured here are not as exotic as was once thought. Indeed, the previously accepted, high metal abundances of these stars caused Bannister et al. (2003) to calculate high mass loss rates for these stars, giving the impression that the stars with circumstellar material all have high mass loss rates; this is not seen in this work.

The examination both of the sample of eight white dwarfs in Chapter 4 and the 23 stars in Chapter 5 throw up some interesting points. The objects observed in Chapter 4 were selected based on their photospheric metal abundances. Chayer & Dupuis (2010) and Dupuis, Chayer & Hénault-Brunet (2010) have reported that, based on radiative levitation calculations and the observed photospheric abundances, accretion is occurring at some of the stars that feature in this thesis. The source of this accretion has not yet been determined, although many studies of the accretion of material at cooler stars suggest that tidally disrupted asteroids and minor planets provide the circumstellar discs detected in the infrared. Were such discs to be found at hotter white dwarfs, it would provide valuable information on how extra solar planetary systems evolve while the white dwarf is young and hot.

The higher temperature of such stars may inhibit the presence of such discs close to the stars, and as the star cools, the disc may be able to persist at ever-closer radii. Detailed models of the evolution and configuration of circumstellar discs at the hotter stars are not available as yet, and would be an interesting avenue of study. Such models may provide quantitative estimates of observables that could be searched for, and allow the detection and charaterisation of circumstellar discs at the hotter DA stars.

This study of circumstellar absorption at hot white dwarfs suggests that a link between the observed features and the ISM may in fact be present. If at least some of the observed high ions are due to ionisation from hot white dwarfs, then conclusions drawn about the morphology and ionisiation of the ISM are at least in part incorrect. Therefore, a thorough further study of all observed, non-photospheric high ions at hot white dwarfs is necessary, to ascertain whether such high ions could be due to the ionisation of the ISM in the locality of the hot white dwarfs. The size of the sample here, with a mere eight stars showing circumstellar absorption, needs to be enlarged to allow more confident conclusions to be drawn; indeed it is difficult to understand the origin of the circumstellar absorption at the eight DAs studied here when the three possible sources of circumstellar material (circumstellar discs, ionised ISM and binary mass loss), or a combination of sources, can be inferred. The observation of more hot DA stars with UV instruments such as *STIS* and *COS* may allow the further identification of circumstellar absorption of hot DAs, allowing a more statistically meaningful sample to be constructed. The re-observation of stars where accretion is required to explain the observed photospheric abundances at higher resolution may allow further circumstellar detections to be made, and will provide a source of the necessary, accreted material.

The applications of this work to our understanding of the morphology of the ISM have been discussed in detail. In 2013, the GAIA observatory is due to launch, and it will map the positions and parallaxes of over a billion stars to an accuracy of 25μ". Given the relative faintness of the white dwarfs studied here, accurate parallaxes are often difficult to measure. However, GAIA's limiting magnitude (20) is sufficiently high enough to allow positions and parallaxes all of the stars here to be measured. When coupled with the better measurements for all other stars, more detailed maps of the ISM structure can be obtained. This will allow a more thorough understanding of how the hot white dwarfs studied here interact with the ISM.

WD 0948+534 remains an enigmatic object. Although preliminary calculations by Barstow et al. (2003b) suggested that stratified, highly abundant photospheric metals may account for the extremely narrow, highly absorbed metal line profiles, this work shows this is not the case. A possible source of the observed line profiles is a previously unaccounted for physical affect that becomes significant for high $T_{eff}$; WD 0948+534 (at 110,000 K) is significantly hotter than the next hottest DA examined in this context (WD 1029+537, $T_{eff}$ = 44,350 K). Investigations into the behaviour of model ions and the ionisation of metals in this high temperature regime will no doubt provide another useful avenue of exploration. Alternatively, a possible secondary component has been seen in the absorption features of the star (Section 5.3.3.3.), and the centroids of the high ions (when fit with one component) are inconsistent with one another, suggesting that the absorption features may be contaminated by another source of high ion absorption along the sight line to the star. Observations at higher resolution may allow the separation of these components, if they are present, and finally determine the mechanism that provides the observed absorption line profiles.

Similarly, the photospheric metal abundance measurements detailed by Barstow et al. (2003b) modelled the complete absorption line profiles, including both the photospheric and circumstellar absorption. The complete charaterisation of the circumstellar absorbing components provided here allows the circumstellar contribution to the high ion metal lines to be accounted for in the line profile modelling. Re-analysing the sample of Barstow et al. (2003b) in this context will allow a better understanding of the patterns between metal abundance and $T_{eff}$. This worthwhile study will inform our understanding of how hot white dwarfs evolve, and the new metal abundances will allow better estimates of $T_{eff}$ to be made, that can then be applied to studies of white dwarf evolution, the initial-final mass relation and the mass-radius relation. Given its proximity, WD 0501+527 is often used as a bench mark hot white dwarf against which other hot DAs are compared; correctly knowing the photospheric abundances of this star is important for all of the other hot white dwarf studies that use the metal abundances of this star in the modelling of other objects.

Indeed, the inconsistency between the $T_{eff}$ values obtained from the Balmer and Lyman line series observed in hot white dwarfs (Barstow et al. 2003a; Good et al. 2004) may be due to uncertainties in the inclusion of metals in the stellar models. Using the better metal abundance estimates, and the updated Stark broadening modelling of Tremblay & Bergeron (2009), may allow this effect to be better understood. The use of Lyman line $T_{eff}$ derivations for white dwarfs in binary systems will benefit greatly from this, as will the mass-radius and initial-final mass relation studies that use white dwarfs in binary systems.

All of these conclusions and suggestions for further investigation show that the field of hot white dwarfs is very much alive, and worthy of considerable interest. The application of this further work to our understanding of the ultimate fate of the overwhelming majority of stars is clear, and from this we can gain a far greater insight into how our Solar System will eventually expire. I greatly look forward to the advances made in this field in the years to come, and playing a part in them.

# *Appendix A.*

## Chapter 6: Tables of DA and solar metal abundances used for mass loss calculations.

**Table A1.** Hot white dwarf metal abundances.

| Star | C III[*] | C IV[*] | N IV[†] | N V[*] | O V[*] | Si IV[†] | P IV[†] |
|---|---|---|---|---|---|---|---|
| WD 0050–335 | 0 | $5.00 \times 10^{-8}$ | $1.00 \times 10^{-8}$ | $6.05 \times 10^{-7,a}$ | 0 | $4.80 \times 10^{-9}$ | |
| WD 0232+035 | $7.64 \times 10^{-8}$ | $4.00 \times 10^{-7}$ | $8.71 \times 10^{-8}$ | $1.60 \times 10^{-7}$ | | $2.31 \times 10^{-6}$ | $6.34 \times 10^{-8}$ |
| WD 0455–282 | 0 | $4.00 \times 10^{-8}$ | $3.76 \times 10^{-7}$ | $1.60 \times 10^{-7}$ | $1.00 \times 10^{-5}$ | $3.00 \times 10^{-8}$ | $1.80 \times 10^{-8}$ |
| WD 0501+527 | $1.99 \times 10^{-7}$ | $4.00 \times 10^{-7}$ | $1.67 \times 10^{-7}$ | $1.60 \times 10^{-7}$ | $3.51 \times 10^{-7}$ | $8.65 \times 10^{-7}$ | $2.08 \times 10^{-8}$ |
| WD 0556–375 | $1.70 \times 10^{-7}$ | $4.00 \times 10^{-7}$ | | $1.86 \times 10^{-7}$ | $2.89 \times 10^{-6}$ | $3.00 \times 10^{-6}$ | |
| WD 0621–376 | $1.59 \times 10^{-7}$ | $4.00 \times 10^{-7}$ | $1.60 \times 10^{-6}$ | $1.60 \times 10^{-7}$ | $2.00 \times 10^{-6}$ | $7.50 \times 10^{-7}$ | $3.07 \times 10^{-8}$ |
| WD 0939+262 | $3.68 \times 10^{-7}$ | $1.71 \times 10^{-7}$ | | $8.38 \times 10^{-8}$ | $1.21 \times 10^{-7}$ | $1.21 \times 10^{-7}$ | |
| WD 1029+537 | $3.00 \times 10^{-7}$ | $1.60 \times 10^{-7}$ | $1.00 \times 10^{-8}$ | $2.20 \times 10^{-7,a}$ | $1.20 \times 10^{-7}$ | $9.50 \times 10^{-8}$ | |
| WD 1057+719 | | 0 | | 0 | | 0 | |
| WD 1123+189 | 0 | | | | | | |
| WD 1254+223 | 0 | 0 | | 0 | 0 | 0 | |
| WD 1314+293 | 0 | 0 | | | | | |
| WD 1337+705 | 0 | 0 | $1.00 \times 10^{-9}$ | 0 | 0 | $2.4 \times 10^{-7}$ | $9.98 \times 10^{-8}$ |
| WD 1611–084 | | $4.00 \times 10^{-7}$ | $1.00 \times 10^{-7}$ | $1.76 \times 10^{-6,a}$ | | $9.50 \times 10^{-9}$ | |
| WD 1738+669 | $9.14 \times 10^{-8}$ | $1.34 \times 10^{-7}$ | | $5.81 \times 10^{-8}$ | $4.84 \times 10^{-7}$ | $4.84 \times 10^{-7}$ | |
| WD 2023+246 | 0 | 0 | | 0 | 0 | $3.80 \times 10^{-9}$ | |
| WD 2111+498 | 0 | 0 | $1.29 \times 10^{-8}$ | 0 | 0 | $2.80 \times 10^{-6}$ | $4.48 \times 10^{-9}$ |
| WD 2152–548 | 0 | $4.00 \times 10^{-9}$ | $1.00 \times 10^{-8}$ | 0 | 0 | $7.50 \times 10^{-10}$ | $2.01 \times 10^{-9}$ |
| WD 2211–495 | $1.00 \times 10^{-7}$ | $8.00 \times 10^{-7}$ | $3.04 \times 10^{-7}$ | $1.60 \times 10^{-7}$ | $2.40 \times 10^{-6}$ | $2.40 \times 10^{-6}$ | $6.07 \times 10^{-9}$ |
| WD 2218+706 | $1.59 \times 10^{-7}$ | $6.70 \times 10^{-7}$ | | $2.72 \times 10^{-7}$ | $9.60 \times 10^{-6}$ | $1.93 \times 10^{-6}$ | |
| WD 2309+105 | 0 | | $3.01 \times 10^{-8}$ | $2.00 \times 10^{-9}$ | $1.60 \times 10^{-7}$ | $1.20 \times 10^{-7}$ | $2.15 \times 10^{-9}$ |
| WD 2331–475 | $8.00 \times 10^{-8}$ | $4.00 \times 10^{-7}$ | $1.60 \times 10^{-8}$ | $2.00 \times 10^{-8}$ | $7.10 \times 10^{-7}$ | $4.80 \times 10^{-7}$ | $3.05 \times 10^{-8}$ |

[*]from Barstow et al. (2003b); [†]from Barstow et al. (*in preparation*); [a]from Chapter 3

**Table A1** - *continued.*

| Star | P V† | S IV† | S VI† | Fe* | Ni* | $Z$ | $\dot{M}$ |
|---|---|---|---|---|---|---|---|
| WD 0050–335 | $8.68\times10^{-9}$ | $3.00\times10^{-9}$ | $2.91\times10^{-7}$ | 0 | 0 | $1.76\times10^{-3}$ | $1.01\times10^{-17}$ |
| WD 0232+035 | $7.44\times10^{-8}$ | $1.53\times10^{-7}$ | $9.85\times10^{-8}$ | $3.56\times10^{-6}$ | $1.16\times10^{-7}$ | $6.05\times10^{-3}$ | $2.08\times10^{-14}$ |
| WD 0455–282 | $6.08\times10^{-8}$ | $3.73\times10^{-8}$ | $9.60\times10^{-8}$ | $3.20\times10^{-6}$ | $1.00\times10^{-6}$ | $1.23\times10^{-2}$ | $4.22\times10^{-15}$ |
| WD 0501+527 | $2.50\times10^{-8}$ | $6.88\times10^{-8}$ | $9.61\times10^{-8}$ | $3.30\times10^{-6}$ | $2.40\times10^{-7}$ | $4.79\times10^{-3}$ | $4.78\times10^{-15}$ |
| WD 0556–375 | | | | $9.93\times10^{-6}$ | $6.71\times10^{-7}$ | $1.49\times10^{-2}$ | $3.16\times10^{-14}$ |
| WD 0621–376 | $6.93\times10^{-8}$ | $1.83\times10^{-7}$ | $9.60\times10^{-8}$ | $1.58\times10^{-5}$ | $1.30\times10^{-6}$ | $1.77\times10^{-2}$ | $2.43\times10^{-13}$ |
| WD 0939+262 | | | | $1.00\times10^{-5}$ | $9.91\times10^{-7}$ | $4.15\times10^{-4}$ | $4.31\times10^{-14}$ |
| WD 1029+537 | $9.36\times10^{-9}$ | $1.22\times10^{-8}$ | $1.67\times10^{-7}$ | 0 | 0 | $1.14\times10^{-3}$ | $8.52\times10^{-17}$ |
| WD 1057+719 | | | | 0 | 0 | | |
| WD 1123+189 | | | | | | | |
| WD 1254+223 | | | | 0 | 0 | | |
| WD 1314+293 | | | | 0 | 0 | | |
| WD 1337+705 | $1.00\times10^{-7}$ | $3.01\times10^{-9}$ | | 0 | 0 | $9.61\times10^{-3}$ | $8.65\times10^{-19}$ |
| WD 1611–084 | | | | 0 | 0 | $5.88\times10^{-3}$ | $1.09\times10^{-16}$ |
| WD 1738+669 | | | | $2.10\times10^{-7}$ | $5.24\times10^{-8}$ | $2.74\times10^{-3}$ | $4.34\times10^{-14}$ |
| WD 2023+246 | | | | 0 | 0 | $1.17\times10^{-4}$ | $1.06\times10^{-20}$ |
| WD 2111+498 | $5.46\times10^{-8}$ | $5.92\times10^{-9}$ | $8.26\times10^{-9}$ | 0 | 0 | $1.14\times10^{-4}$ | $5.10\times10^{-16}$ |
| WD 2152–548 | | $3.00\times10^{-9}$ | $3.12\times10^{-9}$ | 0 | 0 | $3.15\times10^{-4}$ | $3.48\times10^{-18}$ |
| WD 2211–495 | $9.11\times10^{-9}$ | $3.20\times10^{-8}$ | $9.72\times10^{-8}$ | $1.45\times10^{-5}$ | $1.00\times10^{-6}$ | $1.72\times10^{-2}$ | $1.51\times10^{-13}$ |
| WD 2218+706 | | | | $1.00\times10^{-5}$ | $9.91\times10^{-7}$ | $2.03\times10^{-2}$ | $1.89\times10^{-13}$ |
| WD 2309+105 | $1.16\times10^{-8}$ | $9.78\times10^{-9}$ | $9.67\times10^{-9}$ | 0 | 0 | $5.06\times10^{-4}$ | $2.52\times10^{-16}$ |
| WD 2331–475 | $5.27\times10^{-8}$ | $7.34\times10^{-8}$ | $9.58\times10^{-8}$ | $2.50\times10^{-6}$ | $2.00\times10^{-8}$ | $3.72\times10^{-3}$ | $3.08\times10^{-15}$ |

*from Barstow et al. (2003b); †from Barstow et al. (*in preparation*); ᵃfrom Chapter 3

**Table A2.** The solar metal abundances reported by Asplund et al. 2009.

| C | N | O | Si | P | S | Fe | Ni |
|---|---|---|---|---|---|---|---|
| $2.64\times10^{-4}$ | $6.76\times10^{-5}$ | $4.90\times10^{-4}$ | $3.24\times10^{-5}$ | $2.57\times10^{-7}$ | $1.32\times10^{-5}$ | $3.16\times10^{-5}$ | $1.66\times10^{-6}$ |

# Bibliography.

Aannestad P.A., Kenyon S.J., Hammond G.L., Sion E.M., 1993, AJ, **105**, 1033

Abbott D.C., 1982, ApJ, **259**, 282

Adams W.S., 1915, PASP, **27**, 236

Arnaud K., 1996, ASPC, **101**, 17

Asplund M., Grevesse N., Jacques Sauval A., Scott P., ARA&A, **47**, 481

Bannister N. P., Barstow M. A., Holberg J. B., Bruhweiler F. C., 2003, MNRAS, 341, 477

Barstow M.A., Boyce D.D., Welsh B.Y., Lallement R., Barstow J.K., Forbes A.E., Preval S., 2010, ApJ, **723**, 1762

Barstow M.A., Bond H.E., Holberg J.B., Burleigh M.R., Hubeny I., Koester D., 2005, MNRAS, **362**, 1134

Barstow M.A., Cruddace R.G., Kowalski M.P., Bannister N.P., Yentis D., Lapington J.S., Tandy J.A., Hubeny I., Schuh S., Dreizler S., Barbee T.W., 2005, MNRAS, **362**, 1273

Barstow M.A., Dobbie P., Holberg J.B, Hubeny I., Lanz T., 1997, MNRAS, **286**, 58

Barstow M.A., Good S., Burleigh M., Hubeny I., Holberg J., Levan A., 2003a, MNRAS, **344**, 562

Barstow M.A., Good S.A., Holberg J.B., Hubeny I., Bannister N.P., Bruhweiler F.C., Burleigh M.R., Napiwotzki R., 2003b, MNRAS, **341**, 870

Barstow M.A., Holberg J.B., Fleming T.A., Marsh M.C., Koester D., Wonnacott D., 1994, MNRAS, **270**, 499

Barstow M.A., Hubeny I., 1999, MNRAS, **299**, 379

Barstow M.A., Hubeny I., Holberg J.B., 1998, MNRAS, **307**, 884

Barstow M.A., Holberg J.B., Marsh M.C., Tweedy R.W., Burleigh M.R., Fleming T.A., Koester D., Penny A.J., Sansom A., 1994b, MNRAS, **271**, 175

Becklin E.E., Farihi J., Jura M., Song I., Weinberger A.J., Zuckerman B., 2005, ApJ, **632**, 119

Bessel F.W., Herschel J.F.W., 1844, MNRAS, **6**, 136

Bergeron P., Saffer R., Liebert J., 1992, ApJ, **432**, 305

Bergeron P., Wesemael F., Lamontagne R., Fontaine G., Saffer R.A., Allard N.F., 1995, ApJ, **449**, 258

Bohlin R.C., Dickinson M.E., Calzetti D., 2001, AJ, **122**, 2118

Bohlin R.C., Gordon K.D., Rieke G.H., Ardilla D., Carey S., Deustua S., Engelbracht C., Ferguson H.C., 2011, AJ, **141**, 173

Bond G., 1862, AN, **57**, 131

Brinkworth C.S., Gänsicke B.T., Marsh T.R., Hoard D.W., Tappert C., 2009, ApJ, **696**, 1402

Bruhweiler F.C., Kondo Y., 1983, ApJ, **269**, 657

Burleigh M.R., Barstow M.A., Farihi J., Bannister N.P., Dickinson N.J., Steele P.R., Dobbie P.D., Faedi F., Gänsicke B.T., 2010, AIPC, **1273**, 473

Burleigh M.R., Barstow M.A., Farihi J., Bannister N.P., Dickinson N.J., Steele P.R., Dobbie P.D., Faedi F., Gänsicke B.T., 2011, AIPC, **1331**, 289

Casewell S., Dobbie P.D., Napiwotzki R., Burleigh M.R., Jameson R.F., 2009, MNRAS, **395**, 1795

Chayer P., Dupuis J., 2010, AIPC, **1273**, 394

Chayer P., Fontaine G., Wesemael F., 1995, ApJS, **99**, 189

Chayer P., Kruk J.W., Ake T.B., Dupree A.K., Malina R.F., Siegmund O.H., Sonneborn G., Ohl R.G., 2000, ApJ, **538**, 91

Chayer P., Oliveira C., Dupuis J., Moos H.W., Welsh B.Y., 2006, ASPC, **348**, 209

Chayer P., Vennes S, Dupuis J., 2005 ASPC, **334**, 181

Chayer P., Vennes S., Pradhan A.K., Thejl P., Beauchamp A., Fontaine G., Wesemael F., 1995, ApJ, **454**, 429

Chiang E.I., Goldreich P., 1997, ApJ, **490**, 368

Chu Y.H., Su K.Y.L., Bilikova J., Gruendl R.A., DeMarco O., Guerrero M., Updike A.C., Volk K., Rauch T., 2011, AJ, **142**, 75

Davidsson B.J.R., 1999, Icarus, **142**, 525

Debes J.H., Sigurdsson S., 2002, ApJ, **572**, 556

Debes J.H., Sigurdsson S., Hansen B., 2007, AJ, **134**, 1662

Debes J.H., Hoard D.W., Wachter S., Leisawitz D.T., Cohen M., 2011, ApJS, **197**, 38

Dreizler S., Werner K., 1993, A&A, **278**, 199

Dreizler S., Wolff B., 1999, A&A, **348**, 189

Dufour P., Liebert J., Fontaine G., Behara N., 2007a, Nature, **450**, 522

Dufour P., Liebert J., Fontaine G., Behara N., 2007b, ASPC, **391**, 241

Dufour P., Fontaine G., Liebert J., Schmidt G.D., Behara N., 2008, ApJ, **683**, 978

Dufour P., Kilic M., Fontaine G., Bergeron P., Lachapelle F.R., Kleinmann S.J., Legget S.K., 2010, ApJ, **719**, 803

Dupree A.K., Raymond J. C., 1982, ApJ, **263**, 63

Dupree A.K., Raymond J.C., 1983, ApJ, **275**, 71

Dupuis J., Chayer P., Hénault-Brunet V., 2010, AIPC, **1273**, 412

Dupuis J., Chayer P., Hénault-Brunet V., Venes S., Kruk J.W., 2010, AAS, **215**, 45205

Dupuis J., Chayer P., Vennes S., Christian D.J., Kruk J.W., 2000, ApJ, **537**, 127

Dupuis J., Fontaine G., Wesemael F., 1993, ApJS, **87**, 345

Dupuis J., Fontaine G., Pelletier C., Wesemael F., 1992, ApJS, **82**, 505

Dupuis J., Fontaine G., Pelletier C., Wesemael F., 1993, ApJS, **84**, 73

Dupuis J., Oliveira C.M., Hébrard G.H., Woos H.W., Sonnentrucker P., 2009, ApJ, **690**, 1045

Eisenstein D.J., Liebert J., Koester D., Kleinmann S.J., Nitta A., Smith P.S., Barentine J.C., Brewington H.J., Brinkmann J., Harvanek M., Krzesinski J., Neilsen E.H. Jr., Long D., Schneider D.P., Snedden S.A., 2006, AJ, **132**, 676

Farihi J., Barstow M.A., Redfield S., Dufour P., Hambly N.C., 2010, MNRAS, **404**, 2123

Farihi J., Jura M., Zuckerman B., 2009, ApJ, **694**, 805

Farihi J., Zuckerman B., Becklin E.E., 2008, ApJ, **674**, 431

Feldman P.D., Sahnow D.J., Kruk J.W., Murphy E.M., Moos H.W., 2001, JGR, **106**, 8119

Fontaine G., Brassard P., 2005, ASPC, **627**, 404

Fontaine G., Brassard P., Bergeron P., 2001, PASP, **113**, 409

Fowler R.H., 1926, MNRAS, **87**, 114

Frew D., Parker Q., 2006, IAUS, **234**, 49

Frisch P.C., York D.G., 1983, ApJ, **271**, L59

Gänsicke B.T., Koester D., Marsh T.R., Rebassa-Masergas A., Southworth J., 2008, MNRAS, **391**, L103

Gänsicke B.T., Marsh T.R., Southworth J., 2007, MNRAS, **380**, L35

Gänsicke B.T., Marsh T.R., Southworth J., Rebassa-Masergas A., 2006, Science, **314**, 1908

Good S., Barstow M., Holberg J., Sing D., Burleigh M., Dobbie P., 2004, MNRAS, **355**, 1031

Graham J.R., Matthews K., Neugebauer G., Soifer B.T., 1990, ApJ, **357**, 216

Greenstein J.L., 1960, In Stars and Stellar Systems, Vol **6**, Stellar Atmospheres, ed. J.L. Greenstein (Chicago: University of Chicago Press)

Hartmann S., Nagel T., Rauch T., Werner K., 2011, A&A, **530**, 7

Herschel W., 1785, RSPT, **75**, 40

Hoard D.W., Wachter S., Sturch L.K., Widhalm A.M., Weiler K.P., Pretorius M.L., Wellhouse J.W., Gibiansky M., 2007, AJ, **134**, 26

Holberg J.B., Barstow M.A., Bruhweiler F.C, Collins J., 1996, AJ, **111**, 2361

Holberg J.B., Barstow M.A., Bruhweiler F.C., Dobbie P.D., 1999b, ApJ, **517**, 841.

Holberg J.B., Barstow M.A., Bruhweiler F.C., Hubeny I., 2000, BAAS, **197**, 8304.

Holberg J.B., Barstow M.A., Bruhweiler F.C., Hubeny I., Green E.M., 1999a, ApJ, **517**, 850

Holberg J.B., Barstow M.A., Bruhweiler F.C., Sion E., 1995, ApJ, **453**, 313

Holberg J.B., Barstow M.A., Lanz T., Hubeny I., 1997, ApJ, **484**, 871

Holberg J.B., Barstow M.A., Green E.M., 1997, ApJ, **474**, 127

Holberg J.B., Barstow M.A., Sion E.M., 1998, ApJS, **119**, 207

Holberg J.B., Barstow M.A., Sion E.M., 1999, ASPC, **169**, 485

Holberg J. B., Bruhweiler F. C., Anderson J., 1995, ApJ, **443**, 753

Holberg J.B., Oswalt T.D., Sion E.M., 2002, ApJ, **571**, 512

Holberg J., Wesemael F., Wegner G., Bruhweiler F., 1985, ApJ, **484**, 871

Horne K., Marsh T.R., 1986, MNRAS, **218**, 761

Hubeny I., Hummer D.G., Lanz T., 1994, A&A, **262**, 501

Hubeny I., Lanz T., 1992, A&A, **262**, 501

Hubeny I., Lanz T., 1995, ApJ, **439**, 875

Hubeny I., Lanz T., 2003, ASPC, **288**, 51

Indebetouw R., Shull J.M., 2004a, ApJ, **605**, 205

Indebetouw R., Shull J.M., 2004b, ApJ, **607**, 309

Jordan S., Heber U., Engels D., Koester D., 1993 A&A, **273**, 27

Jura M., 2003, ApJ, **584**, 91

Jura M., 2003, ApJ, **653**, 613

Jura M., 2008, AJ, **135**, 1785

Jura M., Farihi J., Becklin E.E., 2007, ApJ, **663**, 1285

Jura M., Farihi J., Zuckerman B., 2009, AJ, **137**, 3191

Jura M., Muno M.P., Farihi J., Zuckerman B., 2009, ApJ, **699**, 1473

Kahn S.M., Wesemael F., Liebert J., Raymond J.C., Steiner J.E., Shipman H.L., 1984, ApJ, **278**, 255

Kawka A., Vennes S., Dupuis J., Chayer P., Lanz T., 2008, ApJ, **675**, 1518

Kilic M., Redfield S., 2007, ApJ, **660**, 641

Kilic M., von Hippel T., Legget S.K., Winget D.E., 2005, ApJ, **632**, 115

Kilic M., von Hippel T., Legget S.K., Winget D.E., 2006, ApJ, **646**, 474

Kimble R.A., Woodgate B.E., Bowers C.W., Kraemer S.B., Kaiser M.E., Gul T.R., Heap S.R., Danks A.C., Boggess A., Green R.F., et al. 1998, ApJ, **492**, 83

Klein B., Jura M., Koester D., Zuckerman B., 2011, ApJ, **741**, 64

Kleinman S.J., Harris H.C., Eisenstein D.J., Liebert J., Nitta A., Krzesinski J., Munn J.A., Dahn C.C. et al., 2004, ApJ, 607, 426

Koester D., 1989, ApJ, **342**, 999

Koester D., Wilken D., 2006, A&A, **453**, 1051

Kudritzki R.P., 2002, ApJ, **577**, 389

Kuiper G.P., 1941, PASP, **53**, 248

Lallement R., Ferlet R., Lagrange A. M., Lemonie M., Vidal-Madjar A., 1995, A&A, **304**, 461

Lallement R., Welsh B.Y., Barstow M.A., Casewell S.L., 2011, A&A, **533**, 140

Lanz T., Barstow M.A., Hubeny I., Holberg J.B., 1996, ApJ, **473**, 1089

Lanz T., Hubeny I., 1995, ApJ, **439**, 905

Lanz T., Hubeny I., 2003a, ASPC, **146**, 417

Lanz T., Hubeny I., 2003a, ASPC, **288**, 117

Laughlin G., Bodenheimer P., Adams F.C., 1997, ApJ, **482**, 420

Liebert J., Bergeron P., Eisenstein D., Harris H.C., Kleinmann S.J., Nitta A., Krzesinski J., 2004, ApJL, **606**, 147

Liebert J., Harris H.C., Dahn C.C., Schmidt G.D., et al., 2003, ApJ, **126**, 2521

Liebert J., Young P.A., Arnett D., Holberg J.B., Williams K.A., 2005, ApJ, **630**, 69

Luyten W.J., 1952, ApJ, **116**, 283

MacDonald J., 1992, ApJ, **396**, 619

Marsh M.C., 1995, PhD Thesis, University of Leicester

Marsh M.C., Barstow M.A., Buckley D.A., Burleigh M.R., Holberg J.B., Koester D., O'Donoghue D., Penny A.C., Sansom A.E., 1997, MNRAS, **286**, 369

Maxted P.F.L., Marsh T.R., North R.C., 2000, MNRAS, **317**, 41

McCook G.P., Sion E.M., 1999, ApJS, **121**, 1

Meeting of the Royal Astronomical Society, 11$^{th}$ January 1935, in The Observatory, 1935, **58**, 33

Melis C., Farihi J., Dufour P., Zuckerman B., Burgasser A.J., Bergeron P., Bochanski J., Simcoe R., 2011, ApJ, **732**, 90

Melis C., Jura M., Albert L., Klein B., Zuckerman B., 2010, ApJ, **722**, 1078

Moos H.W., Cash W.C., Cowie L.L., Davidsen A.F., Dupree A.K., Feldman P.D., Friedman S.D., Green J.C., Green R.F., Gry C., et al. 2000, ApJ, **538**, 1

Mullally F., Kilic M., Reach W.T., Kuchner M.J., von Hippel T., Burrows A., Winget D.E., 2007, ApJS, **171**, 206

Napiwotzki R., 1997, A&A, **322**, 256

Napiwotzki R., Schönberner D., 1995, A&A, **301**, 545

Napiwotzki R., Green P.J., Saffer R.A., 1999, ApJ, **517**, 399

Napiwotzki R., 1999, A&A, 350, 101

Pelletier C., Fontaine G., Wesemael F., Michaud G., Wegner G., 1986, ApJ, **307**, 242

Peters C.A.F., 1851, AN, **32**, 1

Redfield S., Falcon R., 2008, ApJ, **683**, 207

Redfield S., Linsky J.L., 2002, ApJS, **139**, 439

Redfield S., Linsky J.L., 2004a, ApJ, **602**, 776

Redfield S., Linsky J., 2004b, ApJ, **613**, 104

Redfield S., Linsky J.L., 2008, ApJ, **673**, 283

Renzini A., Voli M., 1981, A&A, **94**, 175

Saffer R., Livio M., Yungelson L.R., 1998, ApJ, **502**, 394

Sahu M.S., Landsman W., Bruhweiler F.C., Gull T.R., Bowers C.A., Lindler D., Feggans K., Barstow M.A., Hubeny I., Holberg J.B., 1999, ApJ, **523**, 159

Sahnow D.J., Moos H.W., Ake T.B., Andersen J., Andersson B.-G., Andre M., Artis D., Berman A.F., Bair W.P., Brownsberger K.R., et al., 2000, ApJ, **538**, 7

Savage B.D., Lehner N., 2006 ApJS, **162**, 134

Savage B.D., Sembach K.R., 1991, ApJ, **379**, 245

Schatzmann E., 1958, White Dwarfs (Amsterdam: North Holland Publishing)

Schmidt G.D., Smith P.S., 1995, ApJ, **448**, 305

Schönberg M., & Chandrasekhar S., 1942, ApJ, **96**, 161

Schuh S., Barstow M.A., Dreizler S., 2005, ASPC, **334**, 237

Schuh, S., Dreizler S., Wolff B., 2002, A&A, **382**, 164

Seaton M., 1996, Ap&SS, **237**, 107

Sfeir D.M., Lallement R., Crifo F., Welsh B.Y., 1999, A&A, **346**, 785

Shaprio P.R., Field G.B., 1976, ApJ, **205**, 762

Sion E.M., Bohlin R.C., Tweedy R.W., Vauclair G.P., 1992, ApJ, **391**, 29

Sion E.M., Greenstein J.L., Landstreet J.D., Liebert J., Shipman H.L., Wegner G.A., 1983, ApJ, **269**, 253

Strömgren B., 1939, ApJ, **89**, 526

Struve O., 1866, MNRAS, **26**, 268

Su K.Y.L., Chu Y.-H., Rieke G.H., Huggins P.J., Gruendl R., Napiwotzki R., Rauch T., Latter W.B., Volk K., 2007, ApJ, **657**, 41

Tat H., Terzian Y., 1999, PASP, **111**, 1258

Tremblay P.E. & Bergeron P, 2009, ApJ, 696, 1755

Tweedy R. W., Kwitter K. B., 1994, ApJ, **433**, 93

Unglaub K., 2007, ASPC, **372**, 201

Unglaub K., 2008, A&A, **486**, 923

Unglaub K, Bues I., 2000, A&A, **359**, 1042

Vallegra J., 1998, ApJ, **497**, 921

Vallerga J., Vedder P., Craig N., Welsh B., 1993, ApJ, **411**, 729

Van Maanen A., 1917, PASP, **29**, 258

Vauclair G., Vauclair S., Greenstein J.L., 1979, A&A, **80**, 79

Vennes S., Chayer P., Dupuis J., 2005, APJ, **622**, L121

Vennes S., Chayer P., Dupuis J., Lanz T., 2006, ApJ, **652**, 1554

Vennes S., Lanz T., 2001, ApJ, **553**, 399

Vennes S., Pelletier C., Fontaine G., Wesemael F., 1988, ApJ, **331**, 876

Vennes S., Thorstensen J.R., 1994, AJ, **108**, 1881

Vink J., de Koeter A., Lamers H., 2001, A&A, **369**, 574

von Hippel T., Kuchner M.J., Kilic M., Mullally F., Reach W.T., 2007, ApJ, **662**, 544

Votruba V., Feldmeier A., Krticka J., Kubát J., 2010, Ap&SS, **329**, 159

Weidemann V., 1987, A&A, **188**, 74

Welsh B.Y., 1991, ApJ, **373**, 556

Welsh B.Y., Lallement R., 2005, A&A, **436**, 615

Welsh B.Y., Lallement R., 2010, PASP, **122**, 1320

Welsh B.Y., Lallement R., Vergely J.-L., Raimond S., 2010a, A&A, **510**, 54

Welsh B.Y., Wheatley J., Siegmund O.H.W., Lallement R., 2010b, ApJL, **712**, 199

Welsh B.Y., Sfeir D., Sirk M., Lallement R., 1999, A&A, **352**, 308

Welsh B.Y., Vallegra P.W., Vedder J.V., 1990, ApJ, **358**, 473

Welsh B.Y., Wheatley J., Siegmund O.H.W., Lallement R., 2010, ApJL, **712**, 199

Werner K., Rauch T., Kruk J.W., 2007, A&A, **466**, 317

Wolff B., Koester D., Dreizler S., Haas S., 1998, A&A, **329**, 1045

Wolff B., Kruk J.W., Koester D., Allard N.F., Ferlet R., Vidal-Madjar A., 2001 A&A, **373**, 674

Wood M., 1992, ApJ, **386**, 539

Wood M., 1995, LNP, **443**, 41

Woodgate B.E., Kimble R.A., Bowers C.W., Kraemer S., Kaiser M.E., Danks A.C., Grady J.F., Loicaono J.J., Brumfield M., Feinberg L., et al. 1998, PASP, **110**, 1183

Young P., Schneider D.P., Shectman S.A., 1981, ApJ, **245**, 1035

Zuckerman B., Koester D., Dufour P., Melis C., Klein B., Jura M., 2011, ApJ, **739**, 101

Zuckerman B., Koester D., Melis C., Hansen B.M., Jura M., 2007, ApJ, **671**, 872

Zuckerman B., Koester D., Reid I.N., Hünsch M., 2003, ApJ, **596**, 477

Zuckerman B, Reid I.N., 1998, ApJ, **505**, 143